\documentclass[11pt,a4paper]{article}
\usepackage{jheppub}
\usepackage[normalem]{ulem}
\usepackage{capt-of}
\newcommand{\tr}{\text{tr}}
\newcommand{\bx}{{\boldsymbol x}}

\newcommand{\bp}{{\boldsymbol p}}

\newcommand{\bk}{{\boldsymbol k}}

\newcommand{\bu}{{\boldsymbol u}}

\newcommand{\bnabla}{{\boldsymbol \nabla}}
\newcommand{\bJ}{{\boldsymbol J}}

\newcommand{\bomega}{{\boldsymbol \omega}}

\newcommand{\bzero}{{\boldsymbol 0}}

\newcommand{\bcalB}{\boldsymbol{\mathcal{B}}}
\newcommand{\tF}{\tilde{F}}

\newcommand{\tcalR}{\widetilde{\calR}}

\newcommand{\tf}{\tilde{f}}
\newcommand{\hp}{{\hat p}}
\newcommand{\hk}{{\hat k}}
\newcommand{\hbp}{{\hat{\boldsymbol p}}}
\newcommand{\hbk}{{\hat{\boldsymbol k}}}

\newcommand{\hzero}{{\hat{0}}}

\newcommand{\lD}{\overleftarrow{D}}

\newcommand{\calD}{\mathcal{D}}

\newcommand{\calI}{\mathcal{I}}

\newcommand{\calR}{\mathcal{R}}

\newcommand{\calE}{\mathcal{E}}
\newcommand{\calB}{\mathcal{B}}
\newcommand{\zero}{ {(0)} }
\newcommand{\one}{ {(1)} }
\newcommand{\two}{ {(2)} }

\newcommand{\dis}{\displaystyle}
\renewcommand{\flat}{\text{flat}}

\newcommand{\ind}{\text{ind}}

\newcommand{\eq}{\text{eq}}
\newcommand{\dyn}{\text{dyn}}
\newcommand{\st}{\text{st}}
\newcommand{\nonst}{\text{nonst}}

\renewcommand{\c}{@{\hspace{4pt}}c@{\hspace{4pt}}}
\newcommand{\rout}{\bgroup \color{red} \ULdepth=-.5ex \ULset}
\newcommand{\bout}{\bgroup \color{blue} \ULdepth=-.5ex \ULset}

\begin{document}

\title{Second order chiral kinetic theory under gravity and antiparallel charge-energy flow}
\preprint{KEK-TH-2279, J-PARC-TH-0234, RIKEN-QHP-488}
\author[a]{Tomoya Hayata,}
\affiliation[a]{
	Departments of Physics, Keio University, 4-1-1 Hiyoshi, Kanagawa 223-8521, Japan
}

\author[b]{Yoshimasa Hidaka}
\affiliation[b]{KEK Theory Center, Tsukuba 305-0801, Japan}
\affiliation{Graduate University for Advanced Studies (Sokendai), 
Tsukuba 305-0801, Japan}
\affiliation{RIKEN iTHEMS, RIKEN, Wako 351-0198, Japan}

\author[c]{and Kazuya Mameda}
\affiliation[c]{
Theoretical Research Division, Nishina Center, RIKEN, Wako, Saitama 351-0198, Japan
}
\emailAdd{hayata@keio.jp}\emailAdd{hidaka@post.kek.jp}\emailAdd{kazuya.mameda@riken.jp}

\abstract{%
We derive the chiral kinetic theory under the presence of a gravitational Riemann curvature.
It is well-known that in the chiral kinetic theory there inevitably appears a redundant ambiguous vector corresponding to the choice of the Lorentz frame.
We reveal that on top of this conventional frame choosing vector, higher-order quantum correction to the chiral kinetic theory brings an additional degrees of freedom to specify the distribution function. 
Based on this framework, we derive new types of fermionic transport, that is, the charge current and energy-momentum tensor induced by the gravitational Riemann curvature.
%\red{It is also shown that the CVE vanishes when the fluid vorticity is dynamical.}
Such novel phenomena arise not only under genuine gravity but also in a (pseudo-)relativistic fluid, for which inhomogeneous vorticity or temperature are effectively represented by spacetime metric tensor.
It is especially found that the charge and energy currents are antiparallelly induced by an inhomogeneous fluid vorticity (more generally, by the Ricci tensor ${R_0}^i$), as a consequence of the spin-curvature coupling.
We also briefly discuss possible applications to Weyl/Dirac semimetals and heavy-ion collision experiments.%
}

\maketitle
%\pacs{03.65.Vf,73.43.-f,03.65.Sq,72.10.bg}
\section{Introduction and Summary}
Transport phenomena are a pivotal subject in modern quantum field theories.
Similar to external electromagnetic fields, (effective) background gravitational fields are intriguing sources to generate various currents.
First, the most widely well-known example is fluid vorticity;
the fluid velocity can be described by a metric tensor of the comoving frame with fluid. 
The corresponding transport phenomenon, i.e., the chiral vortical effect (CVE)~\cite{Vilenkin:1979ui} can be regarded as the gravitational counterpart of the chiral magnetic effect (CME)~\cite{Vilenkin:1980fu,Nielsen:1983rb,Fukushima:2008xe}, as is clear from the gravitoelectromagnetism~\cite{Landsteiner:2011cp}.
The CVE is not only a theoretically interesting phenomenon in the sense that it is originated from quantum anomaly~\cite{Son:2009tf,Landsteiner:2011cp}, but also an important experimental probe to study the rotation of quark-gluon plasmas created in relativistic heavy-ion collision experiments~\cite{STAR:2017ckg}.
Second, the mechanical strain plays a role of an effective $U(1)$ or axial $U(1)$ magnetic field, and accordingly yields a charge current~\cite{PhysRevLett.115.177202,Cortijo:2016wnf,PhysRevX.6.041046}.
Third, spacetime torsion is recently under active investigation, as it can bring novel currents, which are referred to as the chiral torsional effect~\cite{Sumiyoshi:2015eda,Khaidukov:2018oat,Ferreiros:2018udw,Imaki:2019ite,Imaki:2020csc,Gao:2020gcf}.

In contrast to the aforementioned effects from the spacetime geometry, we do not fully understand the effect of the gravitational Riemann curvature in the quantum transport theory.
Even at the classical level, however, its importance has been known; 
the trajectory of a spinning particle is modified by the Riemann curvature~\cite{Mathisson:1937zz,Dixon:1970zza,Papapetrou:1951pa}.
In the context of quantum transport theory, this knowledge suggests that the Riemann curvature can be the trigger of a characteristic transport of the fermion chirality (or spin, more generally). 
In cosmological systems, such spacetime distortions may become dominant contributions to determine the fermionic transport rather than background electromagnetic fields.
In laboratory environments, a fluid motion and temperature gradient can be described by effective gravities leading to non-vanishing Riemann curvatures.
Therefore, the curvature-induced transport phenomena could be relevant in a wide range of physics from table-top experiments to the Universe.

%For the nonequilibrium dynamics of chiral fermions, one of the promising theoretical implements would be the so-called CKT (CKT)~\cite{Stephanov:2012ki,Son:2012wh,Son:2012zy,Chen:2012ca,Manuel:2013zaa,Chen:2014cla,Chen:2015gta,Hidaka:2016yjf,Hidaka:2017auj,Mueller:2017lzw,Mueller:2017arw,Huang:2018wdl,Carignano:2018gqt,Dayi:2018xdy,Liu:2018xip,Lin:2019ytz,Carignano:2019zsh}.
For the nonequilibrium dynamics in the weak interaction regime, one of the promising theoretical implements would be the kinetic theory.
In particular, the so-called chiral kinetic theory (CKT)~\cite{Stephanov:2012ki,Son:2012wh},
 which nicely reproduces the chiral anomaly, plays a pivotal role in the development of various studies of the chiral transport phenomena~\cite{Son:2012zy,Chen:2012ca,Manuel:2013zaa,Chen:2014cla,Chen:2015gta,Hidaka:2016yjf,Hidaka:2017auj,Mueller:2017lzw,Mueller:2017arw,Huang:2018wdl,Carignano:2018gqt,Dayi:2018xdy,Liu:2018xip,Lin:2019ytz,Carignano:2019zsh} in the context of heavy-ion collision, condensed matter and neutrino physics; 
although the kinetic theory is inapplicable to strongly-coupled quark-gluon plasmas, the early stage of heavy-ion collisions is described well by the Boltzmann transport theory~\cite{Baym:1984np,Mueller:1999pi,Baier:2000sb}.
The CKT conventionally involves only the leading order quantum correction so that the anomalous aspects can be taken into account as the Berry curvature.
However, the leading order CKT is insufficient to capture the gravitational curvature contributions to the transport coefficients, although the kinetic equation involves the spin-curvature coupling~\cite{Liu:2018xip}.
As is readily expected, higher-order corrections make the theory much more complicated, and an intuitive deduction would not avail.
This fact can be found from the equilibrium distribution function.
The $O(\hbar)$ contribution to enter the distribution is anticipated to be the spin-vorticity coupling, if we recall the conservation of the total angular momentum~\cite{Chen:2015gta}.
On the other hand, this intuition is inapplicable to the $O(\hbar^2)$ contribution, particularly, under a background gravitational field, as it is nontrivial to identify how the total angular momentum is modified at this order.
%Unlike the effective formalisms relied on the Berry curvature, the Wigner function approach works well against such a complication and systematically derives the CKT from quantum field theory~\cite{Hidaka:2016yjf,Liu:2018xip}.
Unlike the effective formalisms that relied on the Berry curvature, the derivation from quantum field theory works well against such a complication.
In this case, the semiclassical (or weak coupling) dynamics is described by the Wigner transformation of the fermion propagator, which is a quantum-extended quantity of the classical distribution function.
The Wigner function approach systematically involves quantum corrections appearing in the CKT, and definitely keeps the covariance of the fundamental theory~\cite{Hidaka:2016yjf} even in the general coordinate system~\cite{Liu:2018xip}.
In this paper, thus we study the semiclassical transport theory with gravitational Riemann curvatures, based on the CKT derived with the Wigner function.

In the following, we present a summary of the findings in this paper.
First, we solve the collisionless CKT in general coordinate and derive the Wigner function of Weyl fermions up to $O(\hbar^2)$.
This is a contrast to the conventional works, where only the $O(\hbar)$ quantum correction is taken into account.
It is intriguing that the CKT in curved spacetime is systematically solvable even with the $O(\hbar^2)$ contributions, while the $O(\hbar^2)$ electromagnetic effect is not so tamable; the theory suffers from severe infrared divergence, for which so far no correct prescription is found.

The analysis of the higher-order quantum corrections reveals new aspects of the ambiguity underlying the CKT. 
It is well-known that due to degrees of freedom in terms of the Lorentz frame, the distribution function of chiral fermions cannot be uniquely determined~\cite{Chen:2014cla}.
As a result, a frame vector representing such an ambiguity is inevitably introduced~\cite{Chen:2015gta}.
From the Wigner function up to $O(\hbar^2)$, we find that on top of the conventional frame vector, there emerges a different frame vector to define the distribution function.
These extra degrees of freedom should be irrelevant to the Wigner function and thus physical quantities, as so is the conventional one.
%This concept is helpful to identity an equilibrium distribution function.
This is one of the guiding principles to identify an equilibrium distribution function.
Indeed we find that there is no equilibrium solution for the kinetic equation in general curved spacetime.
In other words, in general, the $O(\hbar^2)$ CKT under gravity does not reach a global equilibrium without collisions.
However, we elucidate that an equilibrium solution is admitted for several gravitational fields, such as the stationary weak one.

The remaining parts are devoted to the evaluation of physical quantities from the Wigner function that we derived before.
For instance, the charge current and energy-momentum tensor are given by the momentum integrals, as follows:
\begin{equation}
\begin{split}
\label{eq:JT_intro}
 J^\mu (x)
% &= \int_p \tr\biggl[\gamma^\mu \frac{1+\gamma^5}{2} W(x,p)\biggr]
 = 2\int_p \calR^\mu (x,p)\,, 
 \quad
 T^{\mu\nu}(x) 
% &= \int_p \tr\biggl[ \frac{i\hbar}{2} \gamma^{(\mu} \overleftrightarrow{D}^{\nu)} \frac{1+\gamma^5}{2} W(x,p)\biggr]
 = 2\int_p p^{(\mu}\calR^{\nu)} (x,p) + O(\hbar^3)
\end{split}
\end{equation}
with $\int_p := \int \frac{d^4 p}{(2\pi)^4\sqrt{-g(x)}}$ and $X^{(\mu} Y^{\nu)} = \frac{1}{2}(X^\mu Y^\nu + Y^\mu X^\nu)$.
Here $\calR^\mu (x,p)$ is the Wigner function of the right-handed Weyl fermions.
Under a static weak gravity, the $O(\hbar)$ part of Eq.~\eqref{eq:JT_intro} correctly reproduces the CVE.
We find that under a time-dependent gravity, the CVE totally vanishes in the dynamical limit of the background gravitational field.
Although this fact is originally derived with the diagrammatic computation~\cite{Landsteiner:2013aba}, our calculation is its first verification based on the CKT.

\begin{table}
\begin{minipage}{1\textwidth}
\begin{equation*}
{\renewcommand{\arraystretch}{2}
 \begin{array}{|\c||\c|\c|} \hline
  & \text{charge current $C_0 j^\mu$} & \text{energy-momentum tensor  $C_1 t^{\mu\nu}$}  \\  \hline\hline
\text{static} 
	& {\renewcommand{\arraystretch}{1.5}
		\begin{array}{l}
		j^\mu=\frac{1}{12} {R^\mu}_{\alpha}\xi^\alpha
 		- \frac{1}{24} \xi^\mu R \\
 		\qquad\ 
 		+ \frac{1}{6} \xi^\mu  R_{\alpha\beta}\xi^\alpha\xi^\beta
		\end{array}
		}
 	& {\renewcommand{\arraystretch}{1.5}
	\begin{array}{l}
	 	t^{\mu\nu}
	 	=  - \frac{1}{12} R^{\mu\nu} 
 		 - \frac{1}{12} R \xi^\mu \xi^\nu
 		 + \frac{1}{24} R \eta^{\mu\nu} \\
   \qquad\quad
 		 - \frac{1}{6}R^{\alpha(\mu} \xi^{\nu)} \xi_\alpha  
  		 +\frac{1}{6} R^{\alpha\beta} \xi_\alpha \xi_\beta
  		 	(4\xi^\mu\xi^\nu-\eta^{\mu\nu}) \\
  \qquad\quad
  		 + \frac{1}{6} R^{\mu\alpha\nu\beta} \xi_\alpha \xi_\beta
 	\end{array} } \\ \hline
\text{dynamical} 
	& j^\mu=\frac{1}{20}{R^\mu}_\alpha\xi^\alpha
 	  - \frac{1}{40}\xi^\mu R
 	& {\renewcommand{\arraystretch}{1.5}
	\begin{array}{l}
	 	t^{\mu\nu}
	 	= - \frac{1}{12} R^{\mu\nu} 
 		 + \frac{1}{105} R \xi^\mu \xi^\nu
 		 + \frac{13}{840} R \eta^{\mu\nu} \\
		\qquad\ \ 
 		 + \frac{1}{15}R^{\alpha(\mu} \xi^{\nu)} \xi_\alpha  
  		-\frac{2}{105} R^{\alpha\beta} \xi_\alpha \xi_\beta \xi^\mu\xi^\nu\\
		\qquad\ \   		 
		-\frac{1}{70} R^{\alpha\beta} \xi_\alpha \xi_\beta \eta^{\mu\nu} 
  		+ \frac{1}{30} R^{\mu\alpha\nu\beta} \xi_\alpha \xi_\beta 
 	\end{array} }   \\ \hline
 \end{array}
} 
\end{equation*}
\vspace{-1.5em}

\caption{Transport phenomena induced by general weak gravity.
The static and dynamical parts are derived in Eqs.~\eqref{eq:JT2_eq} and~\eqref{eq:JT2_dyn}, respectively.
}
\label{table:general}
\end{minipage}
\vspace{1em}

\begin{minipage}{1\textwidth}
\begin{equation*}
{\renewcommand{\arraystretch}{2}
 \begin{array}{|c||c|c|} \hline
  	& \text{charge current $C_0 j^\mu$}  & \text{energy-momentum tensor $C_1 t^{\mu\nu}$}
  	  \\  \hline  \hline 
\text{static} 
	&
	\ j^0 = \frac{\bnabla^2 T}{6\bar T} \,,\quad j^i = \frac{1}{12} (\bnabla\times\bomega)^i \
 	& 
 	\begin{array}{c}
 	 	t^{00}=\frac{\bnabla^2 T}{6\bar T} \,, 
 	 	\quad
 	 	t^{0i}=-\frac{1}{6} (\bnabla\times\bomega)^i \,, \\ 
		t^{ij} = -\frac{1}{12\bar{T}} (\partial^i\partial^j + \eta^{ij}\bnabla^2 )T  \quad\ \ 	 
	\end{array}  \\ \hline
\text{dynamical} 
	& 
	 j^0=0 \,,\quad j^i = \frac{1}{20} (\bnabla\times\bomega)^i 
 	& 
 	\begin{array}{c}
 		t^{00} = 0 \,,\quad
 		t^{0i} = - \frac{1}{20} (\bnabla\times\bomega)^i\,, \\ 
 		\ t^{ij}
 		=\frac{1}{20\bar{T}} (\partial^i\partial^j + \eta^{ij}\bnabla^2 )T 
 			+ \frac{1}{20}\partial_0\sigma^{ij} \ 
	  \end{array}  \\
 	 \hline
 \end{array}
}
\end{equation*}
\vspace{-1.5em}

\caption{Transport phenomena induced by weak fluid vorticity and temperature gradient.
The static and dynamical parts are derived in Eqs.~\eqref{eq:gravito_eq} and~\eqref{eq:gravito_dyn}, respectively.
}
\label{table:fluid}
\end{minipage}
\vspace{1.5em}

\end{table}

The $O(\hbar^2)$ contribution of Eq.~\eqref{eq:JT_intro} corresponds to the novel transport phenomena induced by Riemann curvatures, which is also our main finding.
They are represented as the following charge current and energy-momentum tensor:
\begin{equation}
J^\mu= C_1 j^\mu\,,\quad
T^{\mu\nu}= C_0 t^{\mu\nu} \,.
\end{equation}
Here $C_0 = \mu/(2\pi^2)$ and $C_1 = \mu^2/(4\pi^2)+T^2/12$ are the metric-independent coefficients determined by temperature $T$ and chemical potential for right-handed fermions $\mu$.
Also $j^\mu$ and $t^{\mu\nu}$ are functions of Riemann curvature $R_{\mu\nu\rho\sigma}$, Ricci tensor $R_{\mu\nu}$ and Ricci scalar $R$.
In Table~\ref{table:general}, we summarize the analytic forms of $j^\mu$ and $t^{\mu\nu}$.
The `static' (`dynamical') implies that a time-independent (time-dependent) metric tensor is used.
We denote $\xi^\mu=\delta^\mu_0$ and $\eta_{\mu\nu}$ being the Minkowski metric tensor.
It is demonstrated in Appendix~\ref{app:alternative} that the same charge current under a static gravity is also derived from different field-theoretical approaches.
This consistency apparently guarantees the validity of our formalism in this paper.

The fermionic system under a background fluid is a pedagogical and informative environment to demonstrate the aforementioned novel phenomena.
In general the fluid effect is translated into an effective curved spacetime described by the following metric tensor:
\begin{equation}
 \label{eq:metric_intro}
  g_{00} = 1 + h_{00}(t,\bx)\,,\quad
  g_{0i} = h_{0i}(t,\bx) \,,\quad
  g_{ij} = \eta_{ij} \,,
\end{equation}
where $h_{\mu\nu}$ is the fluctuation around the flat spacetime.
With this metric, temperature gradient and vorticity are given by
\begin{equation}
 \partial_i T/ \bar{T} = - \frac{1}{2} \partial_i h_{00}\,,
 \quad
 \omega^i = -\frac{1}{2} \varepsilon^{0ijk}\partial_j h_{k0} \,.
\end{equation}
Applying Eq.~\eqref{eq:metric_intro} to $j^\mu$ and $t^{\mu\nu}$ in Table~\ref{table:general}, we get the transport phenomena induced by temperature gradient or the inhomogeneity of vorticity. The results are summarized in Table~\ref{table:fluid}.
Here we define the shear tensor as $\sigma^{ij} = \epsilon^{ij} - \frac{1}{3} \eta^{ij} {\epsilon_k}^k $ with $\epsilon_{ij} = \partial_{(i} h_{j)0}$.

There are two crucial features of the novel transport phenomena induced by gravity (or equivalently inhomogeneous fluid profiles).
One is that even if the collisionless kinetic equation holds, these transport phenomena are induced, as so are the CME and CVE.
Therefore these do not generate any entropy, and thus be nondissipative.
This fact motivates us to analyze the relation of Tables~\ref{table:general} and~\ref{table:fluid}  to the quantum anomaly, from different approaches, such as hydrodynamics.
This is an interesting open question that will be revisited in the future.
The other is the antiparallel flows of the charge current $j^i$ and energy current $t^{0i}$.
Tables~\ref{table:general} and~\ref{table:fluid} for $\mu>0$ show the coefficients of these currents have opposite signs whether the metric is static or dynamical and whatever the metric tensor is.
This is never explained by the classical particle motions; both charge and energy are carried along the classical particle momenta.
The essential ingredient of the antiparallel flow is the spin-Riemann-curvature coupling.
To find a more intuitive explanation of such curious flows is also a fascinating task.

In these respects, the novel gravity-induced transport phenomena should involve a lot of implications, e.g., in heavy-ion collisions or Dirac/Weyl semimetals (some of them are discussed in this paper).
The latter systems may provide a good playground to study our novel phenomena and can be complementary environments to the former.
For them, we need more detailed analysis based on the hydrodynamic model calculation, and quantitative comparison between theory and experiments.
%Although our analysis ignores the collisional integral resulting from fermionic self-energy, the general formalism we obtain here can be applied to the transport dynamics of chiral fermion under general gravity, and can be extended to massive fermion system.
Besides, the CKT in curved spacetime and the resulting curvature-induced transport phenomena could play a more crucial role under genuine gravity, although we do not discuss it in this paper.
For example, we can discuss the geodesics deviation of chiral fermions due to the spin-curvature coupling.
Such a deviation may lead to some correction to the gravitational lensing of neutrinos~\cite{liebes1964gravitational}.
Also, the present work could be applicable to the physics of core-collapse supernova explosions and neutron star formations~\cite{Yamamoto:2015gzz}.
In this direction, we need to take the collisional effects into account~\cite{Hidaka:2016yjf,Hidaka:2017auj,Yamamoto:2020zrs}, based on the Kadanoff-Baym equation in curved spacetime, which respects the diffeomorphism covariance~\cite{Hohenegger:2008zk}.

This paper is organized as follows.
In Sec.~\ref{sec:ckt}, solving the constraint equations, we obtain the general solution of the Wigner function up to $O(\hbar^2)$ under general curved spacetime.
In Sec.~\ref{sec:equilibrium}, we determine an equilibrium distribution function involving $O(\hbar^2)$ corrections, based on the frame-independence of the Wigner function.
In Sec.~\ref{sec:stationary}, we obtain curvature-induced charge current and energy-momentum tensor in equilibrium, which is consistent with different field-theoretical approaches in Appendix~\ref{app:alternative}.
%It is demonstrated in Sec.~\ref{app:alternative} that the resulting expression of the charge current is also derived from different field-theoretical approaches.
%This consistency guarantees the relevance of the general Wigner function derived in Sec.~\ref{sec:ckt} and equilibrium distribution function found in Sec.~\ref{sec:equilibrium}.
In Secs.~\ref{sec:dynamical} and~\ref{sec:dynamical_JT}, we analyze the dynamical response from background gravitational fields.
As a practical application, in Sec.~\ref{sec:fluid} we argue the transport phenomena in a fluid with inhomogeneous vorticity and temperature, which yields effective gravitational curvatures.
In particular, we observe the antiparallel flows of charge and energy due to an inhomogeneous vorticity (or the Ricci tensor ${R_0}^i$).
Through this paper, the convention follows from Ref.~\cite{Liu:2018xip}.

\section{Chiral kinetic theory at $O(\hbar^2)$}\label{sec:ckt}
We first show the brief outline of the derivation of the CKT in the Wigner function approach.
The kinetic theory is a perturbative effective theory in the infrared momentum regime.
In quantum field theory, this perturbation is equivalent to the semiclassical truncation~\cite{Elze:1986qd}.
The CKT is obtained from the semiclassical expansions of the equation of motions for the Wigner functions, that is, the Wigner transformed Dyson-Schwinger equation.
The Wigner function of chiral fermions obeys three equations, which correspond to Eqs.~\eqref{eq:R1-curved}-\eqref{eq:R3-curved} in this paper.
Two of them are the constraints to the Wigner function, and the other becomes the kinetic equation eventually.
In the classical limit, the equation for the Wigner function reduces to the Boltzmann equation~\eqref{eq:Vlasov}. 
The Wigner function formalism gives the quantum generalization of the classical kinetic theory.

Let us start from the Wigner function for the right-handed Weyl fermions, which is defined as~\cite{Fonarev:1993ht}
\begin{eqnarray}
 &\dis \calR^\mu (x,p)
  = \frac{1}{2}\tr\biggl[
 		\gamma^\mu\frac{1+\gamma^5}{2}  W (x,p)
     \biggr]\,,\\\
 & \dis W_{ab}(x,p) 
  = \int d^4y\,\sqrt{-g(x)}\,e^{-ip\cdot y/\hbar} 
  	\langle\bar\psi_b(x,y/2)\psi_a(x,-y/2) \rangle \,,\ \ \ 
\end{eqnarray}
with $g(x)=\det(g_{\mu\nu})$, $\bar\psi(x):= \psi^\dag(x)\gamma^\hzero$, $\psi(x,y) =  \exp(y\cdot D)\psi(x)$, $\bar \psi(x,y) = \bar\psi(x)\exp(y\cdot\lD)$, and $\bar\psi\overleftarrow{O}:= [O \psi]^\dag \gamma^\hzero$.
The above $W(x,p)$ is the general relativistic extension of the one in gauge theory~\cite{Elze:1986qd}.
Here $D_\mu$ is called the horizontal lift;
for a function on $(x^\mu,y^\mu)$ and $(x^\mu,p_\mu)$, the horizontal lift is represented as
\begin{equation}
 D_\mu 
 = 
 \begin{cases}
  \nabla_\mu - \Gamma_{\mu\nu}^\rho y^\nu \partial_\rho^y \,, \\
  \nabla_\mu + \Gamma_{\mu\nu}^\rho p_\rho \partial_p^\nu \,,
 \end{cases} 
\end{equation}
where $\nabla_\mu$ is the covariant derivative in terms of diffeomorphism and the local Lorentz transformation.
The most beneficial property of $D_\mu$ is that it commutes with both $y^\mu$ and $p_\mu$, while $\nabla_\mu$ does not.

Hereafter we consider the Dirac theory under an external torsionless gravitational field.
In this paper, we focus on the collisionless kinetic theory.
The Dirac equation is thus given by $\gamma^\mu \nabla_\mu \psi(x) = 0$, which brings the dynamical equation that the Wigner function $W(x,p)$ obeys.
After a long computation, the set of equations for $\calR^\mu(x,p)$ is up to $O(\hbar^2)$ given by~\cite{Liu:2018xip}
\begin{eqnarray}
 \label{eq:R1-curved}
 & (D_\mu + \hbar^2 P_\mu)\calR^\mu = 0 \,, \\ 
 \label{eq:R2-curved}
 & (p_\mu + \hbar^2 Q_\mu )\calR^\mu = 0 \,, \\
 \label{eq:R3-curved}
 & \dis \hbar\varepsilon_{\mu\nu\rho\sigma}D^{\rho}\calR^{\sigma}
 + 4 \Bigl[(p_{[\mu} 
 + \hbar^2 T_{[\mu})\calR_{\nu]} + \hbar^2 S_{\alpha\mu\nu}\calR^\alpha\Bigr]
  = 0 \,,
\end{eqnarray}
where we introduce the following notations:
\begin{eqnarray}
  \label{eq:P}
 &\dis P_\mu   
  = 	- \frac{1}{8} \nabla_\lambda R_{\mu\nu}\partial_p^\lambda\partial_p^\nu
  		-\frac{1}{24} \nabla_\lambda {R^\rho}_{\sigma\mu\nu} \partial_p^\lambda
  		 	\partial_p^\nu \partial_p^\sigma p_\rho
  		+\frac{1}{8}{R^\rho}_{\sigma\mu\nu} \partial_p^\nu \partial_p^\sigma D_\rho \,,\\
 &\dis Q_\mu 
  = \frac{1}{8} R_{\mu\nu}\partial_p^\nu 
  + \frac{1}{24}{R^\rho}_{\sigma\mu\nu}\partial^\nu_p \partial_p^\sigma p_\rho
  = 3 A_\mu + B_\mu \,, \\
  \label{eq:T}
 &\dis T_\mu 
  = \frac{1}{4} R_{\mu\nu}\partial_p^\nu 
  +\frac{1}{24}{R^\rho}_{\sigma\mu\nu}\partial^\nu_p \partial_p^\sigma p_\rho  
  = 6 A_\mu + B_\mu\,,\\
  \label{eq:ABS}  
 &\dis A_\mu 
  = \frac{1}{24} R_{\mu\nu}\partial_p^\nu \,, \quad
 B_\mu 
  = \frac{1}{24} {R^\rho}_{\sigma\mu\nu}  \partial_p^\nu \partial_p^\sigma p_\rho\,,\quad
 S_{\alpha\mu\nu}
  = -\frac{1}{16} R_{\lambda\alpha\mu\nu}\partial^\lambda_p \,.
\end{eqnarray}
In the above equations, we denote $X^{[\mu}Y^{\nu]} = (X^\mu Y^\nu - X^\nu Y^\mu)/2$, the Riemann tensor is defined as ${R^\rho}_{\sigma\mu\nu} = 2\bigl(\partial_{[\nu} \Gamma^\rho_{\mu]\sigma} + \Gamma_{\lambda[\nu}^\rho \Gamma_{\mu]\sigma}^\lambda \bigr)$ with $\Gamma^\rho_{\mu\nu} =g^{\rho\lambda}(\partial_\mu g_{\lambda\nu} + \partial_\nu g_{\lambda\mu} - \partial_\lambda g_{\mu\nu})/2$, and the Ricci tensor is $R_{\mu\nu} = {R^\lambda}_{\mu\lambda\nu}$.
For left-handed Weyl fermions,  similar equations are derived, but only the sign in front of $\varepsilon_{\mu\nu\rho\sigma}$ is flipped, as is parity-odd.
The first equation~\eqref{eq:R1-curved} corresponds to the kinetic equation while the others~\eqref{eq:R2-curved} and~\eqref{eq:R3-curved} are constraints that 
determine the functional form of $\calR^\mu$.
It is worthwhile to mention that Eqs.~\eqref{eq:R1-curved}-\eqref{eq:R3-curved} are the Ward identities in terms of the symmetries that Weyl fermions respect in a given coordinate;
the $U(1)$ gauge symmetry, the conformal symmetry, and the Lorentz symmetry, respectively~\cite{Liu:2020flb}.

Let us parametrize the solution for Eqs.~\eqref{eq:R2-curved} and~\eqref{eq:R3-curved} as
\begin{equation}
 \calR^\mu = \calR^\mu_\zero + \hbar\calR^\mu_\one+ \hbar^2 \calR_\two^\mu \,.
\end{equation}
Contracting Eq.~\eqref{eq:R3-curved} with $p^\nu$, we find
\begin{equation}
 p^2 \calR_\mu 
 = p_\mu p\cdot\calR
 + \frac{\hbar}{2} \varepsilon_{\mu\nu\rho\sigma}p^\nu D^\rho \calR^\sigma
 + 2\hbar^2  p^\nu 
  \Bigl(
	  T_{[\mu} \calR_{\nu]} + S_{\alpha\mu\nu} \calR^\alpha
  \Bigr)\,.
\end{equation}
Combined with Eq.~\eqref{eq:R2-curved}, this equation is decomposed into
\begin{eqnarray}
 \label{eq:0theq-1}
 & p^2\calR_\mu^\zero = 0 \,, \\
 \label{eq:1steq-1}
 & \dis p^2\calR_\mu^\one
 =   \frac{1}{2}\varepsilon_{\mu\nu\rho\sigma} p^\nu D^\rho \calR^\sigma_\zero \,,\\ 
 \label{eq:2ndeq-1}
 & \dis p^2\calR_\mu^\two 
 = - p_\mu Q\cdot\calR_\zero 
  + \frac{1}{2}\varepsilon_{\mu\nu\rho\sigma} p^\nu D^\rho \calR^\sigma_\one 
  + 2 p^\nu 
  \Bigl(
	  T_{[\mu} \calR_{\nu]}^\zero  + S_{\alpha\mu\nu}\calR^\alpha_\zero
  \Bigr) \,.
\end{eqnarray}
Also Eqs.~\eqref{eq:R2-curved} and~\eqref{eq:R3-curved} yield
\begin{eqnarray}
 \label{eq:0theq-2}
 & p\cdot \calR_\zero = 0 \,,\\
 \label{eq:1steq-2}
 & p\cdot \calR_\one = 0 \,,\\
 \label{eq:2ndeq-2}
 & p\cdot \calR_\two + Q\cdot \calR_\zero = 0 \,,\\
 \label{eq:0theq-3}
 & 4 p_{[\mu} \calR^\zero_{\nu]} 
   = 0 \,, \\
 \label{eq:1steq-3}
 & 4 p_{[\mu} \calR^\one_{\nu]} 
   + \varepsilon_{\mu\nu\rho\sigma}D^\rho\calR^\sigma_\zero 
   = 0 \,, \\
 \label{eq:2ndeq-3}
 & 4 p_{[\mu} \calR_{\nu]}^\two
   + \varepsilon_{\mu\nu\rho\sigma}D^{\rho}\calR_\one^{\sigma}
   + 4 \Bigl(
   		T_{[\mu}\calR_{\nu]}^\zero 
 	 	+ S_{\alpha\mu\nu} \calR_\zero^\alpha
 	   \Bigr)
   = 0 \,.
\end{eqnarray}
In the following, we look for $\calR^\mu_\zero$, $\calR^\mu_\one$ and $\calR^\mu_\two$ that satisfy Eqs.~\eqref{eq:0theq-1}-\eqref{eq:2ndeq-3}.

First, let us solve the zeroth and first-order parts.
Equations~\eqref{eq:0theq-1} and~\eqref{eq:0theq-2} imply
\begin{equation}
\label{eq:Rmu0}
 \calR^\mu_\zero = 2\pi\delta(p^2) p^\mu f_\zero \,,
\end{equation}
where $f_\zero$ is a scalar function that satisfies $\delta(p^2) p^2 f_\zero = 0$.
From Eq.~\eqref{eq:0theq-3}, we can check that there does not appear any other term in $\calR_\mu^\zero$.
Equation~\eqref{eq:R1-curved} in the zeroth order, $D_\mu  \calR^\mu_\zero=0$, gives the collisionless Boltzmann equation,
\begin{equation}
2\pi\delta(p^2) (p^\mu \nabla_\mu + \Gamma_{\mu\nu}^\rho p^\mu p_\rho \partial_p^\nu ) f_\zero=0\,.
\label{eq:Vlasov}
\end{equation}
Higher-order terms give quantum corrections to the Boltzmann equation.

From Eq.~\eqref{eq:1steq-1} and the above $\calR^\mu_\zero$, we find
\begin{equation}
\label{eq:p2Rmu1}
 p^2 \calR^\mu_\one = 0 \,.
\end{equation}
This does not necessarily mean that $\calR^\mu_\one$ itself vanishes for arbitrary $p_\mu$.
Indeed if $\calR^\mu_\one$ involves $\delta(p^2)$, it fulfils Eq.~\eqref{eq:p2Rmu1}.
Therefore, the first-order correction is generally written as
\begin{equation}
 \calR^\mu_\one = 2\pi\delta(p^2) \tcalR^\mu_\one \,.
\end{equation}
Here the undetermined part $\tcalR_\mu^\one$ satisfies $\delta(p^2) p^2 \tcalR^\mu_\one = 0$ so that Eq.~\eqref{eq:p2Rmu1} holds.
Plugging this $\calR^\mu_\one$ and $\calR^\mu_\zero$ into Eq.~\eqref{eq:1steq-3}, we obtain
\begin{equation}
 \label{eq:tcalR1_eq}
 \delta(p^2)\Bigl[ \varepsilon_{\mu\nu\rho\sigma}p^{\rho} D^{\sigma} f_\zero
 - 4 p_{[\mu} \tcalR_{\nu]}^\one \Bigr]
 = 0 \,.
\end{equation}
We contract this with $n^\nu/(2\,p\cdot n)$, where $n^\mu(x)$ is a vector field independent of $p_\mu$.
Then we get
\begin{equation}
 \begin{split}
  \tcalR_\mu^\one\delta(p^2)
  = \delta(p^2)
 	\biggl[
 		p_\mu \frac{n\cdot \tcalR_\one}{p\cdot n} 
 		+ \frac{\varepsilon_{\mu\nu\rho\sigma}p^\rho n^\sigma}{2p\cdot n} 
 			D^\nu f_\zero
 	\biggr] \,.
 \end{split}
\end{equation}
Thus the first-order correction is given by
\begin{equation}
 \label{eq:Rmu1}
 \calR^\mu_\one 
 = 2\pi\delta(p^2)\Bigl[
  p^\mu f_\one + \Sigma_n^{\mu\nu}D_\nu f_\zero
 \Bigr] \,,
\end{equation}
where we define
\begin{equation}
  f_\one
  = \frac{n\cdot\tcalR^\one}{p\cdot n} \,,
   \quad 
  \Sigma_n^{\mu\nu} 
   = \frac{\varepsilon^{\mu\nu\rho\sigma} p_\rho n_\sigma}{2p\cdot n} \,.
\end{equation}
In the above $\calR^\mu_{\one}$, an arbitrary vector $n^\mu$ emerges through $\Sigma^{\mu\nu}_{n}$.
This ambiguity is related to the (local) Lorentz transformation~\cite{Chen:2014cla}, and thus $\Sigma^{\mu\nu}_n$ is regarded as the spin tensor defined in the frame $n^\mu$~\cite{Chen:2015gta}.
In particular, at $n^\mu = (1,\bzero)$, we have $f_\one = \tcalR^\one_0/p_0$, i.e., the charge density divided by the particle energy.
In this sense, $f_\one$ can be regarded as the quantum correction to the distribution function. 
Note that the solution $\calR^\mu_\one$ fulfils Eqs.~\eqref{eq:1steq-1} and~\eqref{eq:1steq-2} as long as $\delta(p^2)p^2 f_\one = 0$ holds.

Now we solve the second-order correction.
By plugging the above $\calR_\zero^\mu$ and $\calR_\one^\mu$ into Eq.~\eqref{eq:2ndeq-1}, we obtain
\begin{equation}
 \begin{split}
  p^2\calR_\mu^\two 
   & = 2\pi
 		\Bigl(
  			- p_\mu Q\cdot p 
      		+ p^\nu\calD_{\mu\nu}
		\Bigr) \delta(p^2) f_\zero  \,,
 \end{split}
\end{equation}
where the derivative operator $\calD_{\mu\nu}$ is defined as
\begin{equation}
  \calD_{\mu\nu} 
 = 2\Bigl(
 		T_{[\mu}p_{\nu]}
 		+ S_{\alpha\mu\nu}p^\alpha
 	\Bigr)
 	+ \frac{1}{2}\varepsilon_{\mu\nu\rho\sigma}D^\rho\Sigma_n^{\sigma\lambda}D_\lambda \,.
\end{equation}
The general form of the second-order correction then reads
\begin{equation}
 \calR_\mu^\two
 = 2\pi\delta(p^2) \tcalR_\mu^\two 
    + \frac{2\pi}{p^2} 
		\biggl[
      		- p_\mu Q\cdot p 
      		+ p^\nu\calD_{\mu\nu}
		\biggr] \delta(p^2) f_\zero \,.
\end{equation}
Here we again introduced the undetermined part $\tcalR_\mu^\two$, which satisfies $\delta(p^2) p^2 \tcalR^\mu_\two = 0$.
Plugging $\calR_\mu^\zero$, $\calR_\mu^\one$, and $\calR_\mu^\two$ into Eq.~\eqref{eq:2ndeq-3}, we obtain
\begin{equation}
 \begin{split}
  0
  & = 4p_{[\mu}\calR^\two_{\nu]}
  	  + (2\pi)\varepsilon_{\mu\nu\rho\sigma} D^\rho
  			\Bigl(
  				p^\sigma f_\one
  				+ \Sigma_n^{\sigma\lambda} D_\lambda f_\zero
  			\Bigr) \delta(p^2) 
  	  + 4(2\pi)\Bigl(
 			T_{[\mu}p_{\nu]}
 			+ S_{\alpha\mu\nu}p^\alpha
 		\Bigr) f_\zero \delta(p^2) \\
  & = 2\pi
  		\biggl[
  			4p_{[\mu}\tcalR^\two_{\nu]}
  			+ \varepsilon_{\mu\nu\rho\sigma}D^\rho p^\sigma f_\one 
  			+ \frac{4}{p^2} p_{[\mu}p^\rho\calD_{\nu]\rho} f_\zero 
  			+ 2 \calD_{\mu\nu} f_\zero
  		\Bigr] \delta(p^2) \\
  & = 2\pi \biggl[
  			4p_{[\mu}\tcalR^\two_{\nu]}
  			- \varepsilon_{\mu\nu\rho\sigma}p^\rho D^\sigma f_\one 
  			- \frac{1}{p^2} \varepsilon_{\mu\nu\rho\sigma} 
  				p^\rho\varepsilon^{\alpha\beta\gamma\sigma} p_\alpha \calD_{\beta\gamma}
  				f_\zero
  		   \biggr] \delta(p^2) \,. 					
 \end{split}
\end{equation}
Similarly to Eq.~\eqref{eq:tcalR1_eq}, we solve the above equation by introducing a vector $u^\mu$ (this is in general different from $n^\mu$), as follows:
\begin{equation}
\label{eq:tRmu2}
 \begin{split}
   \tcalR_\mu^\two \delta(p^2) 
 	& = \delta(p^2) 
  	 \Bigl[
 		 p_\mu f_\two + \Sigma_{\mu\nu}^u D^\nu f_\one
 	  \Bigr] 
  	 + \frac{1}{p^2} \varepsilon^{\alpha\beta\gamma\nu} \Sigma_{\mu\nu}^u
  	 		p_\alpha\calD_{\beta\gamma} \delta(p^2) f_\zero \\
 	& = \delta(p^2) 
  	 \Bigl[
 		 p_\mu f_\two + \Sigma_{\mu\nu}^u D^\nu f_\one
 	  \Bigr] 
  	 - \frac{\delta(p^2)}{p^2} \Sigma_{\mu\nu}^u 
   			\biggl[
   				\frac{1}{2}\tilde{R}^{\alpha\beta\nu\rho}p^\rho p^\alpha \partial_p^\beta
   				+ p\cdot D\Sigma_n^{\nu\rho} D_\rho
   			\biggr] f_\zero \,,
 \end{split}
\end{equation}
where we defined $\tilde{R}_{\alpha\beta\mu\nu} =  {R_{\alpha\beta}}^{\rho\sigma}\varepsilon_{\rho\sigma\mu\nu}/2$ and
\begin{equation}
 f_\two
 = \frac{u\cdot \tcalR_\two }{p\cdot u} \,.
\end{equation}
In the second line of Eq.~\eqref{eq:tRmu2}, we utilized
\begin{equation}
 \begin{split}
 \label{eq:ABS}
   [A_\mu, p_\nu] 
   	 = \frac{1}{24} R_{\mu\nu} \,, \qquad
   [B_\mu, p_\nu] 
   	 = -\frac{1}{24} R_{\mu\nu}
 	 + \frac{1}{24} 
 		\Bigl(
 			{R^\rho}_{\nu\mu\sigma} 
 			+ {R^\rho}_{\sigma\mu\nu} 
 		\Bigr)
 			p_\rho \partial_p^\sigma \,,
 \end{split}
\end{equation}
which yield
\begin{equation}
 \begin{split}
  2\varepsilon^{\alpha\beta\gamma\nu} p_\alpha
  \Bigl(
 	T_{\beta} p_{\gamma}
 	+ S_{\lambda\beta\gamma} p^\lambda
  \Bigr) 
  & = -\frac{1}{2} \tilde{R}^{\alpha\beta\nu\rho}p_\rho p_\alpha\partial_\beta^p \,.
 \end{split}
\end{equation}
Therefore, the second-order correction reads
\begin{equation}
 \begin{split}
 \label{eq:Rmu2}
  \calR_\mu^\two
   	 & = 2\pi\delta(p^2)
 	  	\biggl[
 	  		p_\mu f_\two
 	  		+ \Sigma_{\mu\nu}^u D^\nu f_\one
	    \biggr]  
   	 	   + 2\pi\frac{1}{p^2}
    		 \biggl[
	 	     - p_\mu Q\cdot p 
    			  + 2 p^\nu \Bigl( T_{[\mu} p_{\nu]} + S_{\alpha\mu\nu} p^\alpha \Bigr)
   	 	 \biggr] \delta(p^2) f_\zero \\
   	 & \quad
	  + 2\pi\frac{\delta(p^2)}{p^2}
	 	\biggl[
      	  	 \frac{1}{2}\varepsilon_{\mu\nu\rho\sigma}p^\nu
  		 	 	D^\rho \Sigma_n^{\sigma\lambda} D_\lambda
      	  	 - \Sigma_{\mu\nu}^u
      	  	 	\biggl(
      	  	 		\frac{1}{2} {\tilde{R}}^{\alpha\beta\nu\rho} 
      	  	  			p_\rho p_\alpha \partial_\beta^p
      	  			 + p\cdot D \Sigma_n^{\nu\rho} D_\rho
      	  		\biggr)
      	  \biggr] f_\zero \,.
 \end{split}
\end{equation}
We mention that Eq.~\eqref{eq:2ndeq-1} is still fulfilled for the above $\calR_\mu^\two$ as long as 
\begin{equation}
 \label{eq:f2_constraint}
 \delta(p^2)p^2 f_\two = 0
\end{equation}
holds.
Indeed we can check
\begin{equation}
 \begin{split}
  \delta(p^2)p^2 \tcalR^\two_\mu
  & = - \delta(p^2) \Sigma_{\mu\nu}^u 
   			\biggl[
   				\frac{1}{2}\tilde{R}^{\alpha\beta\nu\rho}p_\rho p_\alpha \partial^p_\beta
   				+ p\cdot D\Sigma_n^{\nu\rho} D_\rho
   			\biggr] f_\zero \\
  & = - \delta(p^2) \Sigma_{\mu\nu}^u 
   			\biggl[
   				\frac{1}{2}\tilde{R}^{\alpha\beta\nu\rho}p_\rho p_\alpha \partial^p_\beta
   				+ D_\lambda
   					\biggl(
   						\Sigma_n^{\nu\lambda}p^\rho
   						+ \frac{1}{2}\varepsilon^{\nu\lambda\rho\sigma}
   							p_\sigma
   						- \frac{1}{2}\varepsilon^{\nu\lambda\rho\sigma}\frac{p^2 n_\sigma}{p\cdot n}
   					\biggr) D_\rho
   			\biggr] f_\zero \\
  & = - \delta(p^2) \Sigma_{\mu\nu}^u 
   			 D_\lambda \Sigma_n^{\nu\lambda} p\cdot D f_\zero = 0 \,.
 \end{split}
\end{equation}
In the second line, we utilized
\begin{equation}
 \label{eq:Sigma_p}
 \Sigma_{\alpha[\mu}^n p_{\nu]}
 = -\frac{1}{2}\Sigma_{\mu\nu}^n p_\alpha
   - \frac{1}{4}\varepsilon_{\mu\nu\alpha\beta}
 		\biggl(
 			p^\beta
 			- \frac{p^2 n^\beta}{p\cdot n}
 		\biggr) \,,
\end{equation}
which follows from the Schouten identity:
$ p_\mu \varepsilon_{\nu\rho\sigma\lambda} 
 + p_\nu \varepsilon_{\rho\sigma\lambda\mu}
 + p_\rho \varepsilon_{\sigma\lambda\mu\nu}
 + p_\sigma \varepsilon_{\lambda\mu\nu\rho}
 + p_\lambda \varepsilon_{\mu\nu\rho\sigma}
 = 0 $.
Also the last line follows from $[D_\mu,D_\nu] f = - R_{\alpha\beta\mu\nu}p^\alpha\partial^\beta_p f$, and the classical kinetic equation~\eqref{eq:R1-curved}, i.e., $\delta(p^2)\, p\cdot D f_\zero = 0$.
We stress that Eq.~\eqref{eq:f2_constraint} is a crucial constraint to $f_\two$, especially when we determine the equilibrium distribution function (see Sec.~\ref{sec:equilibrium}).

Eventually, the Wigner function up to $O(\hbar^2)$ is derived as
\begin{equation}
 \begin{split}
 \label{eq:Rmu}
  \calR_\mu
 	& = 2\pi\delta(p^2)
 	  	\Bigl[
			p_\mu \Bigl(f_\zero + \hbar f_\one + \hbar^2 f_\two\Bigr)
 	  		+ \hbar\Sigma_{\mu\nu}^n D^\nu f_\zero
 	  		+ \hbar^2\Sigma_{\mu\nu}^u D^\nu f_\one
	    \Bigr] \\
	& \quad
	   + 2\pi\hbar^2\frac{1}{p^2}
    		 \biggl[
	 	     - p_\mu Q\cdot p 
    			  + 2 p^\nu \Bigl( T_{[\mu} p_{\nu]} + S_{\alpha\mu\nu} p^\alpha \Bigr)
   	 	 \biggr] \delta(p^2) f_\zero \\
   	 & \quad
	  + 2\pi\hbar^2\frac{\delta(p^2)}{p^2}
	 	\biggl[
      	  	 \frac{1}{2}\varepsilon_{\mu\nu\rho\sigma}p^\nu
  		 	 	D^\rho \Sigma_n^{\sigma\lambda} D_\lambda
      	  	 - \Sigma_{\mu\nu}^u
      	  	 	\biggl(
      	  	 		\frac{1}{2} {\tilde{R}}^{\alpha\beta\nu\rho} 
      	  	  			p_\rho p_\alpha \partial_\beta^p
      	  			 + p\cdot D \Sigma_n^{\nu\rho} D_\rho
      	  		\biggr)
      	  \biggr] f_\zero \,.
 \end{split}
\end{equation}
%Since the above resulting Wigner function has a highly complicated structure, we need argue its relevance.
%One way to do this is to confirm that a physical quantity from the above $\calR^\mu$ is consistent with the one derived from the different computation.
As the above $\calR^\mu$ is a fermion propagator, performing the momentum integration involving it we evaluate a corresponding quantity of Weyl fermions under a gravitational field.
In particular, the charge current and the symmetric energy-momentum tensor are given by
\begin{equation}
\begin{split}
\label{eq:J-T}
 J^\mu (x)
 &= \int_p \tr\biggl[\gamma^\mu \frac{1+\gamma^5}{2} W(x,p)\biggr]
 = 2\int_p \calR^\mu (x,p)\,, \\
 \quad
 T^{\mu\nu}(x) 
 &= \int_p \tr\biggl[ \frac{i\hbar}{2} \gamma^{(\mu} \overleftrightarrow{D}^{\nu)} \frac{1+\gamma^5}{2} W(x,p)\biggr]
 = 2\int_p p^{(\mu}\calR^{\nu)} (x,p) + O(\hbar^3) \,,
\end{split}
\end{equation}
where we define $\int_p := \int \frac{d^4 p}{(2\pi)^4\sqrt{-g(x)}}$, $X^{(\mu}Y^{\nu)} = (X^\mu Y^\nu + X^\nu Y^\mu)/2$ and $\overleftrightarrow{D_\mu}(\bar\psi_b \psi_a) = \bar\psi_b D_\mu \psi_a - \bar\psi_b \lD_\mu \psi_a$.
Here we used the expansion of the derivative; $D_\mu \psi(x,y) = \partial_\mu^y \psi(x,y) + O(\hbar^2)$ (see Appendix~C in Ref.~\cite{Liu:2018xip}).
In Sec.~\ref{sec:stationary}, we derive the equilibrium Wigner function and show that it yields the gravity-induced parts of $J^\mu$ and $T^{\mu\nu}$. 
We also demonstrate that the different approaches in Appendix~\ref{app:alternative} lead to the same $J^\mu$ [see Eq.~\eqref{eq:JT2_eq} and Eqs.~\eqref{eq:J_kubo} and~\eqref{eq:J_RNC}].

\section{Frame dependence and equilibrium}\label{sec:equilibrium}
In the evaluation of physical quantities such as Eq.~\eqref{eq:J-T}, it is necessary to identify the explicit form of $f_{\zero,\one,\two}$.
For this purpose, the frame (i.e., $n^\mu$ and $u^\mu$) dependences of $\calR^\mu$ are a crucial concept; as shown below, we derive a constraint on the distribution function from the proper transformation law under the shift of the frames.
Combined this with the kinetic equation, we determine $f_{\zero,\one,\two}$ at equilibrium, which are utilized to evaluate equilibrium transport phenomena induced by external gravity in Sec.~\ref{sec:stationary}.
This section is devoted to the analysis of the frame dependence and the equilibrium solution found from it.

In the above derivation of $\calR^\mu$, the frame vectors $n^\mu$ and $u^\mu$ are algebraically introduced.
It is valid to expect that the frame-dependence disappears in $\calR^\mu$, which generates physical quantities.
Indeed, as is well-known in the $O(\hbar)$ CKT, the choice of the frame vector $n^\mu$ corresponds to the Lorentz transformation, and the frame-dependence is totally compensated in physical quantities, due to the shift of $f_\one$.
Hence we may plausibly require that the same is true in the $O(\hbar^2)$ CKT.
That is, we determine the transformation law of $f_\two$ under $n^\mu\to n'^\mu$ and $u^\mu\to u'^\mu$ so that the frame dependence vanishes in $\calR^\mu$.
%\red{These transformation laws are utilized to identify the equilibrium distribution function, and eventually the equilibrium transport phenomena induced by external gravity.}

Let us first take the Lorentz transformation in terms of $n^\mu$, namely, $(x^\mu,p^\mu) \to (x'^\mu,p'^\mu) = {({\Lambda}_n)^\mu}_\nu (x^\nu,p^\nu)$ and $u^\mu\to u'^\mu = {({\Lambda}_n)^\mu}_\nu u^\nu$, where ${({\Lambda}_n)^\mu}_\nu$ is the matrix representation of the local Lorentz transformation.
This transformation is equivalent to the one of the frame vector $n^\mu$ as
\begin{equation}
 n^\mu \to n'^\mu = {{({\Lambda}_n^{-1})}^\mu}_\nu n^\nu \,.
\end{equation}
We also parametrize the transformation of $f$ as
\begin{equation}
 f(x,p)\to f'(x',p') = f(x,p) + \hbar\delta_n f_\one(x,p) + \hbar^2 \delta_n f_\two(x,p) \,.
\end{equation}
Due to the Lorentz covariance of $\calR^\mu$, we have
\begin{equation}
 \begin{split}
 \label{eq:covariance_n}
  0 & = {(\Lambda^{-1}_n)_\mu}^\nu \calR'_\nu(x',p') - \calR_\mu(x,p) \\
  	& = 2\pi\delta(p^2)
  			\biggl[
  				p_\mu \Bigl(
  						\hbar\delta_n f_\one + \hbar^2 \delta_n f_\two
  					  \Bigr)
  				+ \hbar\Bigl(
  						\Sigma^{n'}_{\mu\nu}
  						- \Sigma^{n}_{\mu\nu}
  					   \Bigr) D^\nu f_\zero
  				+ \hbar^2\Sigma^{u}_{\mu\nu} D^\nu \delta_n f_\one \\
  & \qquad\qquad\qquad
  				+\frac{\hbar^2}{p^2}
  					\biggl(
  						\frac{1}{2}\varepsilon_{\mu\nu\rho\sigma}
  							p^\nu D^\rho 
  							\Bigl(
  								\Sigma^{\sigma\lambda}_{n'}
  								- \Sigma^{\sigma\lambda}_n
  							\Bigr)
  								D_\lambda f_\zero
  						- \Sigma_{\mu\nu}^{u} p\cdot D
  						\Bigl(
  							\Sigma_{n'}^{\nu\rho}
  							- \Sigma_{n}^{\nu\rho}
  						  \Bigr)  D_\rho f_\zero
  					\biggr)
  			\biggr] \,.
 \end{split}
\end{equation}
Contracting Eq.~\eqref{eq:covariance_n} with $n^\mu$ and picking up only the $O(\hbar)$ terms, we find
\begin{equation}
 \begin{split}
 \label{eq:delta_f1}
  \delta_n f_\one 
   & = - \frac{n^\mu}{p\cdot n}\Sigma_{\mu\nu}^{n'} D^\nu f_\zero \,.
 \end{split}
\end{equation}
Similarly, contracting Eq.~\eqref{eq:covariance_n} with $u^\mu$, we obtain
\begin{equation}
 \begin{split}
 \label{eq:deltan_f2}
  \delta_n f_\two
   & = \frac{1}{p^2} \Sigma_{\mu\nu}^u D^\mu 
   			\Bigl(
   				\Sigma^{\nu\rho}_{n'}
   				- \Sigma^{\nu\rho}_{n}
   			\Bigr)
   				 D_\rho f_\zero \,.
 \end{split}
\end{equation}
The above $\delta_n f_{\one,\two}$ fulfills $\delta(p^2) p^2 \delta_n f_{\one,\two} = 0$.
Also, we can show that they satisfy Eq.~\eqref{eq:covariance_n}.
%We note that by contracting Eq.~\eqref{eq:covariance_n} with $n^\mu$, we also obtain the above equation.

Let us also perform the Lorentz transformation with
\begin{equation}
u^\mu \to u'^\mu = {{({\Lambda}_u^{-1})}^\mu}_\nu u^\nu \,,
\end{equation}
for which the Lorentz covariance of $\calR_\mu$ requires
\begin{equation}
 \begin{split}
 \label{eq:covariance_u}
  0 & = {(\Lambda^{-1}_u)_\mu}^\nu \calR'_\nu(x',p') - \calR_\mu(x,p) \\
  	& = 2\pi\delta(p^2)
  			\biggl[
  				p_\mu \Bigl(
  						\hbar\delta_u f_\one + \hbar^2 \delta_u f_\two
  					  \Bigr)
  				+ \hbar^2\Bigl(
  						\Sigma^{u'}_{\mu\nu}
  						- \Sigma^{u}_{\mu\nu}
  					   \Bigr)  D^\nu f_\one
  				+ \hbar^2\Sigma^{u'}_{\mu\nu}  D^\nu \delta_u f_\one \\
  	& \qquad\qquad\qquad
  				-\frac{\hbar^2}{p^2}
  					\Bigl(
  						\Sigma^{u'}_{\mu\nu}
  						- \Sigma^u_{\mu\nu}
  					\Bigr)
  					\biggl(
  						\frac{1}{2}\tilde{R}^{\alpha\beta\nu\rho}p_\rho p_\alpha \partial^p_\beta
  						+ p\cdot D \Sigma_{n}^{\nu\rho}  D_\rho
  					\biggr) f_\zero
  			\biggr] \,.
 \end{split}
\end{equation}
From the $O(\hbar)$ part, we readily find
\begin{equation}
\label{eq:deltau_f1}
 \delta_u f_\one = 0 \,.
\end{equation}
By contracting Eq.~\eqref{eq:covariance_u} with $u^\mu$, we find
\begin{equation}
\label{eq:deltau_f2}
 \delta_u f_\two
 = -\frac{u^\mu}{p\cdot u} \Sigma^{u'}_{\mu\nu}
 	\biggl[
 		   D^\nu f_\one
   		- \frac{1}{p^2} 
   					\biggl(
  						\frac{1}{2}\tilde{R}^{\alpha\beta\nu\rho}p_\rho p_\alpha\partial^p_\beta
  						+ p\cdot D \Sigma_{n}^{\nu\rho}  D_\rho
  					\biggr) f_\zero 
   	\biggr] \,.
\end{equation}
We can check that the above $\delta_u f_{\two}$ fulfills $\delta(p^2) p^2 \delta_u f_{\two} = 0$ and Eq.~\eqref{eq:covariance_u}.

In the Wigner function~\eqref{eq:Rmu}, the frame vectors $n^\mu$ and $u^\mu$ are in general chosen independently.
As long as $f_\one$ and $f_\two$ obey the transformation laws~\eqref{eq:delta_f1}, \eqref{eq:deltan_f2}, \eqref{eq:deltau_f1} and~\eqref{eq:deltau_f2}, however, we can always set $u^\mu = n^\mu$ by redefining $f_\two$.
Then, Eq.~\eqref{eq:Rmu} is simplified as
\begin{equation}
 \begin{split}
 \label{eq:Rmu_same_frame}
  \calR_\mu
 	& = 2\pi
 		\Biggl[
 			\delta(p^2)
 			\bigl(
				p_\mu 
 	  			+ \hbar\Sigma_{\mu\nu}^n D^\nu 
 	  		\bigr)
	   		+ \frac{\hbar^2}{p^2}
    		 \Bigl\{
	 	    	 - p_\mu Q\cdot p 
    			  + 2 p^\nu 
    			  	\bigl( 
    			  		T_{[\mu} p_{\nu]} 
    			  		+ S_{\alpha\mu\nu} p^\alpha 
    			  	\bigr) 
    		\Bigr\} \delta(p^2) \\
   	 & \qquad
	  + \frac{\hbar^2\delta(p^2)}{2p^2}
	 	\biggl\{
      	  	 \varepsilon_{\mu\nu\rho\sigma}p^\nu
  		 	 	D^\rho \Sigma_n^{\sigma\lambda} D_\lambda
      	  	 - \Sigma_{\mu\nu}^n
      	  	 	\bigl(
      	  	 		{\tilde{R}}^{\alpha\beta\nu\rho} 
      	  	  			p_\rho p_\alpha \partial_\beta^p
      	  			 + 2 p\cdot D \Sigma_n^{\nu\rho} D_\rho
      	  		\bigr)
      	  \biggr\}
      	 \Biggr] f \,,
 \end{split}
\end{equation}
where we define
\begin{equation}
 f=f_\zero + \hbar f_\one + \hbar^2 f_\two \,.
\end{equation}

The transformation laws under the change of the frames $n^\mu$ and $u^\mu$ are helpful to identify the equilibrium distribution function.
Let us first start from the classical distribution $f_\zero$, which is defined as a function of the collisional conserved quantities:
\begin{eqnarray}
 \label{eq:f0}
 & f_\zero = f_\zero (g_\zero = -\beta\mu +\beta\cdot p) \,,\\
 \label{eq:alphabeta}
 & \nabla_\mu (\beta\mu) = 0 \,, \quad
 \nabla_\mu \beta_\nu + \nabla_\nu \beta_\mu = 0 \,.
\end{eqnarray}
For this $f_\zero$, the transformation law~\eqref{eq:delta_f1} yields
\begin{equation}
 \begin{split}
  \delta_n f_\one 
%  & = - \frac{n_\mu}{p\cdot n}\Sigma_{\mu\nu}^{n'}
%  			D_\nu f_\zero \\ 
  & = -f'_\zero 
  		\frac{n_\mu}{p\cdot n}\Sigma_{\mu\nu}^{n'}p^\rho
  				\nabla_\nu\beta_\rho \\
  & = - f'_\zero 
  		\frac{n_\mu}{p\cdot n}
  				\biggl(
  					-\frac{1}{2}\Sigma^{\nu\rho}_{n'}p^\mu
  					-\frac{1}{4}\varepsilon^{\nu\rho\mu\sigma}
  						p_\sigma
  				\biggr)
  				\nabla_\nu\beta_\rho \\  
  & =   f'_\zero \frac{1}{2}
  		\Bigl(
  			\Sigma^{\nu\rho}_{n'}
 			- \Sigma^{\nu\rho}_{n}
  		\Bigr)
  			\nabla_\nu\beta_\rho \,,
 \end{split}
\end{equation}
where we use Eq.~\eqref{eq:Sigma_p} and define $f'_\zero = df_\zero(g_\zero)/dg_\zero$.
Although the above relation identifies the frame-dependent part involved in $f_\one$ at equilibrium, the frame-independent part is still undetermined.
If we set such an ambiguous part in $f_\one$ to be zero, however, we identify
\begin{equation}
 \label{eq:f1}
 f_{\one} 
 =   f_\zero'\frac{1}{2}\Sigma_n^{\mu\nu}\nabla_\mu\beta_\nu 
% 	+ \phi_{\one}
 \,.
\end{equation}
This is a plausible form in the sense that the spin-vorticity coupling term is correctly reproduced:
$f_\zero + \hbar f_\one \simeq f_\zero (g_\zero + \frac{\hbar}{2} \Sigma_n^{\mu\nu}\nabla_\mu\beta_\nu) + O(\hbar^2)$.
In this case, the first order Wigner function~\eqref{eq:Rmu1} is written as
\begin{equation} 
 \label{eq:Rmu1_eq}
 \calR^\mu_{\one\eq}
 = 2\pi\delta(p^2) f'_\zero \biggl(-\frac{1}{4}\biggr)\varepsilon^{\mu\nu\rho\sigma} 
 	p_\nu \nabla_\rho\beta_\sigma \,.
\end{equation}
This $\calR^\mu_{\one\eq}$ fulfills the kinetic equation at $O(\hbar)$~\cite{Liu:2018xip}.
Subsequently, with the above $f_\zero$ and $f_\one$, the transformation laws~\eqref{eq:deltan_f2} and~\eqref{eq:deltau_f2} lead to
\begin{equation}
 \begin{split}
  \delta_n f_\two 
  & = \frac{1}{p^2} \Sigma_{\mu\nu}^u D^\mu  f'_\zero
   			\Bigl(
   				\Sigma^{\nu\rho}_{n'}
   				- \Sigma^{\nu\rho}_{n}
   			\Bigr)
   				p^\sigma \nabla_\rho \beta_\sigma \\
  & = \Sigma_{\mu\nu}^u D^\mu  
  		\biggl[
  			\frac{1}{4}f'_\zero
   				\varepsilon^{\rho\sigma\nu\lambda}
   					\biggl(
   						\frac{n'_\lambda}{p\cdot n'}
   						- \frac{n_\lambda}{p\cdot n}
   					\biggr)
   				\nabla_\rho\beta_\sigma
   		\biggr] \,, \\
  \delta_u f_\two 
  & = -\frac{u^\mu}{p\cdot u} \Sigma^{u'}_{\mu\nu}
 		\biggl[
 			  D^\nu f_\zero'
  					\frac{1}{2}\Sigma_n^{\rho\sigma}\nabla_\rho\beta_\sigma
   			- \frac{1}{p^2} 
   				\biggl(
   					\frac{1}{2}\tilde{R}^{\alpha\beta\nu\rho}p_\rho p_\alpha\beta_\beta f_\zero'
   					+ p\cdot  D f_\zero' \Sigma^{\nu\rho}_n p^\sigma \nabla_\rho\beta_\sigma 
   				\biggr)
  	 	\biggr] \\
  & =  \Bigl(
 			\Sigma^{u'}_{\mu\nu}
 			- \Sigma^{u}_{\mu\nu}
 		\Bigr)
 			  D^{\mu}
 		\biggl( 
  	 		f_\zero' \frac{ \varepsilon^{\rho\sigma\nu\lambda} n_\lambda}{4p\cdot n}
  			\nabla_\rho\beta_\sigma 
  		\biggr) \,,
 \end{split}
\end{equation}
where we employ Eq.~\eqref{eq:Sigma_p}.
Therefore we find
\begin{equation}
 \label{eq:f2}
 f_\two
 = \Sigma^u_{\mu\nu}  D^\mu 
 	\biggl(
 		f'_\zero
 		\frac{\varepsilon^{\nu\rho\sigma\lambda}}{4\,p\cdot n}
 		n_\rho \nabla_\sigma\beta_\lambda 
 	\biggr) 
% 	+ \phi'_\one \frac{1}{2} \Sigma^{\nu\rho}_{u}
%  			\nabla_\nu\beta_\rho 
  	+ \phi_\two \,.
\end{equation}
Here, unlike the first-order correction $f_\one$, we explicitly keep the frame-independent ambiguity $\phi_\two$.
Importantly, this ambiguous part should be taken into account for the realization of equilibrium, as we elaborate later.
As shown in Appendix~\ref{app:eqRmu2}, inserting the above distribution functions $f_{\zero,\one,\two}$ into Eq.~\eqref{eq:Rmu2}, we reduce the second-order correction part to
\begin{equation}
\begin{split}
\label{eq:Rmu2_eq}
\calR^\mu_{\two\eq}
 & = 2\pi\delta(p^2)
   	\Biggl[
   		\phi_\two p^\mu 
   		+ f_\zero
 	 	\biggl(
 	 		- \frac{1}{2p^2} R^{\mu\alpha} p_\alpha
 	 		- \frac{1}{12p^2} R\, p^\mu
 	 		+ \frac{2}{3(p^2)^2} R^{\alpha\beta} p^\mu p_\alpha p_\beta 
 	 	\biggr) \\
 & \qquad\qquad
 	 	+ f'_\zero
 	 	\biggl(
 	 		-\frac{1}{24} R^{\mu\alpha} \beta_\alpha
 	 		+ \frac{1}{12 p^2} R^{\alpha\beta\gamma\mu} p_\alpha\beta_\beta p_\gamma
 	 	\biggr) \\
 & \qquad\qquad
   		+ f''_\zero
 		\biggl(
 			- \frac{1}{24} R^{\alpha\beta\gamma\mu} p_\alpha\beta_\beta\beta_\gamma
 			-\frac{1}{12 p^2} R_{\alpha\beta\gamma\delta} 
 				p^\mu p^\alpha p^\gamma \beta^\beta \beta^\delta\\
 & \qquad\qquad\qquad\quad
	 		 - \frac{1}{4} \nabla^{[\rho} \beta^{\mu]} p^\nu \nabla_{[\rho}\beta_{\nu]} 
 			 + \frac{p^\mu}{4p^2} p_\nu 
 			 	\nabla^{[\rho} \beta^{\nu]} p^\sigma \nabla_{[\rho} \beta_{\sigma]}
 		\biggr)
 	\Biggr] \,.
\end{split}
\end{equation}
The frame-dependence here vanishes totally, as it should.
Plugging this into the chiral kinetic equation~\eqref{eq:R1-curved}, and after a straightforward calculation in Appendix~\ref{app:kinetic_equation}, we finally arrive at
\begin{equation}
\begin{split}
\label{eq:CKE1}
0 & = \Bigl[D\cdot\calR_\two+P\cdot\calR_\zero\Bigr]/2\pi \\
 & = \delta(p^2)\,p\cdot D\, \phi_\two 
 	+ f_\zero'\delta(p^2)
		\biggl(
			- \frac{1}{8}\beta\cdot \nabla R
			+ \frac{1}{4p^2} \beta\cdot \nabla R^{\alpha\beta} p_\alpha p_\beta
		\biggr) \\
 &\quad
		+ f_\zero''\delta(p^2)
		\biggl(
			- \frac{1}{24} p\cdot\nabla R^{\alpha\beta} \beta_\alpha \beta_\beta
			+ \frac{1}{8} R^{\alpha\beta\mu\nu} p_\alpha\beta_\beta \nabla_\mu \beta_\nu
 		\biggr) \\
 &\quad
  + f_\zero''' \delta(p^2)
  \biggl(
  		- \frac{1}{24}\beta\cdot\nabla R_{\rho\sigma\mu\nu} p^\mu \beta^\nu p^\rho \beta^\sigma
  \biggr) 
 	+ \delta(p^2)
 		\biggl(
 	 		- \frac{1}{2p^2} R^{\mu\alpha} p_\alpha
 	 	\biggr) D_\mu f_\zero \\
 & \quad
 	+ \delta(p^2)
 	 		\biggl(
 	 			\frac{1}{12} R^{\mu\alpha} \beta_\alpha
 	 		\biggr) D_\mu f_\zero' 
 	 +  \delta(p^2)
 	 		\biggl(
 				- \frac{1}{6} R^{\alpha\beta\gamma\mu} p_\alpha\beta_\beta\beta_\gamma
 				- \frac{1}{4} \nabla^{\rho} \beta^{\mu} p^\nu \nabla_{\rho}\beta_{\nu} 
 			\biggr) D_\mu f_\zero'' \,,
\end{split}
\end{equation}
which is the equation to determine $\phi_\two$.
However, there is in general no solution, as is obvious from the constraint~\eqref{eq:f2_constraint};  $\phi_\two$ cannot have $\delta(p^2)/p^2$ terms.
In other words, the collisionless CKT has no global equilibrium solution in general background curved geometry.
We note that an equilibrium distribution function with $\phi_\two = 0$ is realized in the flat spacetime limit $g_{\mu\nu} = \eta_{\mu\nu}$; the Killing equation $\partial_\mu \beta_\nu + \partial_\nu \beta_\mu = 0$ leads to
\begin{equation}
\begin{split}
\delta(p^2) 
 		\biggl( 
 			- \frac{1}{4} \partial^\rho \beta^{\mu} p^\nu \partial_{\rho}\beta_{\nu}
 		\biggr) \partial_\mu f_\zero'' 
& = \delta(p^2) 
 		\biggl( 
 			- \frac{1}{4} \partial^\rho \beta^{\mu} 
 		\biggr) \partial_\rho \partial_\mu f_\zero'
 = 0 \,.
\end{split}
\end{equation}

\section{Stationary weak gravity}\label{sec:stationary}
Although the general curved spacetime does not realize an equilibrium, there may exist a special geometry having a solution for Eq.~\eqref{eq:CKE1}.
One of the simplest cases is the stationary and weak background gravitational field, where the metric tensor is given by
\begin{equation}
\label{eq:stationary_weak}
 g_{\mu\nu}=\eta_{\mu\nu}+h_{\mu\nu}\,,
 \quad
 \partial_0 h_{\mu\nu} = 0\,,
 \quad
 |h_{\mu\nu}|\ll 1 \,.
\end{equation}
In this case, the time-like Killing vector $\beta^\mu \parallel \xi^\mu:=\delta^\mu_0$ is admitted.
%and we neglect the higher order metric fluctuation $O(h_{\mu\nu}^2)$.
Then, the kinetic equation~\eqref{eq:CKE1} is drastically reduced as
\begin{equation}
\begin{split}
\label{eq:CKE2}
 \delta(p^2)
		p\cdot D 
		\biggl(
			\phi_\two
			- \frac{1}{24} R^{\alpha\beta} \beta_\alpha \beta_\beta f_\zero''
 		\biggr) 
 = 0 \,.
\end{split}
\end{equation}
Therefore, for the metric tensor~\eqref{eq:stationary_weak}, we identify
\begin{equation}
\label{eq:phi2}
 \phi_\two
 = \frac{f_\zero''}{24} R^{\alpha\beta} \beta_\alpha \beta_\beta \,.
\end{equation}
Hereafter, we call $f = f_\zero + \hbar f_\one + \hbar^2 f_\two$ with Eqs.~\eqref{eq:f0}, \eqref{eq:f1}, \eqref{eq:f2}~and \eqref{eq:phi2} an equilibrium distribution function.
In this section, we focus on the geometry described by Eq.~\eqref{eq:stationary_weak}.

Let us evaluate the charge current and the symmetric energy-momentum tensor for the equilibrium distribution function.
We employ the classical equilibrium state described by
\begin{eqnarray}
 &\dis f_\zero 
 = \frac{\theta(\beta\cdot p) }{e^{g_\zero}+1} 
 	+ \frac{\theta(-\beta\cdot p)}{e^{-g_\zero}+1} \,,\quad
 g_\zero 
  = -\beta \mu + \beta\cdot p \,,
\end{eqnarray}
where $\beta^\mu$ is a time-like Killing vector $\beta^\mu= \bar\beta\xi^\mu$ with $\bar\beta = \sqrt{\beta\cdot\beta/g_{00}}$.
Also $\bar \beta$ and $\bar\mu$ are the global inverse temperature and chemical potential.
In the flat spacetime limit, the classical charge density becomes
\begin{equation}
J^\mu_{\zero\eq}
 = 2\int_p \calR^\mu_{\zero\eq} 
 = \int_\bp \Bigl[n_F(|\bp|-\mu) - n_F(|\bp|+\mu)\Bigr] \,,
 \quad
  n_F(z) := \frac{1}{e^{\beta z}+1}  
\end{equation}
with $\int_\bp = \int d^3 p/(2\pi)^3$.

From Eq.~\eqref{eq:J-T}, the equilibrium Wigner function $\calR^\mu_{\one\eq}$ yields the CVE~\cite{Chen:2015gta}:
\begin{equation}
 \begin{split}
 \label{eq:JT1_eq}
   J^\mu_{\one\eq}
   = C_1 \omega^\mu \,,
 \quad 
 T_{\one\eq}^{\mu\nu}
  = 2C_2 \xi^{(\mu} \omega^{\nu)}\,,
 \end{split}
\end{equation}
where the vorticity vector is introduced as $\omega^\mu = \varepsilon^{\mu\nu\rho\sigma} \xi_\nu\nabla_\rho \xi_\sigma/2$.
Here the coefficients are defined as
\begin{equation}
 C_{n}
 := \frac{1}{2\pi^2}\int_0^\infty d\rho \rho^n 
 	\Bigl[
 		n_F(\rho-\mu) - (-1)^n n_F(\rho+\mu) 
 	\Bigr]\,,
\end{equation}
and thus we find $C_1 = \mu^2/(4\pi^2)+T^2/12$ and $C_2 =\mu^3/6(\pi^2)+\mu T^2/6$ (see Appendix~\ref{app:int_formulas}).
%Also we defined the local temperature $T = 1/(\bar\beta\sqrt{g_{00}})$, and the local chemical potential $\mu = \bar\mu /\sqrt{g_{00}}$, and the vorticity vector is $\omega^\mu = \varepsilon^{\mu\nu\rho\sigma} \beta_\nu\nabla_\rho \beta_\sigma/(2 \beta^\lambda\beta_\lambda) $.
The above charge current is conserved, namely, $\nabla_\mu J_{\one\eq}^\mu = 0$.
This reflects the absence of the gravitational contribution to the $U(1)$ anomaly in the CKT up to $O(\hbar)$.
Besides, the energy-momentum conservation holds, as we check $\nabla_\mu T_{\one\eq}^{\mu\nu} = 0$.

From Eq.~\eqref{eq:Rmu2_eq} and~\eqref{eq:phi2}, the second order equilibrium Wigner function reads
\begin{equation}
\begin{split}
\calR^\mu_{\two\eq}
 & = 2\pi\delta(p^2)
   	\Biggl[
   		f_\zero
 	 	\biggl(
 	 		- \frac{1}{2p^2} R^{\mu\alpha} p_\alpha
 	 		- \frac{1}{12p^2} R\, p^\mu
 	 		+ \frac{2}{3(p^2)^2} R^{\alpha\beta} p^\mu p_\alpha p_\beta 
 	 	\biggr) \\
 & \qquad\qquad
 	 	+ f'_\zero
 	 	\biggl(
 	 		-\frac{1}{24} R^{\mu\alpha} \beta_\alpha
 	 		+ \frac{1}{12 p^2} R^{\alpha\beta\gamma\mu} p_\alpha\beta_\beta p_\gamma
 	 	\biggr) \\
 & \qquad\qquad
   		+ f''_\zero
 		\biggl(
 			- \frac{1}{24} R^{\alpha\beta\gamma\mu} p_\alpha\beta_\beta\beta_\gamma
 			-\frac{1}{12 p^2} R_{\alpha\beta\gamma\delta} 
 				p^\mu p^\alpha p^\gamma \beta^\beta \beta^\delta
 			 + \frac{1}{24} R^{\alpha\beta} \beta_\alpha\beta_\beta
 		\biggr)
 	\Biggr] \,,
\end{split}
\end{equation}
where $R=g^{\mu\nu} R_{\mu\nu}$ is the Ricci scalar.
Here we dropped the terms including $\nabla_\rho\beta_\nu$ because they are of order $O(h_{\mu\nu}^2)$.
In the momentum integral, the $1/p^2$ terms can be rewritten as  
\begin{eqnarray}
 & \dis 
 \int_p \frac{\delta(p^2)}{p^2} p^\mu F(p) = \int_p \frac{1}{2} \delta(p^2)\partial^\mu_p F(p)\,,
\end{eqnarray}
which follows from $\delta'(x)=-\delta(x)/x$.
The integral with $1/(p^2)^2$ is also computed in a similar manner with $\delta''(x)=2\delta(x)/x^2$.
With the help of several formulas in Appendix~\ref{app:int_formulas}, we eventually derive
\begin{equation}
 \begin{split}
 \label{eq:JT2_eq}
  J^\mu_{\two\eq}
    & = C_0
 	\biggl[
 		\frac{1}{12} {R^\mu}_{\alpha}\xi^\alpha
 		- \frac{1}{24} \xi^\mu R
 		+ \frac{1}{6} \xi^\mu  R_{\alpha\beta}\xi^\alpha\xi^\beta
 	\biggr] \,, \\
 T_{\two\eq}^{\mu\nu}
  & = C_1
 	\biggl[
 		 - \frac{1}{12} R^{\mu\nu} 
 		 - \frac{1}{12} R \xi^\mu \xi^\nu
 		 + \frac{1}{24} R \eta^{\mu\nu} \\
  & \qquad\quad
 		 - \frac{1}{6}R^{\alpha(\mu} \xi^{\nu)} \xi_\alpha  
  		 +\frac{1}{6} R^{\alpha\beta} \xi_\alpha \xi_\beta
  		 	(4\xi^\mu\xi^\nu-\eta^{\mu\nu})
  		 + \frac{1}{6} R^{\mu\alpha\nu\beta} \xi_\alpha \xi_\beta
  	\biggr]
\end{split}
\end{equation}
with $C_0  = \mu/(2\pi^2)$.
Therefore, based on the CKT, we reveal that the nonvanishing fermion transport is induced by the background gravitational field even in equilibrium.
This implies that these phenomena are nondissipative, as so are the CME, and CVE.
This is one of the main findings of this paper.
We can also derive the same current $J^\mu_{\two\eq}$ from the diagrammatic computation (see Appendix~\ref{sec:diagram}) and with the Riemann normal coordinate expansion (see Appendix~\ref{sec:Riemann}).
It is worthwhile to mention that $g_{0i}$ enters in Eqs.~\eqref{eq:JT1_eq} and~\eqref{eq:JT2_eq} only through the field strength $f_{ij} = \partial_i g_{0j} - \partial_j g_{0i}$.
This is a consequence of the Kaluza-Klein gauge symmetry~\cite{Banerjee:2012iz}.

For the left-handed Weyl fermion, $J^\mu_{\two\eq}$ and $T^{\mu\nu}_{\two\eq}$ are written as the same form, while the sign of $J^\mu_{\one\eq}$ and $T^{\mu\nu}_{\one\eq}$ flipped; 
the former does not involve $\varepsilon^{\mu\nu\rho\sigma}$ while the latter does.
As a result, the vector and axial parts are written as
\begin{equation}
 \begin{split}
\label{eq:JT2_eq_VA}
  J^\mu_{\two\eq\,V/A}
    & = C_{0,V/A}
 	\biggl[
 		\frac{1}{12} {R^\mu}_{\alpha}\xi^\alpha
 		- \frac{1}{24} \xi^\mu R
 		+ \frac{1}{6} \xi^\mu  R_{\alpha\beta}\xi^\alpha\xi^\beta
 	\biggr] \,, \\
 T_{\two\eq\,V/A}^{\mu\nu}
  & = C_{1,V/A}
 	\biggl[
 		 - \frac{1}{12} R^{\mu\nu} 
 		 - \frac{1}{12} R \xi^\mu \xi^\nu
 		 + \frac{1}{24} R \eta^{\mu\nu} \\
  & \qquad\quad
 		 - \frac{1}{6}R^{\alpha(\mu} \xi^{\nu)} \xi_\alpha  
  		 +\frac{1}{6} R^{\alpha\beta} \xi_\alpha \xi_\beta
  		 	(4\xi^\mu\xi^\nu-\eta^{\mu\nu})
  		 + \frac{1}{6} R^{\mu\alpha\nu\beta} \xi_\alpha \xi_\beta
  	\biggr]
\end{split}
\end{equation}
where $C_{0,\,V/A} = \mu_{V/A}/\pi^2$, $C_{1,\,V}=(\mu_V^2 + \mu_A^2)/2\pi^2 + T^2/6$ and $C_{1,\,A} = \mu_V\mu_A/\pi^2$, with $\mu_V$ and $\mu_A$ being the vector and chiral chemical potential, respectively.
In Sec.~\ref{sec:fluid}, we argue the novelty and some implications of Eq.~\eqref{eq:JT2_eq} and~\eqref{eq:JT2_eq_VA}.

\section{Dynamical weak gravity}\label{sec:dynamical}
While so far we have focused on the equilibrium state, this section is dedicated to discuss the dynamical response from the time-dependent gravity.
Specifically, we consider a plane-wave weak background gravitational field:
\begin{equation}
\label{eq:dynamical_weak}
 g_{\mu\nu}=\eta_{\mu\nu}+h_{\mu\nu}\,,
 \quad
 h_{\mu\nu} \sim e^{-ik\cdot x} \,,
 \quad
 |h_{\mu\nu}|\ll 1 \,,
\end{equation}
where $k^\mu = (k_0,\bk)$ is the momentum of the gravitational field. 
Let us look for the perturbative distribution function represented as the following form:
\begin{equation}
\label{eq:f-tildef}
 f = f_\text{\flat} + \tf\,,
 \quad
 f_\text{flat}= \frac{\theta(p_0)}{e^{\beta(p_0-\mu)}+1} + \frac{\theta(-p_0)}{e^{-\beta(p_0-\mu)}+1} \,,\quad
 \tf
 = \tf_\zero + \hbar\tf_\one + \hbar^2\tf_\two \,.
\end{equation}
Here $\beta$ and $\mu$ are constant, and thus $f_\text{flat}$ is the static and homogeneous solution of the collisionless Boltzmann equation~\eqref{eq:Vlasov} for $h_{\mu\nu} = 0$.
We define $\tf$ as the fluctuation around $f_\text{flat}$.
For $h_{\mu\nu} \sim e^{-ik\cdot x}$, we may employ the anzatz $\tf \sim e^{-ik\cdot x}$.
For simplicity, we further assume $\partial_\mu n^\nu = 0$.

We first compute the classical and leading order parts.
Plugging the general form of $\calR^\mu_{\zero,\one}$ in Eqs.~\eqref{eq:Rmu0} and~\eqref{eq:Rmu1} into Eq.~\eqref{eq:R1-curved}, we write down the kinetic equation as
\begin{equation}
\begin{split}
 \delta(p^2)
 \biggl[
 	p\cdot D
 	+ \hbar (D_\mu \Sigma_n^{\mu\nu}) D_\nu
 	- \frac{\hbar}{2} \Sigma^{\mu\nu}_n R_{\alpha\beta\mu\nu} p^\alpha \partial_p^\beta
 \biggr]
 	f = 0 \,.
\end{split}
\end{equation}
Expanding the above equation in terms of $h_{\mu\nu}$ and utilizing $\partial_\mu f_\flat = 0$, we obtain
\begin{equation}
 p\cdot \partial \tf
 + 
 	\biggl[
 		\Gamma^\rho_{\mu\nu} p^\mu p_\rho \beta^\nu 
 		- \frac{\hbar}{2} \Sigma^{\mu\nu}_n R_{\alpha\beta\mu\nu} p^\alpha \beta^\beta
 	\biggr] f'_\flat
 = 0 \,,
\end{equation}
where we denote $\beta^\mu = \beta \xi^\mu = \beta\delta^\mu_0$.
Note that after the weak gravitational field expansion, all indices are raised and lowered by $\eta_{\mu\nu}$ and the inner products are defined as $A\cdot B = \eta_{\mu\nu} A^\mu B^\nu$ and $A^2 = \eta_{\mu\nu}A^\mu A^\nu$.
Especially, to get the above equation, we have taken the following replacement:
\begin{eqnarray}
  \dis p^\mu 
 &\to &  g^{\mu\nu}p_\nu \simeq p^\mu - h^{\mu\nu} p_\nu \,, \label{eq:p} \\
  \dis
  \delta(p^2) 
 &\to & \delta(g^{\mu\nu}p_\mu p_\nu)  
 \simeq
%  \delta(p^2) + \delta'(p^2) (-h^{\mu\nu}p_\mu p_\nu)
  \delta(p^2)
  	\biggl(
 		1
 		+ \frac{1}{p^2} h^{\mu\nu} p_\mu p_\nu
 	\biggr) \,,
\end{eqnarray}
which follow from $g^{\mu\nu} \simeq \eta^{\mu\nu} - h^{\mu\nu}$ and $\delta'(x) = - \delta(x)/x$.
For $ h_{\mu\nu}\sim e^{-ik\cdot x}$, the fluctuations $\tf_{\zero,\one} \sim e^{-ik\cdot x}$ are found to be
\begin{eqnarray}
\label{eq:tildef0}
 &\dis \tf_{\zero}
 = \frac{1}{ik\cdot p}
 		\Gamma^\rho_{\mu\nu} p^\mu p_\rho \beta^\nu f'_\flat \,,\\
\label{eq:tildef1}
 &\dis \tf_{\one}
 = -\frac{1}{2ik\cdot p}\Sigma^{\mu\nu}_n {R}_{\alpha\beta\mu\nu} p^\alpha \beta^\beta f'_\flat \,,
\end{eqnarray}
with the linearized Christoffel symbol and Riemann tensor being
\begin{equation}
\begin{split}
 {\Gamma}_{\mu\lambda}^{\rho} 
 &\simeq\frac{-i}{2}
 	(
 		k_{\mu}h_{\lambda}^{\rho}
 		+k_{\lambda}h_{\mu}^{\rho}
 		-k^{\rho}h_{\mu\lambda}
 	)\,,
 	\quad\\
 {R}^{\rho}_{~\lambda\mu\nu}
 &\simeq
	\frac{(-i)^2}{2}
	(
		k_{\nu}k_{\lambda}h_{\mu}^{\rho}
		-k_{\nu}k^{\rho}h_{\mu\lambda} 
		-k_{\mu}k_{\lambda}h_{\nu}^{\rho}
		+k_{\mu}k^{\rho}h_{\nu\lambda} 
	)\,.
\end{split}
\end{equation}
%Then we find
%\begin{equation}
%\label{eq:fdiff1}
% \tf_{\one n'} - \tf_{\one n}
%% = \frac{\hbar}{ik\cdot p}
%% 		(\partial_\mu \Sigma_{n'}^{\mu\nu}-\partial_\mu \Sigma_n^{\mu\nu})
%% 			{\Gamma}^{\lambda}_{\nu\eta} p_\lambda \beta^\eta f_\flat'
% = \frac{i}{2} (\Sigma_{n'}^{\mu\nu} - \Sigma_{n}^{\mu\nu})
% 	\biggl(
% 		- \frac{k\cdot\beta}{k\cdot p} h_{\mu\lambda} k_\nu p^\lambda
% 		+ h_{\mu\lambda} k_\nu \beta^\lambda
% 	\biggr) f_\flat' \,.
%\end{equation}

Plugging the above distribution functions into Eqs.~\eqref{eq:Rmu0} and~\eqref{eq:Rmu1}, we get the Wigner function.
%\begin{equation}
%\begin{split}
%\calR^\mu_{\zero} 
% & = 2\pi\delta(p^2)
% 	\biggl[
% 		 p^\mu \biggl(1+\frac{1}{p^2} h^{\alpha\beta} p_\alpha p_\beta \biggr )f_\flat
% 		 - h^{\mu\nu} p_\nu f_\flat 
% 		+ \frac{1}{i k\cdot p} \Gamma^\rho_{\lambda\nu} p^\mu p^\lambda p_\rho \beta^\nu f_\flat'
% 	\biggr] \,.
%\end{split}
%\end{equation}
It is here informative to decompose the Wigner function into the terms that involve the $k$-dependent pole in the denominator, and the others.
As we show in Sec.~\ref{sec:dynamical_JT}, the momentum integrals of the former vanishes in the static limit $k_0/|\bk|\to 0$, while those of the latter survives.
In this sense, we denote such a (non)static part as $\calR^\mu_{\mathrm{(non)}\st}$.
We note that the static part $\calR^\mu_{\st}$ reproduces the equilibrium Wigner function $\calR^\mu_\eq$ in the previous section, as we show later.

For the classical $O(\hbar^0)$ part, Eq.~\eqref{eq:tildef0} leads to $\calR^\mu_{\zero}=\calR^\mu_{\zero\st}+\calR^\mu_{\zero\nonst}$ with
\begin{eqnarray}
\label{eq:R0_st}
\calR^\mu_{\zero\st}
 &=& 2\pi\delta(p^2)
 	\biggl[
 		 p^\mu \biggl(1+\frac{1}{p^2} h^{\alpha\beta} p_\alpha p_\beta \biggr )f_\flat
 		 - h^{\mu\nu} p_\nu f_\flat 
 	\biggr] \,,\\
\label{eq:R0_nonst}
\calR^\mu_{\zero\nonst}
 &=& 2\pi\delta(p^2)\frac{1}{i k\cdot p} \Gamma^\rho_{\lambda\nu} p^\mu p^\lambda p_\rho \beta^\nu f_\flat'\,,
\end{eqnarray}
where we use $g^{\mu\nu} \simeq \eta^{\mu\nu} - h^{\mu\nu}$ and $\delta(g^{\alpha\beta}p_\alpha p_\beta) \simeq \delta(p^2) (1+h^{\alpha\beta} p_\alpha p_\beta/p^2)$.
Similarly, Eqs.~\eqref{eq:tildef0} and~\eqref{eq:tildef1} yield to the $O(\hbar)$ part as
\begin{equation}
\begin{split}
 \calR^\mu_\one
 & = 2\pi\delta(p^2) 
 		 \Bigl(
 		 	p^\mu \tf_{\one} 
 		 	+ \Sigma_n^{\mu\nu} D_\nu 
 		 		(f_\flat + \tf_\zero)
 		 \Bigr) \\
 & = -\frac{2\pi\delta(p^2) }{4}\frac{1}{ik\cdot p}
 		 \varepsilon^{\mu\eta\nu\lambda} p_\eta R_{\rho\sigma\lambda\nu} p^\rho \beta^\sigma f_\flat' \, ,
\end{split}
\end{equation}
where we used Eq.~\eqref{eq:Sigma_p} to remove $\Sigma_n^{\mu\nu}$.
We again stress that while $\tf_\one$ is the frame-dependent, the Wigner function $\calR^\mu_\one$ is irrelevant to the frame.
Then, the Wigner function is represented as $\calR^\mu_\one = \calR^\mu_{\one\st} + \calR^\mu_{\one\nonst}$ with
\begin{eqnarray}
\label{eq:R1_st}
 \dis
 \calR^\mu_{\one\st}
 &=& -\frac{2\pi\delta(p^2)}{4} \varepsilon^{\mu\nu\rho\sigma} p_\nu (-ik_\rho) 
 		h_{\sigma\lambda} \beta^\lambda f_\flat' \,,\\
\label{eq:R1_nonst}
 \dis
 \calR^\mu_{\one\nonst}
 &=& \frac{2\pi\delta(p^2)}{4} \varepsilon^{\mu\nu\rho\sigma} p_\nu (-ik_\rho) 
 		\frac{k\cdot\beta}{k\cdot p} h_{\sigma\lambda} p^\lambda
 			f_\flat' \,.
\end{eqnarray}
We observe that the above $\calR^\mu_{\one\st}$ is consistent with the equilibrium Wigner function $\calR^\mu_{\one\eq}$ in Eq.~\eqref{eq:Rmu1_eq}.

Applying the totally same manner to the $O(\hbar^2)$ parts, we obtain $\tilde{f}_2$ and eventually the Wigner function as $\calR^{\mu}_{\two} = \calR^{\mu}_{\two\st} + \calR^{\mu}_{\two\nonst}$ with
\begin{equation}
\begin{split}
\label{eq:Rmu2_st}
 \calR^{\mu}_{\two\st}
 & = 2\pi\delta(p^2)
 	 \biggl[
 	 	 -\frac{1}{4p^2} 
  			{R_{\alpha\beta}}^{\mu\nu} p_\nu p^\alpha \beta^\beta f_\flat'  
  		 + \frac{1}{4p^2} 
  			{R_{\beta}}^{\nu} p^\mu  p_\nu \beta^\beta f_\flat'
  	 	- \frac{1}{4} {R_{\beta}}^{\mu}  \beta^\beta f_\flat' \\
 & \quad
 		 + \frac{1}{24}p^\mu R_{\alpha\beta} \beta^\alpha \beta^\beta  f_\flat''
 	 \biggr] 
 	 + \frac{2\pi}{p^2}
    		 \Bigl[
	 	    	 - p^\mu Q\cdot p 
    			  + 2 p_\nu 
    			  	\bigl( 
    			  		T^{[\mu} p^{\nu]} 
    			  		+ S^{\alpha\mu\nu} p_\alpha 
    			  	\bigr) 
    		\Bigr] \delta(p^2) f_\flat \,,
\end{split}
\end{equation}
\begin{equation}
\begin{split}
\label{eq:Rmu2_nonst}
 \calR^{\mu}_{\two\nonst}
 & = 2\pi\delta(p^2)\frac{k\cdot \beta}{k\cdot p} 
 	\biggl[
		- \frac{1}{4p^2} 
  			{R_{\alpha}}^{\nu} p^\mu  p_\nu p^\alpha f_\flat'  
  		+ \frac{1}{4} 
  			{R_{\alpha}}^{\mu} p^\alpha f_\flat' 		
 		+ \frac{1}{24} p^\mu  p^\eta p^\rho \beta^\nu \beta^\sigma 
 			 R_{\rho\sigma\eta\nu}  f_\flat'''	
 	 \biggr] \,,
\end{split}
\end{equation}
for which the precise derivation is shown in Appendix~\ref{app:dynamical_Wigner}.
Again $\calR^\mu_{\two\st}$ is the same as $\calR^\mu_{\two\eq}$ in Eq.~\eqref{eq:Rmu2_eq} up to $O(h_{\mu\nu})$.

\section{Dynamical response}\label{sec:dynamical_JT}
In the following discussion, we evaluate the charge current $J^\mu$ and the energy-momentum tensor $T^{\mu\nu}$ in Eq.~\eqref{eq:J-T} with Eqs.~\eqref{eq:Rmu2_st}, and~\eqref{eq:Rmu2_nonst}.
As the Wigner function $\calR^\mu$ is decomposed into the static ($k$-independent) and nonstatic ($k$-dependent) part, so are $J^\mu$ and $T^{\mu\nu}$, that is, $J^\mu = J^\mu_{\st} + J^\mu_{\nonst}$ and $T^{\mu\nu} = T^{\mu\nu}_{\st} + T^{\mu\nu}_{\nonst}$.
The static part is calculated in the same manner as before.
For instance, using the integral formulas in Appendix~\ref{app:int_formulas}, the momentum integrals of the classical contribution~\eqref{eq:R0_st} yield 
\begin{equation}
\begin{split}
\label{eq:JT0_st}
 J^\mu_{\zero\st}
  & = C_2 \xi^\mu( 1- 2 \xi^\alpha \xi^\beta h_{\alpha\beta} ) \,,\\
 T^{\mu\nu}_{\zero\st}
 & =  C_3  
 		\biggl[
 			 \frac{4}{3} \xi^\mu\xi^\nu 
 			 -\frac{1}{3} \eta^{\mu\nu} 
 	 		 +
 	 		\biggl(
 	 			\frac{1}{3} \eta^{\mu\alpha}\eta^{\nu\beta}
 	 			 - 4 \xi^\mu \xi^\nu \xi^\alpha \xi^\beta
 	  		 	+ \frac{2}{3} \eta^{\mu\nu} \xi^\alpha \xi^\beta
 	  		\biggr) h_{\alpha\beta}
 	 	\biggr],
\end{split}
\end{equation}
with $C_3  = \mu^4/24\pi^2 + \mu^2 T^2/12$.
Here we use $(-g)^{-1/2} \simeq 1 -  h^\mu_\mu/2$, and perform the integration by part.

For the nonstatic part, it is helpful to additionally prepare the following tensor (scalar for $n=0$) function:
\begin{eqnarray}
  I^{j_1\cdots j_n} (x)
  := x \int \frac{d\Omega}{4\pi} 
 	\frac{\hp^{j_1} \cdots \hp^{j_n}}{x-\hbk\cdot\hbp} \,,
\end{eqnarray} 
where we define $x := k_0/|\bk|$ and the integral is over the solid angle of $\bp$.
The evaluations of $I^{j_1\cdots j_n} (x)$'s are summarized in Appendix~\ref{app:angle_int_formulas}.
Here $x$ in the denominators is understood to involve the positive infinitesimal imaginary part $+i\eta$.
The nonstatic part of the classical charge current is from Eq.~\eqref{eq:R0_nonst} evaluated as
\begin{equation}
\begin{split}
 J^\mu_{\zero\nonst}
 & = - k\cdot\beta h_{\lambda\rho} \int_p 2\pi\delta(p^2) 
 	 	\frac{1}{k\cdot p} 
 	 		 p^\mu p^\lambda p^\rho
 	 		 f_\flat' \\
 & = \frac{3}{2} C_2 
 		\biggl[
 	 	 \xi^\mu
 	 		 \Bigl(
 	 		 	 h_{00} I
 	 		 	 + h_{jk} I^{jk}
 	 			 + 2 h_{0j} I^{j}
 	 		 \Bigr)
 		+ \delta^\mu_i 
			\Bigl(
 	 		 	 h_{00} I^i
 	 		 	+  h_{jk} I^{ijk}
 	 			 + 2 h_{0j} I^{ij} 
 	 		\Bigr)
 	 	\biggr]\,.
\end{split}
\end{equation}
To obtain the above second line, we utilized $I^{j_1\cdots j_n}(-x) = (-1)^n I^{j_1\cdots j_n}(x)$ and
\begin{equation}
\begin{split}
 & 2 k\cdot\beta \int \frac{dp_0}{2\pi} 2\pi\delta(p^2) 
 	\frac{(p_0)^n }{k\cdot p} \frac{d^m}{dp_0^m} f_\flat \\
 & = \beta |\bp|^{n-2}
 	\Biggl[
 		\frac{x}{x - \hbk\cdot\hbp + i\eta} n_F^{(m)}(|\bp|-\mu)
 		+ (-1)^{n+m+1} \frac{-x}{-x - \hbk\cdot\hbp - i\eta} n_F^{(m)}(|\bp|+\mu)
 	\Biggr]
\end{split}
\end{equation}
with $n_F({y}) = (e^{\beta {y}} + 1)^{-1}$ and $n_F^{(m)}(y):={d^m n_F(y)/dy^m}$.
Besides, the nonstatic part of the classical energy-momentum tensor is computed as
\begin{equation}
\begin{split}
 T^{\mu\nu}_{\zero\nonst}
 & = -k\cdot\beta h_{\rho\lambda}
 	 \int_p 2\pi\delta(p^2)\frac{1}{k\cdot p}  p^\mu p^\nu p^\rho p^\lambda f_\flat' \\
 & = 2C_3 
 	\biggl[
 		\xi^\mu \xi^\nu
 			\Bigl(
 	 		 	  h_{00} I 
 	 		 	  + 2 h_{0j} I^{j} 
 	 		 	  + h_{jk} I^{jk} 
 	 		\Bigr) 
 		+ 2\delta^{(\mu}_i \delta^{\nu)}_0
			\Bigl(
 	 		 	 h_{00} I^i 
 	 			 + 2 h_{0j} I^{ij} 
 	 		 	 + h_{jk} I^{ijk} 
 	 		\Bigr) \\
 & \qquad\qquad
 		+ \delta^{\mu}_i \delta^{\nu}_j
 			\Bigl(
 	 		 	 h_{00} I^{ij} 
 	 			 + 2 h_{0k} I^{ijk} 
 	 		 	 + h_{kl} I^{ijkl} 
 	 		\Bigr)
 	 \biggr] \,.
\end{split}
\end{equation}

It is worthwhile to notice several properties of the above $I_{j_1\cdots j_n}(x)$.
First we find the following relations:
\begin{eqnarray}
\label{eq:Ix0Ix1}
 &\dis k_0 I  + k_k I^k  = k_0\,, \\
\label{eq:Ix1Ix2}
 &\dis k_0 I^k  + k_j I^{jk}  = 0 \,, \\
\label{eq:Ix2Ix3}
 &\dis k_0 I^{jk}  + k_i I^{ijk}  = k_0\frac{\delta^{jk}}{3} \,, \\
\label{eq:Ix3Ix4}
 &\dis k_0 I^{jkl}  + k_i I^{ijkl}  = 0 \,.
\end{eqnarray}
From these, we can show the charge current and energy-momentum conservation for arbitrary $k^\mu$:
\begin{equation}
\begin{split}
\label{eq:JT0_conserve}
  \nabla_\mu J^\mu_{\zero}  
  &= \nabla_\mu ( J^\mu_{\zero\st} + J^\mu_{\zero\nonst} ) = 0\,, \\
  \nabla_\mu T^{\mu\nu}_\zero
  &= \nabla_\mu ( T^{\mu\nu}_{\zero\st} + T^{\mu\nu}_{\zero\nonst} ) = 0 \,.
\end{split}
\end{equation}
Second, we check that $I^{j_1\cdots j_n}(x)$ fulfills another type of relations:
\begin{eqnarray}
\label{eq:Ix2trace}
 & I  + {I^j}_j   = 0 \,, \\
\label{eq:Ix3trace}
 & I^k  + {I_{j}}^{jk}  = 0 \,, \\ 
\label{eq:Ix4trace}
 & I^{kl}   + {I_{j}}^{jkl}  = 0 \,.
\end{eqnarray}
These bring the dilatation current conservation for arbitrary $k^\mu$:
\begin{equation}
\label{eq:T0traceless}
 g_{\mu\nu} T^{\mu\nu}_\zero
 = g_{\mu\nu}  ( T^{\mu\nu}_{\zero\st} + T^{\mu\nu}_{\zero\nonst} )
 = 0 \,.
\end{equation}

As a particular case, we consider the dynamical limit $x = k_0/|\bk|\gg 1$.
We expand $J^\mu_{\zero\nonst}$ and $T^{\mu\nu}_{\zero\nonst}$ in terms of $1/x$, with the asymptotic forms of $I_{j_1\cdots j_n}(x)$'s, which are derived in Eqs.~\eqref{eq:Ix0_largex}-\eqref{eq:Ix4_largex}.
For later convenience, we here define the total charge current and energy-momentum tensor in the dynamical limit, as follows:
\begin{equation}
\label{eq:JT_dyn}
 J^\mu_\dyn := J^\mu_{\st} + J^\mu_{\nonst}\bigl|_{x\to \infty}\,,
 \quad
 T^{\mu\nu}_\dyn := T^{\mu\nu}_{\st} + T^{\mu\nu}_{\nonst}\bigl|_{x\to \infty} \,.
\end{equation}
Their classical contributions hence become
\begin{equation}
 \begin{split}
 \label{eq:JT0}
 J^\mu_{\zero\dyn}
 &=
 	C_2
  	\biggl[
  		\xi^\mu 
  		\biggl( 
  			1 
  			- \frac{1}{2} {h^\lambda}_\lambda
  		\biggr)
 		- \delta^\mu_i h^i_{0}
 	\biggr] \,,\\
   T^{\mu\nu}_{\zero\dyn}
  & =
   C_3  
 		\biggl[
 			\frac{1}{3} 
 			 \Bigl(
 			  	4\xi^\mu\xi^\nu 
 				-\eta^{\mu\nu} 
 			 \Bigr)
 	 		 +
 	 		\biggl(
 	 			\frac{3}{5} \eta^{\mu\alpha}\eta^{\nu\beta}
 	 			 + \frac{2}{15} \eta^{\mu\nu}\eta^{\alpha\beta}  \\
  & \qquad\quad
 	 			 + \frac{12}{5} \xi^\mu \xi^\nu \xi^\alpha \xi^\beta
 	  		 	- \frac{4}{5}\eta^{\alpha\beta} \xi^\mu\xi^\nu 
 	  		 	-\frac{2}{15} \xi^\alpha\xi^\beta \eta^{\mu\nu}
 	  		 	 - \frac{16}{5} \eta^{\alpha(\mu} \xi^{\nu)} \xi^\beta 
 	  		\biggr) h_{\alpha\beta}
 	 	\biggr] \,.
 \end{split}
\end{equation}

Let us also calculate quantum corrections.
At $O(\hbar)$, the Wigner function~\eqref{eq:R1_nonst} leads to
\begin{equation}
\begin{split}
 J^\mu_{\one\nonst}
 & =
  - C_1  \omega^\mu I
 	 -\frac{1}{2} C_1  \varepsilon^{\mu\nu\rho\sigma} (-ik_\rho) h_{\sigma}^\lambda
 		\biggl[
 				\Bigl(
 					\xi_\nu \delta^k_\lambda 
 					+ \delta^k_\nu \xi_\lambda 
 				\Bigr) I_{k}
 				+ \delta^j_\nu \delta^k_\lambda I_{jk}
 		\biggr]  \,, \\
 T^{\mu\nu}_{\one\nonst}
 & = - \frac{3C_2}{2} \xi^{(\mu}\omega^{\nu)} I
 	-\frac{3C_2}{4} 
 	\biggl[
 		\xi^{(\mu} \varepsilon^{\nu)\eta\rho\sigma}(-ik_\rho) h_{\sigma}^\lambda
 			\Bigl(
 				(\delta_\eta^k \xi_\lambda + \delta_\lambda^k \xi_\eta) I_k
 				+ \delta_\eta^j \delta_\lambda^k I_{jk}
 			\Bigr) \\
 & \qquad
 		+ \eta^{i(\mu}  \varepsilon^{\nu)\eta\rho\sigma}(-ik_\rho) h_{\sigma}^\lambda
 			\Bigl(
 				\xi_\eta \xi_\lambda I_i
 				+ (\delta_\eta^k \xi_\lambda + \delta_\lambda^k \xi_\eta) I_{ik}
 				+ \delta_\eta^j \delta_\lambda^k I_{ijk}
 			\Bigr)
 	\biggr] \,,
\end{split}
\end{equation}
where the vorticity is linearized as
\begin{equation}
\omega^\mu 
	= \frac{1}{2}\varepsilon^{\mu\nu\rho\sigma} \xi_\nu\nabla_\rho \xi_\sigma 
	\simeq  \frac{1}{2}\varepsilon^{\mu\nu\rho\sigma} \xi_\nu (-ik_\rho) h_{\sigma\lambda}\xi^\lambda  \,.
\end{equation}
In particular, taking the dynamical limit, we find
\begin{equation}
\begin{split}
 J^\mu_{\one\nonst}\bigl|_{x\to \infty}
 & = - C_1 \omega^\mu \,, \\
 T^{\mu\nu}_{\one\nonst}\bigl|_{x\to \infty}
 & = - 2 C_2 \xi^{(\mu} \omega^{\nu)} 
 	+ \frac{C_2}{5} \delta^{(\mu}_k \varepsilon^{\nu) 0lm}  (-ik_l) h^k_{m}\,. \\
\end{split}
\end{equation}
These results, combined with the static parts~\eqref{eq:JT1_eq}, yield the $O(\hbar)$ contributions of Eq.~\eqref{eq:JT_dyn}, as follows:
\begin{equation}
\begin{split}
\label{eq:JT1_dyn}
J^\mu_{\one\dyn} 
 = 0\,, \quad
T^{0\mu}_{\one\dyn}
 = 0 \,.
\end{split}
\end{equation}
Therefore we conclude that the CVE vanishes in the dynamical limit.
This is consistent with the diagrammatic calculation in Ref.~\cite{Landsteiner:2013aba}.

At $O(\hbar^2)$, from the Wigner function~\eqref{eq:Rmu2_nonst}, the nonstatic parts are evaluated as
\begin{equation}
\begin{split}
\label{eq:JT2_nonst}
 J^{\mu}_{\two\nonst}
 & = -\frac{C_0}{4}
 	\biggl[
 		\frac{1}{2} {R^{\mu}}_\nu 
 		\biggl(
 			\xi^\nu I  
 			+ \delta^\nu_k I^k 
 		\biggr) 
 	 - \frac{1}{4} R
 		\biggl(
 			\xi^\mu I 
 			+ \delta^\mu_k I^k 
 		\biggr) \\
 & \qquad
 	 + R_{\alpha 0}
 		\biggl(
 			\xi^\mu \xi^\alpha I  
 			+ (\xi^\mu \delta^\alpha_k + \xi^\alpha \delta^\mu_k) I^k 
 			+ \delta^\mu_j \delta^\alpha_k I^{jk} 
 		\biggr) 
 	+ R_{j0k0}
 		\biggl(
 			\xi^\mu I^{jk} 
 			+ \delta^\mu_i I^{ijk} 
 		\biggr)
 	\biggr] \,,\\
 T^{\mu\nu}_{\two\nonst}
 & = (-2C_1)
 	\biggl[
 	- \frac{1}{16} R
 	\biggl(
 		 \xi^\mu \xi^\nu I 
 		+ 2 \xi^{(\mu} \delta^{\nu)}_k  I^k 
 		+ \delta^{\mu}_j \delta^{\nu}_k I^{jk} 
 	\biggr) \\
 & \quad
 	+ \frac{3}{8} R_{0\alpha}
 	\biggl(
 		 \xi^\mu \xi^\nu \xi^\alpha I 
 		+ (2 \xi^{(\mu} \delta^{\nu)}_k \xi^\alpha + \xi^\mu \xi^\nu \delta^\alpha_k)
 		 I^k 
 		+ (2\delta^{(\mu}_j \xi^{\nu)}\delta^{\alpha}_k  + \delta^{\mu}_j \delta^{\nu}_k \xi^\alpha )
 		 I^{jk}  \\
 & \qquad
 		+ \delta^{\mu}_i \delta^{\nu}_j \delta^\alpha_k
 		 I^{ijk} 
 	\biggr)
 	+ \frac{1}{2} R_{k0l0}
 	\biggl(
 		\xi^\mu \xi^\nu I^{kl} 
 		+2 \xi^{(\mu} \delta^{\nu)}_j I^{jkl} 
 		+ \delta^{\mu}_i \delta^{\nu}_jI^{ijkl} 
 	\biggr) 
 	\biggr]\,.
\end{split}
\end{equation}
In the dynamical limit $x\to \infty$, we obtain
\begin{equation}
\begin{split}
 J^\mu_{\two\nonst}\bigl|_{x\to \infty} 
 & = C_0
 	\biggl[
 		 -\frac{1}{6} \xi^\mu R_{00} 
		  + \frac{1}{60}
 	   		\biggl(
 	   			- 2 {R_0}^\mu 
 	   			+ \xi^\mu  R
 	   		\biggr)
 	\biggr] \,, \\
 T^{\mu\nu}_{\two\nonst}\bigl|_{x\to \infty}
 & = C_1
 	\biggl[
 		\xi^\mu\xi^\nu
 		\biggl(
 			\frac{13}{140} R
 			- \frac{24}{35} R_{00}
 		\biggr) 
 		+ \frac{7}{30} \xi^{(\mu} {R_0}^{\nu)}  \\
& \qquad\qquad
 		- \eta^{\mu\nu}
 		\biggl(
 			\frac{11}{420} R
 			- \frac{16}{105} R_{00} 
 		\biggr)
 		- \frac{2}{15} R^{\mu0\nu0}
 	\biggr] \,.
\end{split}
\end{equation}
Here we used
\begin{equation}
 \frac{1}{x}R_{0j0k} \hk^k
 = \frac{k_j}{k_0} R_{00} - R_{j0}\,,
 \quad 
 \frac{1}{x^2} R_{0j0k}\hk^j\hk^k
 = \biggl(1- \frac{1}{x^2}\biggr) R_{00}-\frac{1}{2} R \,,
 \quad
  R 
 = 2 R_{00} + 2 \frac{k_k}{k_0} {R_0}^k  \,,
\end{equation}
which follow from the second Bianchi identity.
Combining these with the static contribution~\eqref{eq:JT2_eq}, we write the $O(\hbar^2)$ contributions of Eq.~\eqref{eq:JT_dyn} as
\begin{equation}
\begin{split}
\label{eq:JT2_dyn}
 J^\mu_{\two\dyn}
 & = \frac{C_0}{20}
 	\biggl[
 		{R^\mu}_\alpha \xi^\alpha
 		- \frac{1}{2}\xi^\mu R
 	\biggr] \,,\\
 T^{\mu\nu}_{\two\dyn}
 & =
 C_1
 	\biggl[
 		 - \frac{1}{12} R^{\mu\nu} 
 		 + \frac{1}{105} R \xi^\mu \xi^\nu
 		 + \frac{13}{840} R \eta^{\mu\nu} 
 		 + \frac{1}{15}R^{\alpha(\mu} \xi^{\nu)} \xi_\alpha  \\
  & \qquad\quad
  		 -\frac{2}{105} R^{\alpha\beta} \xi_\alpha \xi_\beta \xi^\mu\xi^\nu
  		 -\frac{1}{70} R^{\alpha\beta} \xi_\alpha \xi_\beta \eta^{\mu\nu}
  		 + \frac{1}{30} R^{\mu\alpha\nu\beta} \xi_\alpha \xi_\beta
 	\biggr] \,.
\end{split}
\end{equation}
We note that Eqs.~\eqref{eq:Ix0Ix1}-\eqref{eq:Ix3Ix4} and Eqs.~\eqref{eq:Ix2trace}-\eqref{eq:Ix4trace} again result in the conservation laws $\nabla_\mu J^\mu_{\one}=\nabla_\mu J^\mu_{\two}=0$, $\nabla_\mu T^{\mu\nu}_{\one}=\nabla_\mu T^{\mu\nu}_{\two}=0$ and ${T^{\mu}}_{\mu\one} ={T^{\mu}}_{\mu\two} = 0$ for arbitrary $k^\mu$.
In the next section, we discuss some implications of \eqref{eq:JT2_dyn} in the fluid frame [see Eq.~\eqref{eq:gravito_dyn}].

\section{Fluid frame}\label{sec:fluid}
The fermionic system under a background fluid is pedagogical and informative to show the novelty of the gravity-induced transport phenomena given by Eqs.~\eqref{eq:JT2_eq} and~\eqref{eq:JT2_dyn}.
In general, the effect of the fluid can be translated to that of an effective curved geometry with the following metric~\cite{Banerjee:2012iz}:
%Let us now discuss several implications from the results that we have found in the previous sections.
%Here we adopt the following metric tensor:
\begin{equation}
 \label{eq:metric}
  g_{00} = 1 + h_{00}(t,\bx)\,,\quad
  g_{0i} = h_{0i}(t,\bx) \,,\quad
  g_{ij} = \eta_{ij} \,.
\end{equation}
Adopting this metric, we represent the temperature gradient~\cite{Luttinger:1964zz} and the fluid vorticity as
\begin{equation}
 \partial_i T/ \bar{T} = - \frac{1}{2} \partial_i g_{00}\,,
 \quad
 \omega^i = -\frac{1}{2} \varepsilon^{0ijk}\partial_j g_{k0} \,.
\end{equation}
with $\bar{T}$ being the global temperature.
Alternatively, the present coordinate describes the system under the gravitoelectromagnetic fields
$\calE^i = -\frac{1}{2}\partial^i g_{00}$ and $\calB^i = -\frac{1}{2} \varepsilon^{ijk}\partial_j g_{k0}$.
The nonvanishing components of the curvature tensors read
\begin{equation}
\label{eq:tensor_fluid}
 R_{i0j0} = R_{ij}
 = -\frac{\partial_i\partial_j T}{\bar{T}} 
 	- \partial_0 \epsilon_{ij} \,,\quad
 R_{00}
 = \frac{1}{2} R
 = \frac{\bnabla^2 T}{\bar{T}} -\partial_{0}{\epsilon_j}^j \,,\quad
 R_{0i} 
 = (\bnabla\times \bomega)_i \,, 
\end{equation}
with $\epsilon_{ij} = \frac{1}{2}(\partial_i  h_{0 j} + \partial_j h_{0 i})$.

Inserting Eq.~\eqref{eq:tensor_fluid} into Eqs.~\eqref{eq:JT2_eq} and~\eqref{eq:JT2_dyn}, we readily obtain the transport phenomena under the temperature gradient and inhomogeneous vorticity.
In the static limit (or equivalently, for the stationary metric $\partial_0 h_{\mu\nu}=0$), Eq.~\eqref{eq:JT2_eq} is written explicitly as
\begin{eqnarray}
\label{eq:gravito_eq}
 &\dis J^0_{\two\eq}
 = \frac{C_0}{6}
 	\frac{\bnabla^2 T}{\bar T} \,, \quad
 J^i_{\two\eq}
 = \frac{C_0}{12} (\bnabla\times\bomega)^i \,, \notag \\
 &\dis T^{00}_{\two\eq}
 = \frac{C_1}{6} \frac{\bnabla^2 T}{\bar T} \,,\quad
 T^{0i}_{\two\eq}
 = - \frac{C_1}{6} (\bnabla\times\bomega)^i \,,\quad
 T^{ij}_{\two\eq}	
 = -\frac{C_1}{12\bar{T}}
 	(\partial^i\partial^j + \eta^{ij}\bnabla^2 )T \,,
\end{eqnarray}
with $C_0 = \mu/2\pi^2$ and $C_1 = \mu^2/4\pi^2 + T^2/12$.
Similarly, from the expression in the dynamical limit~\eqref{eq:JT2_dyn}, we find
\begin{eqnarray}
\label{eq:gravito_dyn}
 &\dis J^0_{\two\dyn}
 = 0 \,, \quad
 J^i_{\two\dyn}
 = \frac{C_0}{20} (\bnabla\times\bomega)^i \,,\notag \\
 &\dis T^{00}_{\two\dyn}
 = 0 \,,\quad/
 T^{0i}_{\two\dyn}
 = - \frac{C_1}{20} (\bnabla\times\bomega)^i \,,\quad
 T^{ij}_{\two\dyn}	
 = \frac{C_1}{20}
 	\biggl[
 		\frac{1}{\bar{T}}\biggl(\partial^i\partial^j + \frac{1}{3} \eta^{ij} \bnabla^2 \biggr) T
 		+ \partial_0 \sigma^{ij}
 	\biggr] \,,\notag \\
\end{eqnarray}
where we introduce the shear tensor:
\begin{equation}
 \sigma^{ij}
 =  \epsilon^{ij} - \frac{1}{3} \eta^{ij} {\epsilon_k}^k \,.
\end{equation}
The corresponding vector and axial-vector currents are obtained by replacing $C_0$ with $C_{0,V/A} = \mu_{V/A}/\pi^2$, and $C_1$ with $C_{1,V} = (\mu_V^2 +\mu^2_A)/2\pi^2 + T^2/6$ and $C_{1,A} = \mu_{V}\mu_A/\pi^2$ respectively, as we have done to get Eq.~\eqref{eq:JT2_eq_VA}.
Several comments are in order.

We again emphasize that Eq.~\eqref{eq:gravito_eq} represents equilibrium transport phenomena, like the CME and CVE.
It is also intriguing to notice the difference between these and Fourier's law.
For the former, the currents in Eq.~\eqref{eq:gravito_eq} come from the vorticity, namely, the magnetic part $\calB$ of the gravity.
This background source supplies no energy to particles, and thus the currents become finite even in equilibrium.
On the other hand, the latter is the current generation by the temperature gradient.
This electric part $\calE$ of the gravitational field gives an energy to particles.
Therefore the Fourier's law is dissipative and prohibited in equilibrium.

For an static and spatially inhomogeneous vorticity $\bomega(\bx)$, there emerges the nonvanishing charge current $J^i_{\two\eq}$ and the energy current $T^{0i}_{\two\eq}$, on top of the contributions from the CVE~\eqref{eq:JT1_eq}.
Unlike the vector part of the CVE, the curvature-induced currents~\eqref{eq:JT2_eq_VA} or~\eqref{eq:gravito_eq} does not require $\mu_A\neq 0$.
In the system without the chiral imbalance, hence, $J^i_{\two\eq}$ and $T^{0i}_{\two\eq}$ are the leading vortical contributions.
In the dynamical limit, such second-order contributions become more important, since the CVE is washed out as shown in Eq.~\eqref{eq:JT1_dyn} while the currents in Eq.~\eqref{eq:gravito_dyn} are not.

Under the correspondence between magnetic field and vorticity, one would think that the charge current $J^i_{\two\eq/\dyn}$ is the gravitational analogue of Amp\`ere's law: $\bnabla\times \bcalB=\bnabla\times \bomega = \bJ$.
The situation is however not so trivial since $J^i_{\two\eq/\dyn}$ is opposite-signed against the energy current $T^{0i}_{\two\eq/\dyn}$ (for $\mu > 0$).
Namely, Eqs.~\eqref{eq:gravito_eq} and~\eqref{eq:gravito_dyn} cannot be explained based on the naive picture that a particle's momentum carries both charge and energy.
This curious flow dynamics essentially comes from the quantum effects through the spin-curvature coupling.
We should emphasize that such an antiparallel charge-energy flow is not restricted in the present coordinate, but more generally admitted in a lot of curved spacetime;
this phenomenon always takes place as long as ${R_{0}}^i\neq 0$, as shown in Eqs.~\eqref{eq:JT2_eq} and~\eqref{eq:JT2_dyn}.

It is worthwhile to mention the feedback to the gravitational field from Eq.~\eqref{eq:JT2_eq}.
In our sign convention, the Einstein field equation is given by $R_{\mu\nu}-\frac{1}{2} g_{\mu\nu} R = - 8\pi G T_{\mu\nu}$ with the gravitational constant $G$~\cite{birrell1984quantum}.
Following this, the induced Ricci tensor reads $R^\ind_{0i} \sim \alpha R_{0i}$ with a positive coefficient $\alpha>0$.
Hence, the initial gravitational field is enhanced, which evokes the possibility of instability.
We will revisit and analyze more precisely the above brief argument in the future, including the existence of a novel collective dynamics~\cite{Akamatsu:2013pjd} in a gravitational plasma~\cite{Nachbagauer:1995wn,deAlmeida:1993wy}.

One might think that Eq.~\eqref{eq:gravito_eq} is unrelated to an anomaly.
Indeed, Eq.~\eqref{eq:gravito_eq} would be irrelevant to the chiral anomaly, according to the analysis of discrete symmetry~\cite{Kharzeev:2011ds}.
Nevertheless, this fact is not sufficient to conclude the irrelevance to anomaly at all, as for the temperature dependent part of the CVE~\cite{Golkar:2012kb,Golkar:2015oxw,Chowdhury:2016cmh,Glorioso:2017lcn}.
We also mention that the transport coefficients $C_0$ and $C_1$ are time-reversal even quantities, which could be associated with their nondissipative nature similarly to those of the CME and CVE~\cite{Kharzeev:2011ds}.
It should be required to clarify the anomalous aspect of Eq.~\eqref{eq:gravito_eq} from different approaches, such as hydrodynamics.
In the sense that they are of the higher-order of the derivative counting, usual hydrodynamics can neglect Eqs.~\eqref{eq:gravito_eq} and~\eqref{eq:gravito_dyn}.
This would not be the case, however, if these phenomena originate from quantum anomaly like the CME and CVE.

\begin{figure}
\setlength\intextsep{0pt} 
\setlength\textfloatsep{0pt}
\centering
\includegraphics[scale=0.3]{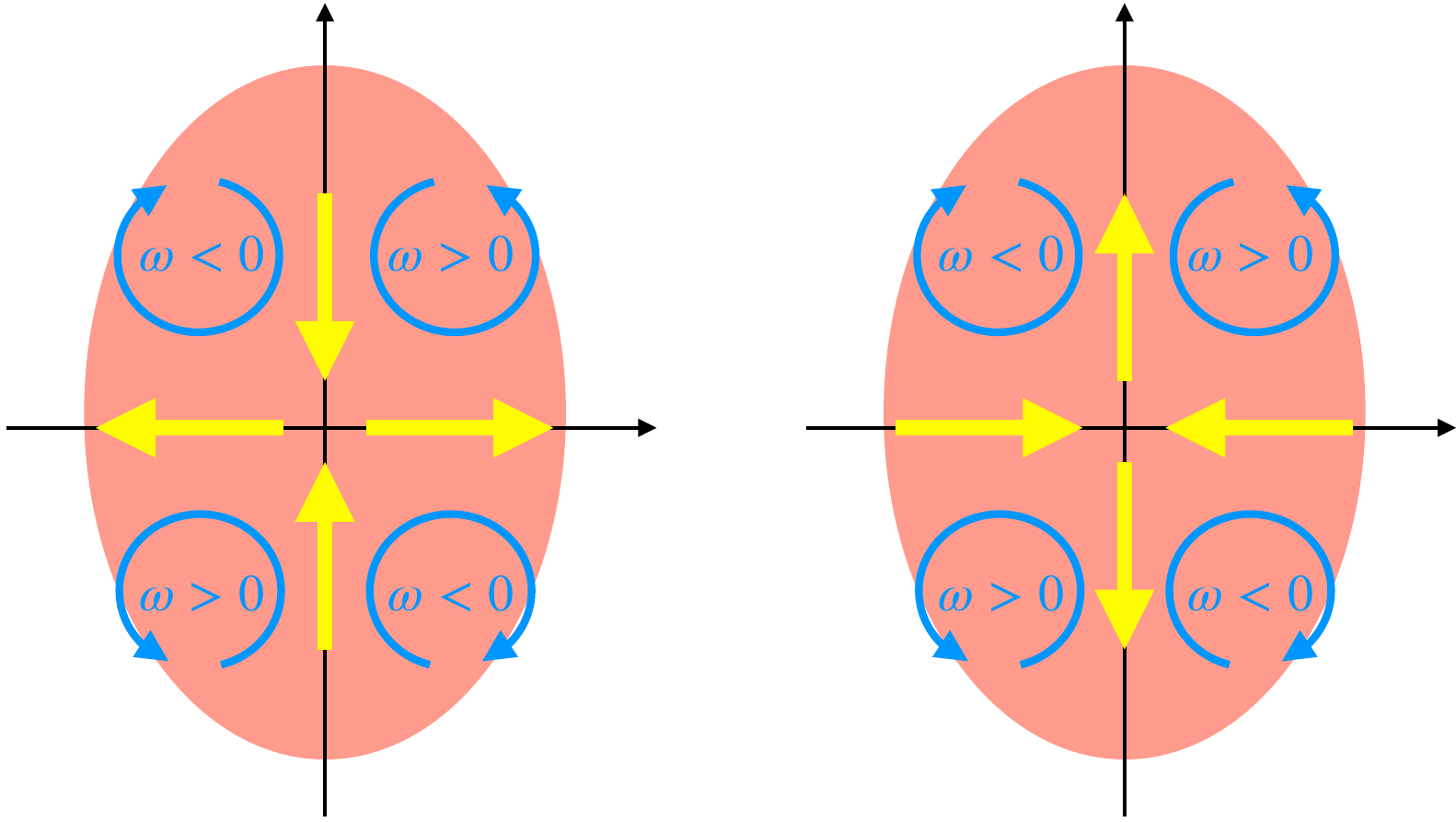}
\caption{Flow directions of $J_\perp$ (left) and $J^\epsilon_\perp$ (right).
The horizontal axis corresponds to the reaction plane of heavy-ion collisions.
The quadrupole vorticity structure is based on the measurement by STAR collaboration~\cite{Adam:2019srw}}
\label{fig:HIC}
\end{figure}

These novel contributions~\eqref{eq:gravito_eq} lead to several implications in relativistic many-body systems where an inhomogeneous fluid vorticity is experimentally generated.
In rotating quark-gluon plasma, there emerges the quadrupole configuration of the vorticity along the beam direction~\cite{Adam:2019srw,Becattini:2017gcx,Wei:2018zfb,Fu:2020oxj}.
Thus, on the transverse plane to the beam direction, the inhomogeneous vorticities generate the charge current $J_\perp$ and the energy current $J^\epsilon_\perp$, as depicted in Fig.~\ref{fig:HIC}. 
As a brief argument, we may estimate the scale of the vorticity gradient to be the inverse of the hot matter size.
Indeed, at the collision energy $\sqrt{s}=19.6\,\text{GeV}$, the gradient of the vorticity is estimated to be $(\bnabla\times\bomega)/\bomega \approx 0.2\,\text{fm} - 0.5\,\text{fm} \approx 40\,\text{MeV}-100\,\text{MeV}$~\cite{Wei:2018zfb}.
Although the whole magnitudes of $J_\perp$ and $J^\epsilon_\perp$ are dependent on the scale of $\bomega$, hence, these are nonnegligible compared with the CVE. 

On top of the charge and energy currents, the stress tensor $T^{ij}_{\two\eq}$ is also induced.
Let us consider a cylindrical system along the $z$ direction with a spatially inhomogeneous temperature $T(z)$.
From the vector part of the energy-momentum tensor in Eq.~\eqref{eq:gravito_eq}, the temperature gradient yields the correction to the transverse pressure $P_\perp (z) = \frac{C_{1,V}}{12} T''(z)/\bar{T}$.
When the temperature takes a Gaussian form $T(z) = \bar{T} e^{-z^2/2\sigma^2}$, we get $P_\perp(z)=\frac{C_{1,V}}{12} e^{-z^2/2\sigma^2} (z^2-\sigma^2)/(3\sigma^2)$, which has the minima $P_\perp(0)=-\frac{C_{1,V}}{12}\sigma^{-2}<0$ and maxima $P_\perp(\sigma)= \frac{C_{1,V}}{12} e^{-3/2}\\ \sigma^{-2} >0$.
%In the context of heavy-ion collision, this would be a nonnegligible contribution.
Such a pressure correction is detectable in Weyl/Dirac semimetal experiments, similarly to the usual thermoelectric transport phenomena~\cite{Lundgren:2014hra,liang2017anomalous}. 

In table-top experiments, an inhomogeneous and dynamical vorticity can be generated by an acoustic surface wave. 
We consider a transverse wave propagating on the $xy$ surface~\cite{9780750626330,PhysRevLett.119.077202,PhysRevB.87.180402} of Weyl/Dirac semimetals.
Also we prepare the wave propagating along the $x$ direction, and its amplitude is normal to the surface, i.e., its displacement vector is given by $\bu = (0,0,u)$ with $u = \bar{u} e^{-ik_0t + ik x -\kappa z}$.
Here $e^{-\kappa z}$ reflects unpenetrating into the material.
Now the response to this surface wave can be evaluated in the coordinate space described by $ g_{\mu\nu} = \eta_{\mu\nu} + h_{\mu\nu}$ with $h_{0z} = -\dot{u} = ik_0 u$, $h_{xz} = -\partial_x{u} = -ik u$, $h_{zz} = -\partial_z{u} = \kappa u$ and other components of $h_{\mu\nu}$ vanishing.
From Eq.~\eqref{eq:gravito_dyn} together with the Wick rotation $\partial_3 \to \kappa$,
we get the charge and energy currents: $ J^x_{\two\dyn} =  \frac{C_0}{20} \frac{1}{2} k_0 k\kappa u $, $ J^z_{\two\dyn} =  \frac{C_0}{20} \frac{1}{2} ik_0 k^2 u $ and $ T^{0x}_{\two\dyn} = -\frac{C_0}{40} \frac{1}{2} k_0 k\kappa u $, $ T^{0z}_{\two\dyn} = -\frac{C_0}{40} \frac{1}{2} ik_0 k^2 u $.
The flows normal to the fluid velocity $\dot{u}$ are induced by the gravitational curvature via quantum effects.
We note that the flows parallel to the fluid velocity are induced from classical contributions.
%The origin of these current along the wave direction is an unclosed path of particles.
%They may thus be regarded as a quantum analogue of the Stokes drift~\cite{stokes1847theory,phillips1966dynamics,lighthill2001waves}.

\acknowledgments
This work was supported by JSPS KAKENHI Grant Numbers~17H06462 and 18H01211.
K.~M. was supported by Special Postdoctoral Researcher (SPDR) Program of RIKEN.

\appendix

\section{Equilibrium Wigner function~\eqref{eq:Rmu2_eq}}\label{app:eqRmu2}
In this appendix, we show the concrete expression of $\calR^\mu_\two$ at equilibrium defined by Eqs.~\eqref{eq:f0}, \eqref{eq:f1}, \eqref{eq:f2}~and \eqref{eq:phi2}.
We decompose $\calR^\mu_\two$ in Eq.~\eqref{eq:Rmu2} into the frame-(in)dependent and the $\phi_\two$ part:
\begin{equation}
 \calR^\mu_\two 
 = \calR^\mu_{\two\text{indep}} + \calR^\mu_{\two\text{dep}} + 2\pi\delta(p^2) p^\mu \phi_\two  \,.
\end{equation}
The first term reads 
\begin{equation}
\begin{split}
\calR^\mu_{\two\text{indep}} 
	& = \frac{2\pi}{p^2}
    		 \biggl[
	 	     	- p^\mu Q\cdot p 
    			+ 2 p_\nu \Bigl( T^{[\mu} p^{\nu]} + S^{\alpha\mu\nu} p_\alpha \Bigr)
   	 	 	\biggr] \delta(p^2) f_\zero \\
 & = \frac{2\pi}{24}
   		\biggl[
  	 		5 R^{\alpha\mu}\partial^p_\alpha
 			- R^{\alpha\beta\gamma\mu} p_\alpha \partial^p_\beta \partial_\gamma^p\
   			-\frac{p^\mu}{p^2}
   			\bigl(
   				2 R
   				+ 6 R^{\alpha\beta}p_\alpha\partial_\beta^p
 				+ 2 R_{\alpha\beta\gamma\delta}
 					p^\alpha p^\gamma \partial_p^\beta \partial^\delta_p
 			\bigr)\\
 & \qquad
 				- \frac{6}{p^2} R^{\alpha\beta\gamma\mu} p_\alpha p_\gamma \partial^p_\beta 
   		\biggr]	
   			\delta(p^2) f_\zero \\
 & = 2\pi\delta(p^2)
 	 \Biggl[
 	 	f_\zero
 	 	\biggl(
 	 		- \frac{1}{2p^2} R^{\mu\alpha} p_\alpha
 	 		- \frac{1}{12p^2} R\, p^\mu
 	 		+ \frac{2}{3(p^2)^2} R^{\alpha\beta} p^\mu p_\alpha p_\beta 
 	 	\biggr)\\
 & \qquad\qquad\quad
 	 	+ f_\zero'
 	 	\biggl(
 	 		\frac{5}{24} R^{\mu\alpha} \beta_\alpha
 	 		- \frac{1}{6 p^2} R^{\alpha\beta\gamma\mu} p_\alpha\beta_\beta p_\gamma
 	 		- \frac{1}{4 p^2} R^{\alpha\beta} p^\mu p_\alpha \beta_\beta
 	 	\biggr)  \\
 & \qquad\qquad\quad
 		+ f_\zero''
 		\biggl(
 			- \frac{1}{24} R^{\alpha\beta\gamma\mu} p_\alpha\beta_\beta\beta_\gamma
 			-\frac{1}{12 p^2} R_{\alpha\beta\gamma\delta} 
 				p^\mu p^\alpha p^\gamma \beta^\beta \beta^\delta
 		\biggr)
 	 \Biggr] \,.
\end{split}
\end{equation}
The frame-dependent part is further decomposed as
\begin{equation}
\begin{split}
 \calR^\mu_{\two\text{dep}} 
 & = 2\pi\delta(p^2)
 	\Bigl(
 		r^\mu_1 + r^\mu_2 + r^\mu_3
 	\Bigr) \,,
\end{split}
\end{equation}
where we define
\begin{equation}
\begin{split}
 r^\mu_1
 & := p^\mu \Sigma^u_{\nu\rho}  D^\nu 
 	\biggl(
 		f'_\zero
 		\frac{\varepsilon^{\rho\sigma\lambda\eta}}{4\,p\cdot n}
 		n_\sigma \nabla_\lambda\beta_\eta 
 	\biggr) 
   + \frac{1}{2}\Sigma_u^{\mu\nu}D_\nu 
 	\Bigl(
 		f'_\zero \Sigma_n^{\rho\sigma}\nabla_\rho\beta_\sigma
 	\Bigr) \,,\\
 r^\mu_2
 & := \frac{1}{2p^2}\varepsilon^{\mu\nu\rho\sigma}p_\nu
  		 	 	D_\rho \Sigma^n_{\sigma\lambda} D^\lambda f_\zero \,, \\
 r^\mu_3
 &: = - \frac{1}{p^2}
    	\Sigma^{\mu\nu}_u 
    		\biggl(
    			\frac{1}{2}{\tilde{R}}_{\alpha\beta\nu\rho} 
      	  	  			p^\rho p^\alpha \partial^\beta_p
      	  	  	+ p\cdot D \Sigma^n_{\nu\rho} D^\rho 
      	  	\biggr) f_\zero \,.
\end{split}
\end{equation}
For the equilibrium distribution $f_\zero$ in Eq.~\eqref{eq:f0}, we reduce $r^\mu_2$ and $r^\mu_3$ to
\begin{equation}
\begin{split}
r^\mu_2
 & = \frac{1}{2p^2} \varepsilon^{\mu\nu\rho\sigma}p_\nu D_\rho
 		\Sigma^n_{\sigma\lambda} p_\eta \nabla^\lambda \beta^\eta f'_\zero \\
 & = -\frac{1}{8p^2} \varepsilon^{\mu\nu\rho\sigma}p_\nu D_\rho
 		\varepsilon_{\lambda\eta\sigma\tau} p^\tau \nabla^\lambda \beta^\eta f'_\zero 
 	 + \frac{1}{8} \varepsilon^{\mu\nu\rho\sigma}p_\nu D_\rho
 		\varepsilon_{\lambda\eta\sigma\tau} \frac{n^\tau}{p\cdot n} 
 			\nabla^\lambda \beta^\eta f'_\zero \,,
 \end{split}
\end{equation}
and
\begin{equation}
\begin{split}
  r^\mu_3
 & = - \frac{1}{p^2}\Sigma_u^{\mu\nu}
 	\biggl[
 		\frac{1}{2} \tilde{R}_{\alpha\beta\nu\rho} 
 			p^\rho p^\alpha \beta^\beta f'_\zero
 		+ p\cdot D\Sigma^n_{\nu\rho} p_\lambda \nabla^\rho \beta^\lambda f'_\zero
 	\biggr] \\
 & = - \frac{1}{4}f'_\zero\Sigma_u^{\mu\nu}
 		\varepsilon_{\rho\lambda\nu\tau} p\cdot D
 			 \frac{n^\tau}{p\cdot n} \nabla^\rho \beta^\lambda\\
 & = - \frac{1}{4}
  		(2\Sigma_u^{\mu[\nu} p^{\sigma]}
 		 + \Sigma_u^{\mu\sigma} p^\nu)
 		\varepsilon_{\rho\lambda\nu\tau} D_\sigma f'_\zero
 			 \frac{n^\tau}{p\cdot n} \nabla^\rho \beta^\lambda\\
 & = - r^\mu_1
 	 + \frac{1}{8}\varepsilon^{\nu\sigma\mu\eta} p_\eta
 			\varepsilon_{\rho\lambda\nu\tau} D_\sigma f'_\zero
 			 \frac{n^\tau}{p\cdot n} \nabla^\rho \beta^\lambda \,,			
\end{split}
\end{equation}
where the $p^2$ term is dropped, and we utilize $\nabla_\mu \nabla_\nu \beta_\rho = - \beta^\lambda R_{\lambda\mu\nu\rho}$ and Eq.~\eqref{eq:Sigma_p}.
The frame-dependent part hence becomes 
\begin{equation}
\begin{split}
 \calR^\mu_{\two\text{dep}}
 & = -\frac{2\pi\delta(p^2)}{8p^2} \varepsilon^{\mu\nu\rho\sigma}p_\nu D_\rho
 		\varepsilon_{\lambda\eta\sigma\tau} p^\tau \nabla^\lambda \beta^\eta f'_\zero \\
 & = 2\pi\delta(p^2)
 	\biggl[
 	f_\zero'
 	\biggl(
 		\frac{1}{4p^2} p_\alpha\beta_\beta p_\gamma R^{\alpha\beta\gamma\mu} 
 	 	- \frac{1}{4} R^{\mu\nu} \beta_\nu 
 	 	+ \frac{p^\mu}{4p^2} R^{\alpha\beta} p_\alpha \beta_\beta
 	\biggr)  \\
 & \quad
 	+ f''_\zero
 	\biggl(
	 	 - \frac{1}{4} \nabla^{\rho} \beta^{\mu} p^\nu \nabla_{\rho}\beta_{\nu}
 		 + \frac{p^\mu}{4p^2} p_\nu \nabla^{\rho} \beta^{\nu} p^\sigma \nabla_{\rho} \beta_{\sigma}
 	\biggr)	
 	\biggr] \,.
\end{split}
\end{equation}
In the above equation, the frame dependence totally vanishes, as it should.
Eventually, $\calR^\mu_\two$ is written as Eq.~\eqref{eq:Rmu2_eq}.

\section{Equilibrium kinetic equation~\eqref{eq:CKE1}}\label{app:kinetic_equation}
In this appendix, we derive the kinetic equation~\eqref{eq:CKE1}.
In later use, we recall the second Bianchi identity for the Riemann tensor:
\begin{equation}
\label{eq:2ndBianchi}
  \nabla_\alpha R_{\mu\nu\beta\gamma}
  + \nabla_\beta R_{\mu\nu\gamma\alpha}
  + \nabla_\gamma R_{\mu\nu\alpha\beta}
  = 0 \,,
\end{equation}
which implies
\begin{equation}
\label{eq:BI2}
 \nabla_\mu R^{\mu\nu} = \frac{1}{2} \nabla^\nu R \,,
 \quad
 \nabla_\mu R^{\rho\sigma\mu\nu}
 = \nabla^\rho R^{\sigma\nu} - \nabla^\sigma R^{\rho\nu} \,.
\end{equation}
Using $\nabla_\mu \beta_\nu = -\nabla_{\nu}\beta_{\mu}$, $R_{\alpha[\mu\nu]\beta} = -\frac{1}{2} R_{\alpha\beta\mu\nu}$ and Eq.~\eqref{eq:BI2}, we evaluate each term in the kinetic equation~\eqref{eq:R1-curved} as follows:
\begin{equation}
\begin{split}
 & - \frac{1}{8}\nabla_\lambda R_{\mu\nu} \partial_p^\lambda \partial_p^\nu p^\mu f_\zero\delta(p^2) \\
 & = 
 	f_\zero
 	\biggl(
 		- \frac{1}{2} \delta'(p^2) p\cdot\nabla R 
 		- \frac{1}{2} \delta''(p^2) p\cdot\nabla R^{\alpha\beta} p_\alpha p_\beta 
 	\biggr) \\
 & \quad
 	+ f_\zero'
 	\biggl(
 		- \frac{3}{16} \delta(p^2) \beta\cdot\nabla R
 		- \frac{1}{4} \delta'(p^2) p\cdot\nabla R^{\alpha\beta} p_\alpha \beta_\beta
 		- \frac{1}{4} \delta'(p^2) \beta\cdot\nabla R^{\alpha\beta} p_\alpha p_\beta
 	\biggr) \\
 & \quad
 	+ f_\zero''
 	\biggl( 
 		- \frac{1}{8} \delta(p^2) \beta\cdot\nabla R^{\alpha\beta} p_\alpha\beta_\beta		
 	\biggr)\,,
\end{split}
\end{equation}
\begin{equation}
\begin{split}
 & - \frac{1}{24}
 	\nabla_\lambda R_{\rho\sigma\mu\nu} \partial_p^\lambda \partial_p^\nu \partial_p^\sigma p^\rho p^\mu 
 	f_\zero \delta(p^2) \\
 & = 
 	f_\zero\biggl(
 		\frac{1}{6} \delta'(p^2) p\cdot\nabla R 
 		+ \frac{1}{6} \delta''(p^2) p\cdot\nabla R^{\alpha\beta} p_\alpha p_\beta
 	\biggr) \\
 & \quad
 	+ f_\zero' 
 	\biggl(
 		\frac{1}{12} \delta(p^2)\beta\cdot\nabla R
 		+ \frac{1}{6} \delta'(p^2) p\cdot\nabla R^{\alpha\beta} p_\alpha \beta_\beta
 		+ \frac{1}{12} \delta'(p^2)\beta\cdot\nabla R^{\alpha\beta} p_\alpha p_\beta
 	\biggr) \\
 & \quad
 	+ f_\zero''
 	\biggl(
 		\frac{1}{6} \delta(p^2) \beta\cdot\nabla R^{\alpha\beta} p_\alpha\beta_\beta
 		-\frac{1}{12} \delta(p^2) p\cdot\nabla R_{\alpha\beta} \beta^\alpha \beta^\beta
 		-\frac{1}{12} \delta'(p^2) p\cdot\nabla R_{\rho\sigma\mu\nu} p^\mu \beta^\nu p^\rho \beta^\sigma
 	\biggr) \\
 & \quad
 	+ f_\zero'''
 	\biggl(
 		-\frac{1}{24} \delta(p^2) 
 			\beta\cdot\nabla R_{\rho\sigma\mu\nu} p^\mu \beta^\nu p^\rho \beta^\sigma
 	\biggr) \,,
\end{split}
\end{equation}
\begin{equation}
\begin{split}
 & \frac{1}{8} {R^\rho}_{\sigma\mu\nu} \partial_p^\nu \partial_p^\sigma D_\rho p^\mu f_\zero \delta(p^2) \\
 & = f_\zero''
 	 \biggl(
 	 	 \frac{3}{16} \delta(p^2) R^{\rho\sigma\mu\nu} p_\mu \beta_\nu \nabla_\rho \beta_\sigma
 	 \biggr)\\
 & \quad
 	 +\biggl(
 		- \frac{1}{8}\delta(p^2) R^{\alpha\rho}\beta_\alpha 
 		+ \frac{1}{4} \delta'(p^2) R^{\rho\sigma\mu\nu} p_\mu \beta_\nu p_\sigma 
 	 \biggr)
 	 	D_\rho f_\zero'
 	 + 
	 \biggl(
		\frac{1}{8} \delta(p^2) R^{\rho\sigma\mu\nu} p_\mu \beta_\nu \beta_\sigma 
 	 \biggr)
 	 	D_\rho f_\zero'' \,,
\end{split}
\end{equation}
\begin{equation}
\begin{split}
 & D_\mu f_\zero\biggl(
 	 		- \frac{1}{2p^2} R^{\mu\alpha} p_\alpha
 	 		- \frac{1}{12p^2} R\, p^\mu
 	 		+ \frac{2}{3(p^2)^2} R^{\alpha\beta} p^\mu p_\alpha p_\beta 
 	 	\biggr) \\
 & = 
  		f_\zero\biggl(
 	 		- \frac{1}{3p^2} p\cdot\nabla R
 	 		+ \frac{2}{3(p^2)^2} p\cdot\nabla R^{\alpha\beta} p_\alpha p_\beta 
 	 	\biggr) 
 	 + \biggl(
 	 		- \frac{1}{2p^2} R^{\mu\alpha} p_\alpha
 	 	\biggr) D_\mu f_\zero \,,
\end{split}
\end{equation}

\begin{equation}
\begin{split}
 & D_\mu f_\zero'
 	 	\biggl(
 	 		-\frac{1}{24} R^{\mu\alpha} \beta_\alpha
 	 		+ \frac{1}{12 p^2} R^{\alpha\beta\gamma\mu} p_\alpha\beta_\beta p_\gamma
 	 	\biggr) \\
 & = 
	 f_\zero'	\biggl(
 	 		-\frac{1}{48} \beta\cdot\nabla R
 	 		-\frac{1}{12 p^2} p\cdot\nabla R^{\alpha\beta}p_\alpha \beta_\beta
 	 		+ \frac{1}{12 p^2} \beta\cdot\nabla R^{\alpha\beta}p_\alpha p_\beta
  	 	\biggr) \\
 & \quad
 	 +  
 	 	\biggl(
 	 		-\frac{1}{24} R^{\mu\alpha} \beta_\alpha
 	 		+ \frac{1}{12 p^2} R^{\alpha\beta\gamma\mu} p_\alpha\beta_\beta p_\gamma
 	 	\biggr)
 	 		D_\mu f_\zero' \,,
\end{split}
\end{equation}
\begin{equation}
\begin{split}
 & D_\mu f_\zero''
 		\biggl(
 			- \frac{1}{24} R^{\alpha\beta\gamma\mu} p_\alpha\beta_\beta\beta_\gamma
 			-\frac{1}{12 p^2} R_{\alpha\beta\gamma\delta} 
 				p^\mu p^\alpha p^\gamma \beta^\beta \beta^\delta
 		\biggr) \\
 & = f_\zero''
  		\biggl(
 			\frac{1}{24} p\cdot\nabla R^{\alpha\beta} \beta_\alpha \beta_\beta
 			- \frac{1}{24} \beta \cdot\nabla R^{\alpha\beta} p_\alpha \beta_\beta
 			- \frac{1}{12p^2} p\cdot\nabla R_{\alpha\beta\gamma\delta}
 				p^\alpha p^\gamma \beta^\beta \beta^\delta \\
 & \qquad
 			+ \frac{1}{16} R^{\alpha\beta\mu\nu} p_\alpha \beta_\beta \nabla_\mu \beta_\nu
 		\biggr) 
 	+ \biggl(
 		\frac{1}{6 p^2} R^{\alpha\beta\gamma\mu} 
 			 p_\alpha p_\gamma \beta_\beta 
 	  \biggr)
 			 D_\mu f_\zero'
 	+ \biggl(
 			- \frac{1}{24} R^{\alpha\beta\gamma\mu} p_\alpha\beta_\beta\beta_\gamma
 	  \biggr) D_\mu f_\zero'' \,,
\end{split}
\end{equation}
\begin{equation}
\begin{split}
 & - \frac{1}{4} D_\mu \nabla^{[\rho} \beta^{\mu]} p^\nu \nabla_{[\rho}\beta_{\nu]} f_\zero '' \\
 & = f_\zero '' 
   	 \biggl(
   	 	- \frac{1}{8} R_{\alpha\beta\mu\nu} p^\alpha \beta^\beta \nabla^{[\mu} \beta^{\nu]}
   	 \biggr) 
   	+ 
   	 \biggl(
   	 	\frac{1}{4} R^{\alpha\beta} \beta_\alpha  
   	 \biggr) D_\beta f_\zero '
  	+ \biggl(
  		- \frac{1}{4} \nabla^{[\rho} \beta^{\mu]} p^\nu \nabla_{[\rho}\beta_{\nu]}
  	 \biggr) D_\mu f_\zero ''\,,
\end{split}
\end{equation}
and
\begin{equation}
\begin{split}
  D_\mu \frac{p^\mu}{4p^2} p_\nu \nabla^{[\rho} \beta^{\nu]}
 	 p^\sigma \nabla_{[\rho} \beta_{\sigma]} f_\zero '' 
 & = 
	\biggl(	 
	 	- \frac{1}{2p^2} R^{\alpha\beta\gamma\mu} p_\alpha \beta_\beta p_\gamma 
	\biggr)
 	 	D_\mu f_\zero ' \,.
\end{split}
\end{equation}
Collecting them, we obtain Eq.~\eqref{eq:CKE1}.

\section{Integration formulas}\label{app:int_formulas}
Here, we present several Integration formulas.
We first define
\begin{equation}
 C_{n}
 := \frac{1}{2\pi^2}\int_0^\infty d\rho \rho^n 
 	\Bigl[
 		n_F(\rho-\mu) - (-1)^n n_F(\rho+\mu) 
 	\Bigr]\,,
 \quad
  n_F (z) := \frac{1}{e^{\beta z} + 1} \,.
\end{equation}
In particular, the first four $C_{n}$'s are 
\begin{eqnarray}
 & \dis
  C_0  = \frac{\mu}{2\pi^2} \,, \\
 & \dis
  C_1 = \frac{\mu^2}{4\pi^2} + \frac{T^2}{12} \,, \\
 & \dis
  C_2  = 2\int_0^\mu d\nu\, C_1(\nu) 
  = \frac{\mu^3}{6\pi^2} + \frac{\mu T^2}{6} \,, \\
 & \dis
  C_3  = 3\int_0^\mu d\nu\, C_2 (\nu) 
  = \frac{\mu^4}{8\pi^2} + \frac{\mu^2 T^2}{4} \,.
\end{eqnarray}
Also in the integral of angular degrees of freedom, we can replace the product of $p_{\mu}$'s in the integral, as follows:
\begin{equation}
\begin{split}
\label{eq:p_replacement}
 p_{\alpha}
	& \to p_{0}\xi_{\alpha}  , \\
 p_{\alpha}p_{\beta} 
	& \to (p_{0})^{2}\xi_{\alpha}\xi_{\beta} 
		+ \frac{\bp^{2}}{3}\Delta_{\alpha\beta},\\
 p_\alpha p_\beta p_\gamma
 &\to (p_0)^3\xi_\alpha\xi_\beta \xi_\gamma
 	 + \frac{p_0\bp^2}{3}
 	 	(\xi_\alpha\Delta_{\beta\gamma}
 	 	 + \xi_\beta\Delta_{\gamma\alpha}
 	 	 + \xi_\gamma\Delta_{\alpha\beta}), \\
p_{\alpha}p_{\beta}p_{\gamma}p_{\delta} 
	& \to (p_{0})^{4}\xi_{\alpha}\xi_{\beta}\xi_{\gamma}\xi_{\delta} \\
	& \quad
		 + \frac{(p_{0})^{2}\bp^{2}}{3}
		 	(\xi_{\alpha}\xi_{\beta}\Delta_{\gamma\delta} 
		 	 + \xi_{\alpha}\xi_{\gamma}\Delta_{\beta\delta} 
		 	 + \xi_{\alpha}\xi_{\delta}\Delta_{\beta\gamma}
		 	 + \xi_{\beta}\xi_{\gamma}\Delta_{\alpha\delta} 
		 	 + \xi_{\beta}\xi_{\delta}\Delta_{\alpha\gamma}
		 	 + \xi_{\gamma}\xi_{\delta}\Delta_{\alpha\beta}) \\
	& \quad
		 + \frac{|\bp|^4}{15} 
		 	(\Delta_{\alpha\beta}\Delta_{\gamma\delta}
		 	 + \Delta_{\alpha\gamma}\Delta_{\beta\delta}
		 	 + \Delta_{\alpha\delta}\Delta_{\beta\gamma}),\\
p_{\alpha}p_{\beta}p_{\gamma}p_{\delta}p_{\lambda} 
	&\to  (p_{0})^{5}\xi_{\alpha}\xi_{\beta}\xi_{\gamma}\xi_{\delta}\xi_{\lambda} \\
	& \quad
		+\frac{(p_{0})^{3}\bp^{2}}{3}  
			(\Delta_{\alpha\beta}\xi_{\gamma}\xi_{\delta}\xi_{\lambda}
			 + \Delta_{\alpha\gamma}\xi_{\beta}\xi_{\delta}\xi_{\lambda}
			 + \Delta_{\alpha\delta}\xi_{\beta}\xi_{\gamma}\xi_{\lambda}
			 + \Delta_{\alpha\lambda}\xi_{\beta}\xi_{\beta}\xi_{\gamma} 
			 + \Delta_{\beta\gamma}\xi_{\alpha}\xi_{\delta}\xi_{\lambda}\\
	& \quad\quad
			 + \Delta_{\beta\delta}\xi_{\alpha}\xi_{\gamma}\xi_{\lambda}
			 + \Delta_{\beta\lambda}\xi_{\alpha}\xi_{\gamma}\xi_{\delta}
			 + \Delta_{\gamma\delta}\xi_{\alpha}\xi_{\beta}\xi_{\lambda}
			 + \Delta_{\gamma\lambda}\xi_{\alpha}\xi_{\beta}\xi_{\delta}
			 + \Delta_{\delta\lambda}\xi_{\alpha}\xi_{\beta}\xi_{\gamma}) \\
	& \quad
		+\frac{1}{15} p_{0}|\bp|^4 
			\Bigl[
				\xi_{\alpha} 
				(\Delta_{\beta\gamma}\Delta_{\delta\lambda}
				 +\Delta_{\beta\delta}\Delta_{\gamma\lambda}
				 +\Delta_{\beta\lambda}\Delta_{\gamma\delta}) \\
	& \quad\quad
				+ \xi_{\beta} 
				(\Delta_{\alpha\gamma}\Delta_{\delta\lambda}
				 +\Delta_{\alpha\delta}\Delta_{\gamma\lambda}
				 +\Delta_{\alpha\lambda}\Delta_{\gamma\delta})
				+ \xi_{\gamma} 
				(\Delta_{\alpha\beta}\Delta_{\delta\lambda}
				 +\Delta_{\alpha\delta}\Delta_{\beta\lambda}
				 +\Delta_{\alpha\lambda}\Delta_{\beta\delta})\\
	& \quad\quad
				+ \xi_{\delta} 
				(\Delta_{\alpha\beta}\Delta_{\gamma\lambda}
				 +\Delta_{\alpha\gamma}\Delta_{\beta\lambda}
				 +\Delta_{\alpha\lambda}\Delta_{\beta\gamma}) 
				+ \xi_{\lambda} 
				(\Delta_{\alpha\beta}\Delta_{\gamma\delta}
				 +\Delta_{\alpha\gamma}\Delta_{\beta\delta}
				 +\Delta_{\alpha\delta}\Delta_{\beta\gamma}) 
			\Bigr] \,,
\end{split}
\end{equation}
where $\xi^{\mu}:=(1,\bzero)$ and $\Delta^{\mu\nu}:=\xi^{\mu}\xi^{\nu}-\eta^{\mu\nu}$.

\section{Alternative derivation of $J^\mu_{\two\eq}$}\label{app:alternative}
%Although the resulting CKT leads to equilibrium physical quantities, these should be derived from the thermodynamics of Weyl fermions in a curved spacetime.
In this appendix, we derive the curvature-induced charge current $J_{\eq\two}^\mu$ in Eq.~\eqref{eq:JT2_eq}, from the thermodynamics of Weyl fermions in a curved spacetime.
At the same time, such alternative derivations leading to the same $J_{\eq\two}^\mu$ ensures the correctness of the Wigner function $\calR^\mu_\two$ in Eq.~\eqref{eq:Rmu} and the equilibrium distribution function $f_\two$ given by Eqs.~\eqref{eq:f2} and~\eqref{eq:phi2}.

\subsection{Diagrammatic computation}\label{sec:diagram}
First, we derive the current of a chiral fluid under a gravitational field, based on the linear response theory.
We consider a Weyl fermion, and the corresponding action is given by
\begin{equation}
\begin{split}
S=\frac{i}{2} \int d^{4}x e \eta^{\dag}\Bigl(\sigma^{{a}}e_{{~a}}^{\mu}(\nabla_\mu +iA_\mu) -(\overleftarrow{\nabla}_{\mu}-iA_\mu)\sigma^{a}e_{{~a}}^{\mu}\Bigr)\eta,
\end{split}
\end{equation}
where we introduce $\sigma^a = ( 1, \sigma^i)$ with the Pauli matrices 
$\sigma^i~(i=1,2,3)$. 
Here $e_\mu^{~a} (e^\mu_{~a})$ denotes (inverse) vierbein satisfying 
$g_{\mu\nu} = e_\mu^{~a} e_\nu^{~b} \eta_{ab},~
\eta^{ab} = e_\mu^{~a} e_\nu^{~b} g^{\mu\nu}$ 
with the spacetime curved metric $g_{\mu\nu}$ and Minkowski metric
$\eta_{ab} = \mathrm{diag} (1,-1,-1,-1)$, and $e:=\det e_{\mu}^{{~a}}$.
The left and right covariant derivatives are defined as 
\begin{equation}
\begin{split}
 \nabla_\mu \eta 
  & := \partial_\mu \eta -  i \mathcal{A}_\mu \eta \,, 
  \quad
 \eta ^{\dag} \overleftarrow{\nabla}_\mu 
   := \partial_\mu \eta^\dag + i  \eta^\dag \mathcal{A}^{\dag}_\mu \,,\\
 \mathcal{A}_\mu 
  & := \frac{1}{2} \omega_\mu^{~ab} \Sigma_{ab} \,,
  \quad
  \Sigma^{ab} :=  \frac{i}{4}  (\bar{\sigma}^a {\sigma}^b - \bar{\sigma}^b {\sigma}^a) 
\end{split}
\end{equation}
with $\bar{\sigma}^a := (1, -\sigma^i)$, which satisfies
$\bar\sigma^a {\sigma}^b + \bar\sigma^b {\sigma}^a=\sigma^a \bar{\sigma}^b + \sigma^b \bar{\sigma}^a = 2 \eta^{ab}$.
Furthermore, employing the torsionless condition, 
we can express the spin connection $\omega_\mu^{~ab} = - \omega_\mu^{~ba}$ 
as
\begin{equation}
\begin{split}
 & \omega_\mu^{~ab} 
  := \frac{1}{2} e^{\nu a}e^{\rho b} 
  (C_{\nu\rho\mu} - C_{\rho\nu\mu} - C_{\mu\nu\rho}) \,,\\
 & C_{\mu\nu\rho} :=
  e_\mu^{~c} (\partial_\nu e_{\rho c} - \partial_\rho e_{\nu c}).
  \label{eq:SpinC}
\end{split}
\end{equation}
The energy-momentum tensor $T^{\mu\nu}$ and $U(1)$ covariant charge current $J^\mu$ are
defined as 
\begin{equation}
\begin{split}
T^{\mu\nu}&=-\frac{1}{e}\frac{\delta S}{\delta e_{\mu}^{~a}} e_{~a}^{\nu} 
  = 
    \frac{ i }{2} \eta^\dag
  ( \sigma^\mu \overrightarrow{\nabla^\nu} 
   - \overleftarrow{\nabla^\nu} \sigma^\mu ) \eta 
   +\frac{1}{4}\nabla_{\rho}(\eta^{\dag}(\sigma^{\mu}\Sigma^{\nu\rho}+\Sigma^{\nu\rho\dag}\sigma^{\mu})\eta)
  - \mathcal{L} g^{\mu\nu} 
  ,\\
J^\mu &=-\frac{1}{e}\frac{\delta S}{\delta A_{\mu}}=  \eta^\dag \sigma^\mu \eta.
  \end{split}
\end{equation}
Note that $T^{\mu\nu}$ is not symmetric, so we introduce the symmetric energy-momentum tensor defined as $T^{\mu\nu}_{S}:=(T^{\mu\nu}+T^{\nu\mu})/2$.
In the following, we consider fluctuation around the flat metric $g_{\mu\nu}=\eta_{\mu\nu}+h_{\mu\nu}$.

In the linear response theory, the current in momentum space can be expressed as 
\begin{equation}
\begin{split}
\label{eq:jmuk}
\langle J^\mu(k) \rangle =  -\frac{1}{2}G^{\mu\nu\rho}(k)h_{\nu\rho}(k)
\end{split}
\end{equation}
with
\begin{equation}
\begin{split}
G^{\mu\nu\rho}(k) 
	:=  T\sum_{n}\int\frac{d^{3}k}{(2\pi)^{3}} e^{- i \bk\cdot\bx} 
		\langle T_{\tau}J^{\mu}(x) T_S^{\nu\rho}(0) \rangle \,,
\end{split}
\end{equation}
where we define $k^{\mu}=(0,\bk)$ and $T_\tau$ denotes the imaginary time ordering.
The two point correlator is computed with the Feynman rule in flat spacetime:
\begin{eqnarray}
& \dis\parbox{1.2cm}{\includegraphics[width=.08\textwidth]{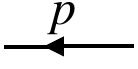} } = \frac{-\bar{\sigma}^{\mu}p_{\mu}}{p^{2}}\,,\\
&\dis \parbox{1.7cm}{\includegraphics[width=.11\textwidth]{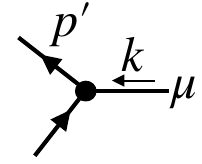} }= \sigma^{\mu} \,,\\
&\dis \parbox{1.8cm}{\includegraphics[width=.12\textwidth]{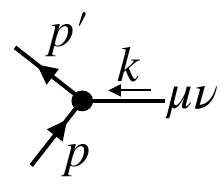} }
= \frac{1}{4}\sigma^{\lambda}\Bigl[\delta_{\lambda}^{\mu}(p^{\nu}+p'^{\nu})+\delta_{\lambda}^{\nu}(p^{\mu}+p'^{\mu}) -2\eta^{\mu\nu}(p_{\lambda}+p'_{\lambda})\Bigr].
\end{eqnarray}
In momentum space, at one-loop order, we get
\begin{equation}
\begin{split}
\label{eq:Gh}
 G^{\mu\nu\rho}(k) h_{\nu\rho} 
 & =  (-1)T\sum_{n}\int_\bp \frac{p_{\alpha}p'_{\beta} }{p^{2}p'^{2}} \,
 	\tr\Bigl[\bar{\sigma}^{\alpha} \sigma^{\mu}\bar{\sigma}^{\beta}\sigma^{\lambda} \Bigr] \\
 & \qquad\qquad\times
  \frac{h_{\nu\rho}}{4}
	\Bigl(
		\delta^{\nu}_{\lambda}(p^{\rho}+p'^{\rho})
		+\delta^{\rho}_{\lambda}(p^{\nu}+p'^{\nu})
		-2\eta^{\nu\rho}(p_{\lambda}+p'_{\lambda})
	\Bigr)  \\
 & = -\calI_{\alpha\beta\gamma}(k) 
 	\Bigl[
 		\eta^{\mu\beta}(h^{\gamma\alpha}-\eta^{\alpha\gamma}h_{\rho}^{\rho})
		-\eta^{\beta\alpha}(h^{\mu\gamma}-\eta^{\mu\gamma}h_{\rho}^{\rho}) \\
 &\qquad\qquad
		+\eta^{\mu\alpha}(h^{\beta\gamma}-\eta^{\beta\gamma}h_{\rho}^{\rho})
		+ i \varepsilon^{\mu\beta\lambda\alpha}
			(h_{\lambda}^{\gamma}-\delta_{\lambda}^{\gamma}h_{\rho}^{\rho})
	\Bigr], 
\end{split}
\end{equation}
where we denote $\int_\bp = \int\frac{d^{3}p}{(2\pi)^{3}}$, $p'=p+k$ and $p^{\mu}=(i\pi T (2n+1)+\mu, \bp)$, and the antisymmetric tensor $\varepsilon^{\mu\nu\rho\sigma}$ is normalized as $\varepsilon^{0123}=+1$.
Also we introduced
\begin{equation}
 \calI_{\alpha\beta\gamma}(k) 
 := T\sum_{n}\int_\bp \frac{p_{\alpha}p'_{\beta}(p_{\gamma}+p'_{\gamma}) }{p^{2}p'^{2}} \,.
\end{equation}
In order to compute the liner response to the gravitational field, we expand $\calI_{\alpha\beta\gamma}(k)$ in terms of $k$ and define $\calI^{(n)}_{\alpha\beta\gamma}(k)$ to be the $O(k^{n})$ contribution of $\calI_{\alpha\beta\gamma}(k)$.
In particular, we find
\begin{equation}
\begin{split}
\calI^\one_{\alpha\beta\gamma}(k)
&= T\sum_{n}\int_\bp \frac{1}{(p^{2})^{2}}
	\Bigl(
 		p_{\alpha}p_{\beta}k_{\gamma}
 		+2p_{\alpha}k_{\beta}p_{\gamma}
		-2p_{\alpha}p_{\beta}p_{\gamma}\frac{2p\cdot k}{p^{2}} 
	\Bigr) \,, \\
\calI^\two_{\alpha\beta\gamma}(k) 
& = T\sum_{n}\int_\bp
	\Bigl(
		-p_{\alpha}p_{\beta}k_{\gamma}\frac{2p\cdot k}{p^{2}}
		-2p_{\alpha}k_{\beta}p_{\gamma}\frac{2p\cdot k}{p^{2}}\\
&\qquad\qquad\qquad\qquad
		+p_{\alpha}k_{\beta}k_{\gamma}
		-2p_{\alpha}p_{\beta}p_{\gamma}\frac{k^{2}}{p^{2}}
		+8p_{\beta}p_{\gamma}p_{\alpha}\frac{(p\cdot k)^{2}}{(p^{2})^{2}}
	\Bigr).
\label{eq:integral}
\end{split}
\end{equation}

There are two steps to compute the momentum integrals.
First, the radial integral is systematically evaluated with the following formulas:
\begin{eqnarray}
\dis
F_{n,m}&:=&T\sum_{l}\int_\bp\frac{|\bp|^{2n-2m}p_{0}^{2m+1} }{(p^{2})^{n+2}} 
 = -F_{0,0}\frac{2\Gamma(m+1/2)}{\Gamma(m-n-1/2)\Gamma(n+2)}\,,\\
\dis
\tF_{n,m}&:=&T\sum_{l}\int_\bp\frac{|\bp|^{2n-2m}(p_{0})^{2m}}{(p^{2})^{n+1}}
 = \tF_{0,0}\frac{\Gamma(m-1/2)}{\Gamma(m-n-1/2)\Gamma(n+1)}
\end{eqnarray}
with
\begin{eqnarray}
&\dis F_{0,0}=-\frac{1}{8\pi^{2}}\mu, \\
&\dis \tF_{0,0}=\frac{1}{8\pi^{2}}\Bigl(\mu^{2}+\frac{\pi^{2}}{3} T^{2} \Bigr) \,.
\end{eqnarray}
The above formulas are proved in Appendix~\ref{app:F-tF}.
%The derivation is shown in Tomoya's note.
Second, for the angle integrals in Eq.~\eqref{eq:integral}, we replace the momentum products $p_{\mu_1}\cdots p_{\mu_j}$, as shown in Eq.~\eqref{eq:p_replacement}.
Then we obtain
\begin{equation}
\begin{split}
\label{eq:integral_k}
 \calI^\one_{\alpha\beta\gamma}
&=
\Bigl(\tF_{1,1}+4\frac{\tF_{2,1}}{3}\Bigr)\xi_{\alpha}\xi_{\beta}k_{\gamma}
+\Bigl(2\tF_{1,1}+4\frac{\tF_{2,1}}{3}\Bigr)\xi_{\alpha}\xi_{\gamma}k_{\beta}
+4\frac{\tF_{2,1}}{3}\xi_{\beta}\xi_{\gamma}k_{\alpha}\\
&\qquad + \Bigl(\frac{\tF_{1,0}}{3}+4\frac{\tF_{2,0}}{15}\Bigr)\Delta_{\alpha\beta}k_{\gamma}
+\Bigl(2\frac{\tF_{1,0}}{3}+4\frac{\tF_{2,0}}{15}\Bigr)\Delta_{\alpha\gamma}k_{\beta}
+4\frac{\tF_{2,0}}{15}k_{\alpha}\Delta_{\gamma\beta}\\
&=
\frac{\tF_{0,0}}{2}(-\xi_{\alpha}\xi_{\gamma}k_{\beta}+\xi_{\beta}\xi_{\gamma}k_{\alpha}-\Delta_{\alpha\gamma}k_{\beta}+\Delta_{\gamma\beta}k_{\alpha}) \,,\\
 \calI^\two_{\alpha\beta\gamma}
&=
\Bigl(2F_{1,0}+F_{0,0}+\frac{16F_{2,0}}{15} \Bigr)\xi_{\alpha}k_{\beta}k_{\gamma}
 +\Bigl(\frac{2}{3}F_{1,0}+ \frac{16F_{2,0}}{15} \Bigr)\xi_{\beta}k_{\alpha}k_{\gamma} 
 + \Bigl(\frac{4}{3}F_{1,0}+ \frac{16F_{2,0}}{15}\Bigr)\xi_{\gamma}k_{\alpha}k_{\beta} \\
&\qquad -\Bigr(2F_{1,1}+\frac{8F_{2,1}}{3}\Bigl)k^{2}\xi_{\alpha}\xi_{\beta}\xi_{\gamma}
-\Bigl(\frac{2}{3}F_{1,0}+\frac{8F_{2,0}}{15}\Bigr)k^{2}\bigl( \xi_{\alpha}\Delta_{\beta\gamma}+ \xi_{\beta}\Delta_{\gamma\alpha}+ \xi_{\gamma}\Delta_{\alpha\beta} \bigr)\\
&=
\frac{F_{0,0}}{6}\bigl(\xi_{\alpha}k_{\beta}k_{\gamma}
 +\xi_{\beta}k_{\alpha}k_{\gamma} 
 -2\xi_{\gamma}k_{\alpha}k_{\beta}-k^{2}\xi_{\alpha}\xi_{\beta}\xi_{\gamma} 
 + k^{2}\xi_{\alpha}\Delta_{\beta\gamma}+ k^{2}\xi_{\beta}\Delta_{\gamma\alpha}+k^{2} \xi_{\gamma}\Delta_{\alpha\beta} 
  \bigr)\,,
\end{split}
\end{equation}
where we denote $\Delta_{\mu\nu} = \xi_\mu\xi_\nu - \eta_{\mu\nu}$.
As a result, the $O(k)$ contribution in Eq.~\eqref{eq:Gh} is written as
\begin{equation}
\begin{split}
& G_{(1)}^{\mu\nu\rho}(k)h_{\nu\rho} \\
&=-\frac{1}{2}\tF_{0,0}(-\xi_{\alpha}\xi_{\gamma}k_{\beta}+\xi_{\beta}\xi_{\gamma}k_{\alpha}-\Delta_{\alpha\gamma}k_{\beta}+\Delta_{\gamma\beta}k_{\alpha})\\
&\quad\times \bigl(\eta^{\mu\beta}(h^{\gamma\alpha}-\eta^{\alpha\gamma}h_{\rho}^{\rho})
-\eta^{\beta\alpha}(h^{\mu\gamma}-\eta^{\mu\gamma}h_{\rho}^{\rho})
+\eta^{\mu\alpha}(h^{\beta\gamma}-\eta^{\beta\gamma}h_{\rho}^{\rho})
+ i  \varepsilon^{\mu\beta\lambda\alpha}(h_{\lambda}^{\gamma}-\delta_{\lambda}^{\gamma}h_{\rho}^{\rho})
\bigr)\\
&=-2 i  \varepsilon^{0\mu jk}\tF_{0,0}
h_{k}^{0}k_{j} \,,
\end{split}
\end{equation}
which reproduces the CVE:
\begin{equation}
 \langle J^\mu_\one \rangle 
 = - \frac{1}{8\pi^{2}}\biggl(\mu^{2}+\frac{\pi^{2}}{3} T^{2} \biggr)
 	\varepsilon^{0\mu jk} \partial_{j}h_{k}^{0} \\
 = \frac{1}{4\pi^2} \biggl(\mu^{2}+\frac{\pi^{2}}{3} T^{2} \biggr) \omega^\mu 
\end{equation}
with $\omega^\mu = \varepsilon^{\mu\nu\rho\sigma}\xi_\nu\partial_\rho h_{\rho\lambda}\xi^\lambda/2$.
Similarly, the $O(k^2)$ parts are computed as
\begin{equation}
\begin{split}
G_{(2)}^{\mu\nu\rho}(k)h_{\nu\rho} 
&=-
\frac{1}{3}F_{0,0}\Bigl[
 -  h^{0\alpha}k_{\alpha}k^{\mu} +  h^{\mu0} k^{2} 
 + \xi^{\mu}( h^{\gamma\alpha}k_{\alpha}k_{\gamma} 
 +   2h^{00}k^{2}
  -   h^{\alpha}_{\alpha}k^{2} )
\Bigr]\,.
\end{split}
\end{equation}
For the stationary gravitational field ($\partial_0 h_{\mu\nu} = 0$), we eventually derive
\begin{equation}
\begin{split}
\label{eq:J_kubo}
 \langle J_{(2)}^{\mu}\rangle
 & = -\frac{\mu}{48\pi^2}\Bigl[
   \partial_{\alpha}\partial^{\mu}h^{0\alpha}-\partial^{2} h^{\mu0} 
 - \xi^{\mu}( \partial_{\alpha}\partial_{\gamma}h^{\gamma\alpha}
   -  \partial^{2} h^{\alpha}_{\alpha}
 +  2\partial^{2}h^{00}
 )\Bigr] \\
 & \simeq \frac{\mu}{24\pi^{2}} 
 	\biggl[
 		R^{0\mu}
  		-\frac{1}{2} \xi^{\mu}R
 		+ 2 \xi^{\mu}  R^{00} 
 	\biggr] \,,
\end{split}
\end{equation}
where we employ
\begin{equation}
\begin{split}
 & R_{\mu\nu}\simeq
\frac{1}{2}(\partial_{\nu}\partial_{\mu}h_{\rho}^{\rho}-\partial_{\nu}\partial^{\rho}h_{\rho\mu} -\partial_{\rho}\partial_{\mu}h_{\nu}^{\rho}+\partial_{\rho}\partial^{\rho}h_{\nu\mu} ) \,, \\
 & R\simeq \partial^{2}h_{\rho}^{\rho}-\partial^{\mu}\partial^{\rho}h_{\rho\mu} \,.
\end{split}
\end{equation}
The above current $\langle J_\two^\mu\rangle$ is consistent with $J^\mu_{\two\eq}$ in Eq.~\eqref{eq:JT2_eq}.

\subsection{Riemann normal coordinate expansion}\label{sec:Riemann}
We reproduce the fermionic current in Eq.~\eqref{eq:JT2_eq}, by employing the Riemann normal coordinate (RNC) expansion~\cite{parker2009quantum}.
We first look for the propagator that satisfies
\begin{equation}
 \label{eq:S}
 i\gamma^\mu\nabla^{x}_\mu S(x,x') = {|-g(x)|}^{-1/2}\delta(x-x') \,,
\end{equation}
where we denote $g = \det(g_{\mu\nu})$ and $S_{ab}(x,x') = - i \langle T \psi_a (x) \bar\psi_b (x')\rangle$.
Here $\nabla^{x}_\mu$ is the diffeomorphic and local Lorentz covariant derivative with respect to $x$, and the spin connection is defined as
\begin{equation}
\label{eq:nabla}
 \nabla_\mu \psi
 = \biggl(
 	\partial_\mu
 	- \frac{i}{4}\omega_{\mu ab }\sigma^{ab}
   \biggr) \psi\,,
  \quad
  \sigma^{ab} = \frac{i}{2}[\gamma^a,\gamma^b]\,,
  \quad
  \omega_{\mu ab} = e_{\nu a}(\partial_\mu e_{~b}^\nu + \Gamma^\nu_{\rho\mu}e_{~b}^\rho ) \,.
\end{equation}
Further we introduce the following bispinor (not scalar) propagator:
\begin{equation}
 \label{eq:GS}
 i\gamma^\mu\nabla^x_\mu G(x,x') = S(x,x') \,.
\end{equation}
From Eqs.~\eqref{eq:S} and~\eqref{eq:GS}, we find
\begin{equation}
 \label{eq:G}
-|-g(x)|^{1/2}  
 \biggl(
 	\nabla^\mu\nabla_\mu
 	+ \frac{1}{4} R
 \biggr)
 	G(x,x')
 = \delta(x-x') \,.
\end{equation}

Let us now introduce the RNC.
We define the normal coordinate $y$ and the origin is at $x'$, that is, we replace $x\to y$ and $x' \to 0$.
In order to evaluate above Green's function, we perform the RNC expansion, as follows:
\begin{eqnarray}
 \label{eq:eta}
 & \dis g_{\mu\nu}(x) 
  = \eta_{\mu\nu} + \frac{1}{3} R_{\mu\alpha\nu\beta} y^\alpha y^\beta + \cdots\,, \\
 & \dis |-g(x)|
 = 1 + \frac{1}{3} R_{\alpha\beta} y^\alpha y^\beta + \cdots\,,\\
 & \dis \Gamma^\rho_{\mu\nu} (x)
 = \frac{2}{3} 
 		{R^\rho}_{(\mu\nu)\alpha} y^\alpha
    + \cdots  \,,\\ 
 & \dis
 	e^a_{~\mu} (x)
 	= e^a_{~\lambda}
 		\biggl(
 			\delta^\lambda_\mu
 			+ \frac{1}{6} {R^\lambda}_{\nu\mu\rho} y^\nu y^\rho
 		\biggr) + \cdots \,, \\   
 & \dis \omega_{\mu \alpha\beta}(x) 
 = \frac{1}{2}  R_{\alpha\beta\mu\nu} y^\nu + \cdots \,,
\end{eqnarray}
where $\cdots$ denotes the $O(R^2)$ or $O(\partial R)$ contribution.
Note that all of the above curvature tensors are evaluated at $y=0$.
We thus reduce Eq.~\eqref{eq:G}, as follows:
\begin{equation}
\begin{split}
\label{eq:G_RNCE}
 \delta(y) 
 & = \biggl[
 		-\eta^{\mu\nu}\partial^y_\mu\partial^y_\nu
 		- \frac{1}{4} R 
 	 	- \frac{1}{6} R_{\alpha\beta} y^\alpha y^\beta \partial_y^2
 	 	+ \frac{1}{3} R_{\mu\alpha\nu\beta} y^\alpha y^\beta \partial_y^\mu\partial_y^\nu \\
 & \qquad
 	 	- \frac{2}{3} R_{\alpha\beta} y^\alpha \partial_y^\beta
 	 	- \frac{i}{4} R_{\mu\nu\alpha\beta}\sigma^{\alpha\beta}y^\mu\partial_y^\nu
 	 \biggr]
 	 	G(x,x') + \cdots \,.
\end{split}
\end{equation}
Now we perform the Fourier transformation:
\begin{equation}
 G(x,x') = \int_p e^{ip\cdot y} G(p)
\end{equation}
with $\int_p = \int \frac{d^4p}{(2\pi)^4}$.
Then $G(p)$ obeys
\begin{equation}
\begin{split}
 1
 &=\biggl[
 	\eta^{\mu\nu}p_\mu p_\nu
 	- \frac{1}{4} R
 	- \frac{1}{6} R_{\alpha\beta}\partial^\alpha_p\partial^\beta_p p^2
 	+ \frac{1}{3} R_{\mu\alpha\nu\beta} \partial^\alpha_p\partial_p^\beta p^\mu p^\nu \\
 & \qquad
 	+ \frac{2}{3} R_{\alpha\beta}\partial_p^\alpha p^\beta
 	- \frac{i}{4} R_{\mu\nu\alpha\beta}\sigma^{\alpha\beta}\partial_p^\mu p^\nu
 	+ \cdots
 \biggr]
 	G(p) \\
 & :=
 	\Bigl(
 		p^2
 		+ \calD 
 	\Bigr) G(p) \,,
\end{split}
\end{equation}
where we denote $p^2 = \eta^{\mu\nu}p_\mu p_\nu$ and $\calD$ is the derivative operators of $O(R)$.
The above equation is solved sequentially, as follows:
\begin{equation}
\begin{split}
 G(p)
 & = \frac{1}{p^2}
 	\Bigl[
 		1 - \calD  G(p)
 	\Bigr]
 	+ \cdots 
  = \frac{1}{p^2}
 	\biggl[
 		1 - \calD \frac{1}{p^2}
 	\biggr] 
 	+ \cdots \\
 & = \frac{1}{p^2}
 	 - \frac{1}{12 (p^2)^2} R
 	 + \frac{2}{3 (p^2)^3 } R_{\alpha\beta} p^\alpha p^\beta 
 	+ \cdots \,.
\end{split}
\end{equation}
Thus we obtain
\begin{equation}
\begin{split}
 S(x,x')
 & = i\gamma^\mu (x) \nabla_\mu \int_p e^{ip\cdot y} G(p) \\
 & = \int_p e^{ip\cdot y} \, 
 		\biggl(
 			- \frac{\gamma\cdot p}{p^2}
 			+ \frac{1}{2 (p^2)^2}R_{\mu\nu}\gamma^\mu p^\nu
 	 		+ \frac{\gamma\cdot p}{12 (p^2)^2} R
 	 		- \frac{2\gamma\cdot p}{3 (p^2)^3 } R_{\alpha\beta} p^\alpha p^\beta 
 	 	\biggr)
 		+ \cdots \,.
\end{split}
\end{equation}

Performing the Wick rotation, we obtain the vector current as 
\begin{equation}
\begin{split}
 J^\mu
 & = - \tr\Bigl[S(x,x)\gamma^\mu\Bigr] \\
 & = T\sum_n \int_\bp
 	\biggl[
 		\frac{4p^\mu}{p^2}
 		- \frac{2p^\nu}{(p^2)^2} {R_{\nu}}^\mu 
 		- \frac{p^\mu}{3 (p^2)^2} R
 		+ \frac{8p^\mu p^\alpha p^\beta }{3 (p^2)^3 } R_{\alpha\beta} 
 	\biggr] 
\end{split}
\end{equation}
with $p^\mu = (i\pi T (2n+1) + \mu, \bp)$.
The above current is evaluated with Eq.~\eqref{eq:p_replacement}.
The first term in the above integrand gives the ordinary charge density.
The other terms are linear in the curvature tensor, and thus the curvature-induced current $J^\mu_\text{curv}$ is calculated as
\begin{equation}
\begin{split}
\label{eq:J_RNC}
 J^\mu_\text{curv}
 & = 
 	 - 2 {R^\mu}_0 F_{0,0}
 	 - \frac{1}{3}\xi^\mu  R F_{0,0} 
 	 + \frac{8}{3} \xi^\mu R_{00} F_{1,1}
 	 + \xi^\mu \frac{8}{3} R_{00} F_{1,0}
 	 - \frac{16}{9} {R^\mu}_0 F_{1,0} 
 	 - \frac{8}{9} \xi^\mu R F_{1,0}\\
 & = 2\cdot \frac{\mu}{24\pi^{2}} 
 	\biggl[
 		R^{0\mu}
  		-\frac{1}{2} \xi^{\mu}R
 		+ 2 \xi^{\mu}  R^{00} 
 	\biggr] \,.
\end{split}
\end{equation}
This is again the same as Eq.~\eqref{eq:JT2_eq} up to the factor $2$, which comes from the right- and left-handed contributions.

\section{Evaluation of $F_{n,m}$ and $\tF_{n,m}$}\label{app:F-tF}
In this appendix, we derive the formulas of the momentum integrals in Euclidean spacetime, which are applied in Appendix~\ref{app:alternative}.
We first compute the following integral:
\begin{eqnarray}
F_{n,m}&:=&T\sum_{l}\int_\bp\frac{|\bp|^{2n-2m}p_{0}^{2m+1} }{(p^{2})^{n+2}} \,.
\end{eqnarray}
This obeys the recursion relation $F_{n,m} = F_{n-1,m-1}+F_{n,m-1}$, and the solutions are given by
\begin{equation}
\begin{split}
F_{n,m} 
 & = \sum_{j=0}^{m} \frac{m!}{j!(m-j)!}F_{n-j,0} \,. 
\label{eq:nm}
\end{split}
\end{equation}
We calculate $F_{n,0}$ as
\begin{equation}
\begin{split}
 F_{n,0}&= T\sum_{l}\int_\bp\frac{|\bp|^{2n} p_0}{(p^2)^{n+2}}\\
&=  T\sum_{l}\int\frac{d\Omega d\bp^{2}}{(2\pi)^{3}}\frac{1}{2}(\bp^2)^{n+1/2}\frac{1}{(n+1)!}\Bigl(\frac{\partial}{\partial\bp^{2}}\Bigr)^{n}\ \frac{p_0}{(p^2)^{2}}\\
&= \frac{(-1)^n \Gamma(n+3/2)}{(n+1)!\Gamma(3/2)} T\sum_{l}\int\frac{d\Omega d\bp^{2}}{(2\pi)^{3}}\frac{1}{2}(\bp^2)^{1/2} \frac{p_0}{(p^2)^{2}}\\
 &=\frac{(-1)^{n}\Gamma(n + 3/2 )}{\Gamma(3/2)(n+1)!} F_{0,0} \,.
\end{split}
\end{equation}
Therefore, we obtain
\begin{equation}
\begin{split}
\label{eq:Fnm}
F_{n,m} 
&=F_{0,0} \sum_{j=0}^{m} \frac{m!}{j!(m-j)!}\frac{(-1)^{n-j} \Gamma(n-j + 3/2 )}{\Gamma(3/2)(n-j+1)!}\\
&=-F_{0,0} \frac{2\Gamma(m+1/2)}{\Gamma(m-n-1/2)\Gamma(n+2)} \,.
\end{split}
\end{equation}
One can check that this solution satisfies the recursion relation:
\begin{equation}
\begin{split}
&F_{n-1,m-1} +F_{n,m-1}- F_{n,m}\\
 &= \sum_{j=0}^{m-1} \frac{(m-1)!}{j!(m-1-j)!}F_{n-1-j,0}+\sum_{j=0}^{m-1} \frac{(m-1)!}{j!(m-1-j)!}F_{n-j,0}-\sum_{j=0}^{m} \frac{m!}{j!(m-j)!}F_{n-j,0}\\
&= \sum_{j=1}^{m} \frac{m!}{j!(m-j)!}\frac{j}{m}F_{n-j,0}+\sum_{j=0}^{m-1} \frac{m!}{j!(m-j)!}\frac{m-j}{m}F_{n-j,0}-\sum_{j=0}^{m} \frac{m!}{j!(m-j)!}F_{n-j,0}\\
&=0.
\end{split}
\end{equation}
The overall factor $F_{0,0}$ in Eq.~\eqref{eq:Fnm} is computed as
\begin{equation}
\begin{split}
F_{0,0}&=T\sum_{l}\int_\bp \frac{1}{2|\bp|}\frac{\partial}{\partial|\bp|}\frac{p_0}{( p^2 )}\\
&=-\frac{1}{4\pi^{2}}T\sum_{l}\int_{0}^{\infty} d|\bp|\frac{p_0}{ p^2 }\\
&=-\frac{1}{4\pi^{2}}T\sum_{l}\int_{0}^{\infty} d|\bp|\frac{1}{2}\biggl(\frac{1}{p_0-|\bp|}+\frac{1}{p_0+|\bp|}\biggr)\\
&=-\frac{1}{8\pi^{2}}\int_{0}^{\infty} d|\bp|\biggl(\frac{1}{e^{\beta(|\bp|-\mu)}+1}-\frac{1}{e^{\beta(|\bp|+\mu)}+1}\biggr)\\
&=-\frac{\mu}{8\pi^{2}}.
\end{split}
\end{equation}

Also we evaluate
\begin{equation}
\begin{split}
\tilde{F}_{n,m}: =  T\sum_{l}\int_\bp\frac{(\bp^{2})^{n-m}(p_0)^{2m} }{( p^2 )^{1+n}},
\end{split}
\end{equation}
which obeys the same recursion relation $\tilde{F}_{n,m} = \tilde{F}_{n-1,m-1}+\tilde{F}_{n,m-1}$. 
In the same manner for $F_{n,m}$, we get
\begin{eqnarray}
\tilde{F}_{n,m} &=& \sum_{j=0}^{m} \frac{m!}{j!(m-j)!}\tilde{F}_{n-j,0} , 
\label{eq:nm}
\\
\tilde{ F}_{n,0} &=&\frac{(-1)^{n}\Gamma(n + 3/2 )}{\Gamma(3/2)n!} \tilde{F}_{0,0},
\\
\tilde{F}_{n,m} &=&\tilde{F}_{0,0} \sum_{j=0}^{m}(-1)^{n-j} \frac{m!}{j!(m-j)!}\frac{\Gamma(n-j + 3/2 )}{\Gamma(3/2)(n-j)!}
\notag
\\
&=&\tilde{F}_{0,0} \frac{\Gamma(m-1/2)}{\Gamma(m-n-1/2)\Gamma(n+1)} .
\end{eqnarray}
The overall factor $\tF_{0,0}$ is calculated as
\begin{equation}
\begin{split}
\tilde{F}_{0,0}
&=\frac{1}{2\pi^{2}}T\sum_{l}\int_{0}^{\infty} d|\bp|\frac{|\bp|^2}{ p^2 }\\
&=\frac{1}{2\pi^{2}}T\sum_{l}\int_{0}^{\infty} d|\bp|\frac{|\bp|}{2}\biggl(\frac{1}{p_0-|\bp|}-\frac{1}{p_0+|\bp|}\biggr)\\
&=\frac{1}{4\pi^{2}}\int_{0}^{\infty} d|\bp| |\bp|\biggl(\frac{1}{e^{\beta(|\bp|-\mu)}+1}+\frac{1}{e^{\beta(|\bp|+\mu)}+1}-1\biggr)\\
&=\frac{1}{8\pi^{2}}\left(\mu^2+\frac{\pi^2}{3}T^2\right) {\rm +(const)}.
\end{split}
\end{equation}
Here, (const) denotes the divergent term that is independent of $T$ and $\mu$.

\section{Wigner function under the dynamical gravity~\eqref{eq:Rmu2_st} and~\eqref{eq:Rmu2_nonst}}
\label{app:dynamical_Wigner}
In this appendix, we derive the Wigner function under a time-dependent gravitational field, Eqs.~\eqref{eq:Rmu2_st} and~\eqref{eq:Rmu2_nonst}.
Plugging $\calR^\mu$ and $P_\mu$ given by Eqs.~\eqref{eq:Rmu} and~\eqref{eq:P}, we write down the kinetic equation~\eqref{eq:R1-curved} as
\begin{equation}
\begin{split}
\label{eq:kinetic_eq}
 0 &=
 	 \delta(p^2)  
 	 \Bigl[
 	 	p\cdot D 
 	 	+ \hbar D_\mu \Sigma^{\mu\nu}_n D_\nu
 	 \Bigr] f
 	 +\delta(p^2)   \hbar^2 D_\mu (\Sigma^{\mu\nu}_u - \Sigma^{\mu\nu}_n) D_\nu \tf_{\one} \\
 & \quad
	   		+ \frac{\hbar^2}{p^2}D^\mu
    		 \Bigl[
	 	    	 - p_\mu Q\cdot p 
    			  + 2 p^\nu 
    			  	\bigl( 
    			  		T_{[\mu} p_{\nu]} 
    			  		+ S_{\alpha\mu\nu} p^\alpha 
    			  	\bigr) 
    		\Bigr] \delta(p^2) f \\
   	 & \quad
	  + \frac{\hbar^2\delta(p^2)}{2p^2}D^\mu
	 	\biggl[
      	  	 \varepsilon_{\mu\nu\rho\sigma}p^\nu
  		 	 	D^\rho \Sigma_n^{\sigma\lambda} D_\lambda
      	  	 - \Sigma_{\mu\nu}^u
      	  	 	\bigl(
      	  	 		{\tilde{R}}^{\alpha\beta\nu\rho} 
      	  	  			p_\rho p_\alpha \partial_\beta^p
      	  			 + 2 p\cdot D \Sigma_n^{\nu\rho} D_\rho
      	  		\bigr)
      	  \biggr] f \\
  	& \quad
	+ \hbar^2
	\biggl(
 		- \frac{1}{8} \nabla_\lambda R_{\mu\nu}\partial_p^\lambda\partial_p^\nu
  		-\frac{1}{24} \nabla_\lambda {R^\rho}_{\sigma\mu\nu} \partial_p^\lambda
  		 	\partial_p^\nu \partial_p^\sigma p_\rho
  		+\frac{1}{8}{R^\rho}_{\sigma\mu\nu} \partial_p^\nu \partial_p^\sigma D_\rho
 	\biggr) p^\mu \delta(p^2) f \,.
\end{split}
\end{equation}
Here $\tilde f_\zero$ and $\tf_\one$ involved in $f$ have already been obtained in Eqs.~\eqref{eq:tildef0} and~\eqref{eq:tildef1}.
After some computation keeping $O(h_{\mu\nu})$ together with $\hbar^2 p\cdot D f \sim O(\hbar^3)$ and $D_\mu f \sim O(h_{\mu\nu})$, we reduce the kinetic equation~\eqref{eq:kinetic_eq} to
\begin{equation}
\begin{split}
\label{eq:kineticeq_eq_weak}
 &\delta(p^2)
 \biggl[
 	\biggl(
 		1
 		+ \frac{1}{p^2} h^{\mu\nu} p_\mu p_\nu
 	\biggr)
 		p\cdot \partial 
 	- h^{\mu\nu} p_\mu \partial_\nu 
 	+ \Gamma^\rho_{\mu\nu} p^\mu p_\rho \partial_p^\nu 
 	+ \hbar 
 		\biggl(
 			- \frac{1}{2} \Sigma^{\mu\nu}_n R_{\alpha\beta\mu\nu} p^\alpha\partial^\beta_p
 		\biggr) \\
 & \ 
 	+ \hbar^2
 		\biggl(
 			- \frac{1}{24} p\cdot \nabla R_{\alpha\beta} \partial_p^\alpha \partial_p^\beta
 			- \frac{1}{24} p^\mu p^\rho \partial_p^\nu \partial_p^\sigma 
 					\partial_p \cdot \nabla R_{\rho\sigma\mu\nu} 
 			- \Sigma_{\mu\nu}^u \frac{n_\lambda}{2p\cdot n}
   	 		\nabla^\mu \tilde{R}^{\alpha\beta\nu\lambda} p_\alpha\partial^p_\beta
   	 	\biggr)
 \biggr] f
 = 0 \,,
\end{split}
\end{equation}
which yields the second order fluctuation as
\begin{equation}
\begin{split}
\label{eq:deltaf2}
 \tf_{\two}
 & = \Sigma_{\mu\nu}^u \frac{n_\lambda}{2p\cdot n}\frac{k^\mu}{k\cdot p} 
   	 		\tilde{R}^{\alpha\beta\nu\lambda} p_\alpha \beta_\beta f_\flat'
 	+ \frac{1}{24} R_{\alpha\beta} \beta^\alpha \beta^\beta  f_\flat''
 	+ \frac{k\cdot\beta}{24k\cdot p} p^\mu p^\rho \beta^\nu \beta^\sigma 
 			 R_{\rho\sigma\mu\nu}  f_\flat''' \,.
\end{split}
\end{equation}
Collecting Eqs.~\eqref{eq:tildef0} and~\eqref{eq:tildef1}, the above $\tf_{\zero,\one,\two}$ and the general form of the Wigner function~\eqref{eq:Rmu2}, we find
\begin{equation}
\begin{split}
\label{eq:Rmu2_dyn_app}
 & \calR^\mu_\two/(2\pi) \\
 & = \delta(p^2)
  \biggl[
  		p^\mu 
 		\biggl(
 			\Sigma_{\eta\nu}^u \frac{n_\lambda}{2p\cdot n}\frac{k^\eta}{k\cdot p} 
   	 		\tilde{R}^{\alpha\beta\nu\lambda} p_\alpha \beta_\beta f_\flat'
 			+ \frac{1}{24} R_{\alpha\beta} \beta^\alpha \beta^\beta  f_\flat''
 			+ \frac{k\cdot\beta}{24k\cdot p} p^\eta p^\rho \beta^\nu \beta^\sigma 
 			 R_{\rho\sigma\eta\nu}  f_\flat'''	
   	 	\biggr) \\
  & \quad
   	 	- \frac{1}{2ik\cdot p} \Sigma^{\mu\nu}_u
   	 		\Sigma^{\lambda\eta}_n (-ik_\nu)
   	 		R_{\alpha\beta\lambda\eta} p^\alpha \beta^\beta 
   	 		f_\flat' 
		 + \frac{1}{2p^2} \varepsilon^{\mu\nu\rho\sigma}
		 p_\nu \Sigma_{\sigma\lambda}^n 
		 	 D_\rho D^\lambda (f_\flat + \tf_\zero)\\
 & \quad
 		 - \Sigma^{\mu\nu}_u
   	 		\frac{n^\sigma}{2p\cdot n}
   	 		\tilde{R}_{\alpha\beta\nu\sigma} p^\alpha\beta^\beta f_\flat'
   	 \biggr] 
   	 + \frac{1}{p^2}
    		 \Bigl[
	 	    	 - p^\mu Q\cdot p 
    			  + 2 p_\nu 
    			  	\bigl( 
    			  		T^{[\mu} p^{\nu]} 
    			  		+ S^{\alpha\mu\nu} p_\alpha 
    			  	\bigr) 
    		\Bigr] \delta(p^2) f_\flat \,.
%  & =: (\calR^\mu_{\two\,\text{dep}} + \calR^\mu_{\two\,\text{indep}})/(2\pi) \,.
\end{split}
\end{equation}
In the above equation, there are the four frame-dependent terms.
However, the dependence is totally cancelled out, as shown in the following.
These are rewritten as
\begin{eqnarray}
 p^\mu \Sigma_{\eta\nu}^u \frac{n_\lambda}{2p\cdot n}\frac{k^\eta}{k\cdot p} 
   	\tilde{R}^{\alpha\beta\nu\lambda} p_\alpha \beta_\beta f_\flat'
 &=& \biggl(
 	- \frac{1}{2} p^\mu \Sigma_{\nu\alpha}^u
 		\frac{n_\lambda}{p\cdot n}
   	 		\tilde{R}^{\alpha\beta\nu\lambda}  \beta_\beta 
   	 + \frac{1}{2} p^\mu \frac{k^\eta}{k\cdot p}\Sigma_{\eta\alpha}^u  \Sigma_{\nu\lambda}^n
   	 		{R}^{\alpha\beta\nu\lambda} \beta_\beta \notag\\
 && 
 	+ \frac{1}{4} \varepsilon^{\tau\mu\nu\alpha} p_\tau
 		\frac{n^\lambda}{p\cdot n} 
   	 		\tilde{R}_{\alpha\beta\nu\lambda}  \beta^\beta 
   	 +  \frac{1}{4} p^\alpha \varepsilon^{\eta\tau\mu\nu} p_\tau
 		\frac{n^\lambda}{p\cdot n}\frac{k_\eta}{k\cdot p} 
   	 		\tilde{R}_{\alpha\beta\nu\lambda}  \beta^\beta\notag\\
 && 
   	 +  \frac{1}{4} \varepsilon^{\alpha\eta\tau\mu} p_\tau
 		\frac{k_\eta}{k\cdot p} \Sigma^{\nu\lambda}_n
   	 		R_{\alpha\beta\nu\lambda}  \beta^\beta
   	 \biggr) f_\flat' \,,\\
 \frac{1}{2k\cdot p} \Sigma^{\mu\nu}_u
   	 		\Sigma^{\lambda\eta}_n k_\nu
   	 		R_{\alpha\beta\lambda\eta} p^\alpha \beta^\beta f_\flat'
 &=& \biggl(
 	- \frac{1}{2k\cdot p} \Sigma_u^{\nu\alpha} p^\mu
   	 		\Sigma^{\lambda\eta}_n k_\nu
   	 		R_{\alpha\beta\lambda\eta} \beta^\beta 
   	 + \frac{1}{2} \Sigma^{\mu\alpha}_u
   	 		\Sigma^{\lambda\eta}_n
   	 		R_{\alpha\beta\lambda\eta} \beta^\beta \notag\\
 && 
 	- \frac{1}{4k\cdot p} \varepsilon^{\nu\alpha\mu\rho} p_\rho
   	 		\Sigma^{\lambda\eta}_n k_\nu
   	 		R_{\alpha\beta\lambda\eta} \beta^\beta 
   	\biggr) f_\flat',\qquad \\
 - \Sigma^{\mu\nu}_u
   	 		\frac{n^\sigma}{2p\cdot n}
   	 		\tilde{R}_{\alpha\beta\nu\sigma} p^\alpha\beta^\beta f_\flat'
 &=& \biggl(
 		p^\mu \Sigma_u^{\nu\alpha}
   	 		\frac{n^\sigma}{2p\cdot n}
   	 		\tilde{R}_{\alpha\beta\nu\sigma} \beta^\beta
	 	+ \frac{1}{4} \varepsilon^{\nu\alpha\mu\rho} p_\rho 
   	 		\frac{n^\sigma}{p\cdot n} 
   	 		\tilde{R}_{\alpha\beta\nu\sigma} \beta^\beta\notag\\
 && 
 	- \frac{1}{2} \Sigma^{\mu\alpha}_u
   	 		\Sigma^{\lambda\eta}_n
   	 		R_{\alpha\beta\lambda\eta} \beta^\beta 
   	\biggr) f_\flat' \,,\\
 \frac{\varepsilon^{\mu\nu\rho\sigma}}{2p^2}
		 p_\nu \Sigma_{\sigma\lambda}^n 
 	 D_\rho D^\lambda (f_\flat + \tf_\zero)  
 &=& 
 	\biggl(
 		-\frac{1}{4p^2} \varepsilon^{\mu\nu\rho\sigma}
		 p_\nu p^\eta\frac{1}{k\cdot p} 
  		\tilde{R}_{\alpha\beta\sigma\eta}k_{\rho}p^\alpha \beta^\beta \notag \\
 &&
   	+ \frac{1}{4} \varepsilon^{\mu\nu\rho\sigma}
		 p_\nu \frac{n^\eta}{p\cdot n}
  	\frac{1}{k\cdot p}
  		\tilde{R}_{\alpha\beta\sigma\eta}k_{\rho}
   	p^\alpha \beta^\beta 
   	\biggr) f_\flat' \,,
\end{eqnarray}
where we use Eq.~\eqref{eq:Sigma_p}, $D_\rho D_\lambda f \simeq (-ik_\rho)(-ik_\lambda) \tf_\zero  +  (-ik_\rho) \Gamma^{\tau}_{\lambda\kappa} p_\tau \beta^\kappa f_\flat'$ and the second Bianchi identity~\eqref{eq:2ndBianchi} for $R_{\alpha\beta\rho[\lambda}k_{\tau]}$.
Hence, the four frame-dependent terms in Eq.~\eqref{eq:Rmu2_dyn_app} are recast into 
\begin{equation}
\begin{split}
&\quad 
 \biggl(
 	p^\mu \Sigma_{\eta\nu}^u \frac{n_\lambda}{2p\cdot n}\frac{k^\eta}{k\cdot p} 
  		\tilde{R}^{\alpha\beta\nu\lambda} p_\alpha \beta_\beta 
 	 + \frac{1}{2k\cdot p} \Sigma^{\mu\nu}_u \Sigma^{\lambda\eta}_n k_\nu
   	 		R_{\alpha\beta\lambda\eta} p^\alpha \beta^\beta 
 	 - \Sigma^{\mu\nu}_u\frac{n^\sigma}{2p\cdot n}
   		 \tilde{R}_{\alpha\beta\nu\sigma} p^\alpha\beta^\beta 
  \biggr) f_\flat' \\
& \quad
  + \frac{1}{2p^2} \varepsilon^{\mu\nu\rho\sigma}
		 p_\nu \Sigma_{\sigma\lambda}^n 
 	 D_\rho D^\lambda (f_\flat + \tf_\zero) \\
 & =  \frac{1}{4p^2} \varepsilon^{\mu\nu\rho\sigma}
		 p^\eta p_\nu \frac{k_{\rho} }{k\cdot p} 
  		\tilde{R}_{\alpha\beta\eta\sigma}
   	p^\alpha \beta^\beta f_\flat' \\  
  & = -\frac{1}{4p^2} 
  			{R_{\alpha\beta}}^{\mu\nu} p_\nu p^\alpha \beta^\beta f_\flat'  
  	 + \frac{1}{4p^2} 
  			{R_{\beta}}^{\nu} p^\mu  p_\nu \beta^\beta f_\flat'
  	 - \frac{1}{4} {R_{\beta}}^{\mu}  \beta^\beta f_\flat' \\
 & \quad
  	 - \frac{1}{4p^2} \frac{k\cdot \beta}{k\cdot p} 
  			{R_{\alpha}}^{\nu} p^\mu  p_\nu p^\alpha f_\flat'  
  	 + \frac{1}{4} \frac{k\cdot\beta }{k\cdot p} 
  			{R_{\alpha}}^{\mu} p^\alpha f_\flat'  \,.
\end{split}
\end{equation}
Inserting this into Eq.~\eqref{eq:Rmu2_dyn_app}, we finally derive Eqs.~\eqref{eq:Rmu2_st} and~\eqref{eq:Rmu2_nonst}.

\section{Angle integrals}\label{app:angle_int_formulas}
In this appendix, we derive the integral formulas in terms of the momentum angle valuables.
We introduce the following function for the angular integral:
\begin{eqnarray}
 & \dis
  I^{j_1\cdots,j_n} (x)
  = x \int \frac{d\Omega}{4\pi} 
 	\frac{\hp^{j_1} \cdots \hp^{j_n}}{x-\hbk\cdot\hbp} \,,
\end{eqnarray}
where $x$ involves the positive infinitesimal imaginary part $+i\eta$ and we define $x=k_0/|\bk|$.
By definition, we can readily show Eqs.~\eqref{eq:Ix0Ix1}-\eqref{eq:Ix3Ix4} and Eqs.~\eqref{eq:Ix2trace}-\eqref{eq:Ix4trace}.
Let us now evaluate the angle integrals.
We define $\theta$ and $\phi$ as the polar and azimuthal angles when the polar axis is along $\hbk$.
First, we compute
\begin{equation}
\begin{split}
 I (x)
 = \frac{x}{2} \int_{-1}^1 \frac{dy}{x-y} 
  & = \frac{x}{2} \ln \frac{x+1}{x-1} 
   = \frac{x}{2} 
 	\ln \biggl|\frac{x+1}{x-1}\biggr| 
   	 - x\frac{i\pi}{2} \theta(1-|x|) 
\end{split}
\end{equation}
with $y=\cos\theta$.
In order to evaluate the other integrals, we prepare the following formulas:
\begin{equation}
\begin{split}
 \int_0^{2\pi} \frac{d\phi}{2\pi}\hp^k
 & = \hk^k y \,, \\
 \int_0^{2\pi} \frac{d\phi}{2\pi} \hp^j \hp^k
 & = \hk^j \hk^k y^2
 	 + \tilde{\Delta}^{jk} \frac{1}{2}(1-y^2) \,, \\
 \int_0^{2\pi} \frac{d\phi}{2\pi} \hp^i \hp^j \hp^k 
 & = \hk^i \hk^j \hk^k y^3
 	 + \Bigl(
 	 		\hk^i \tilde{\Delta}^{jk} 
 	 		+ \hk^j \tilde{\Delta}^{ki}
 	 		+ \hk^k \tilde{\Delta}^{ij}
 	   \Bigr)
 	   	\frac{1}{2} y(1-y^2) \,, \\
 \int_0^{2\pi} \frac{d\phi}{2\pi} \hp^i \hp^j \hp^k \hp^l
 & = \hk^i \hk^j \hk^k \hk^l y^4\\
 & \quad
 	 + \Bigl(
 			\tilde{\Delta}^{ij} \hk^k \hk^l
 			+ \tilde{\Delta}^{jk} \hk^l \hk^i
 			+ \tilde{\Delta}^{kl} \hk^i \hk^j 
 			+ \tilde{\Delta}^{il} \hk^j \hk^k
 			+ \tilde{\Delta}^{ik} \hk^j \hk^l
 			+ \tilde{\Delta}^{jl} \hk^i \hk^k
 		\Bigr) \\
 & \qquad \times
 			\frac{1}{2} y^2(1-y^2) \\
 & \quad
	 + \Bigl(
 			\tilde{\Delta}^{ij} \tilde{\Delta}^{kl}
 			+ \tilde{\Delta}^{ik} \tilde{\Delta}^{jl}
 			+ \tilde{\Delta}^{il} \tilde{\Delta}^{jk}
 		\Bigr) 
	 	\frac{1}{8} (1-y^2)^2  \,,
\end{split}
\end{equation}
where we introduce $\tilde{\Delta}_{ij} = \delta_{ij} - \hk_i \hk_j$ (we note $\Delta_{\mu\nu} = \xi_\mu\xi_\nu - \eta_{\mu\nu}$).
We prepare the following integrals:
\begin{equation}
  \dis
 \frac{x}{2} \int_{-1}^1 dy \frac{y^n}{x-y}
 =
 \begin{cases}
  \dis \,x(I-1) \quad &(n=1) \vspace{1em}\\
  \dis \,x^2(I-1) \quad &(n=2) \vspace{1em}\\
  \dis - \frac{x}{3} + x^3(I-1) \quad &(n=3) \vspace{0.5em}\\
  \dis - \frac{x^2}{3} + x^4(I-1) \quad &(n=4)
 \end{cases}\quad .
\end{equation}
These yield  
\begin{eqnarray}
\label{eq:Ix1}
I^j (x)
 & = & \hk^j x (I-1)  \,, \\
\label{eq:Ix2}
I^{jk}(x)
 & = & \hk^j\hk^k x^2 (I-1)
 	 + \frac{1}{2} \tilde{\Delta}^{jk} 
 	 	\biggl(
 	 		I
 	 		- x^2(I-1)
 	 	\biggr) \,,\\
\label{eq:Ix3}
I^{ijk}(x)
 & = & \hk^i\hk^j\hk^k 
 		\Bigl(
 			- \frac{x}{3}
 			+ x^3(I-1)
 		\Bigr) \notag \\
 & &
 	 + \frac{1}{2}
 			\Bigl(
 	 			\hk^i \tilde{\Delta}^{jk} 
 	 			+ \hk^j \tilde{\Delta}^{ki}
 	 			+ \hk^k \tilde{\Delta}^{ij}
 	   		\Bigr)
 	   		\Bigl(
 	   			x (I-1)
 	   			+ \frac{x}{3}
 	   			- x^3(I-1) 
 	   		\Bigr) \,, \\
\label{eq:Ix4}
 I^{ijkl}(x)
 & = & \biggl(
 		-\frac{x^2}{3} 
 		+ x^4(I-1) 
 	 \biggr) \hk^i \hk^j \hk^k \hk^l \notag \\
 & &
 	 + \frac{1}{2} 
 	 	\biggl(
 	 		x^2(I-1)
 	 		+ \frac{x^2}{3} 
 			- x^4(I-1) 
 	 	\biggr) \notag \\
 & &\qquad \times
 	 	\Bigl(
 			\tilde{\Delta}^{ij} \hk^k \hk^l
 			+ \tilde{\Delta}^{jk} \hk^l \hk^i
 			+ \tilde{\Delta}^{kl} \hk^i \hk^j 
 			+ \tilde{\Delta}^{il} \hk^j \hk^k
 			+ \tilde{\Delta}^{ik} \hk^j \hk^l
 			+ \tilde{\Delta}^{jl} \hk^i \hk^k
 		\Bigr)  \notag \\
 & &
	 + \frac{1}{8} 
	 	\biggl(
	 		I
	 		- 2 x^2(I-1)
	 		-\frac{x^2}{3} 
 			+ x^4(I-1) 
	 	\biggr)
	 	\Bigl(
 			\tilde{\Delta}^{ij} \tilde{\Delta}^{kl}
 			+ \tilde{\Delta}^{jk} \tilde{\Delta}^{li}
 			+ \tilde{\Delta}^{il} \tilde{\Delta}^{jk}
 		\Bigr) \,.\quad
\end{eqnarray}
In particular, the asymptotic forms of $I_{j_1\cdots j_n} $ in the dynamical limit $x\gg 1$ are
\begin{eqnarray}
\label{eq:Ix0_largex}
 I 
 & \simeq  &
  1 + \frac{1}{3x^2} + \frac{1}{5x^4} + \frac{1}{7x^6} + O(x^{-8}) \,,  \\
\label{eq:Ix1_largex}
 I^j  
  & \simeq  &
   \frac{\hk^j}{3x} + O(x^{-3}) \,, \\
\label{eq:Ix2_largex}
 I^{jk} 
 & \simeq &
  \biggl(\frac{1}{3} + \frac{1}{15x^2}\biggr) \delta^{jk}
  		+ \frac{2}{15x^2}\hk^j\hk^k + O(x^{-4}) \,, \\
\label{eq:Ix3_largex}
 I^{ijk} 
  & \simeq &
 		\frac{1}{15x}
 			\Bigl(
 	 			\hk^i \delta^{jk} 
 	 			+ \hk^j \delta^{ki}
 	 			+ \hk^k \delta^{ij}
 	   		\Bigr) 
 	  + O(x^{-3}) \,, \\
\label{eq:Ix4_largex}
 I^{ijkl} 
  & \simeq &
		\biggl(
	 		\frac{1}{15}
	 		+ \frac{1}{105x^2}
	 	\biggr)
	 	\Bigl(
 			\delta^{ij} \delta^{kl}
 			+ \delta^{ik} \delta^{il}
 			+ \delta^{il} \delta^{jk}
 		\Bigr) \notag \\
  & &
 		+ \frac{2}{105x^2}
 	 	\Bigl(
 			\delta^{ij} \hk^k \hk^l
 			+ \delta^{jk} \hk^l \hk^i
 			+ \delta^{kl} \hk^i \hk^j 
 			+ \delta^{il} \hk^j \hk^k
 			+ \delta^{ik} \hk^j \hk^l
 			+ \delta^{jl} \hk^i \hk^k
 		\Bigr) 
  		+ O(x^{-4}) .\qquad\quad 
\end{eqnarray}
On the other hand, in the static limit $x \ll 1$, we find $I_{j_1\cdots j_n} \simeq O(x)$.

%%--- Bibliography ---%
\bibliographystyle{JHEP}
\bibliography{bibfile}
%--- Bibliography ---%

\end{document}